\documentclass[
tightenlines,
superscriptaddress,
 preprint,
 amsmath,amssymb,
 aps,
]{revtex4-1}

\usepackage{natbib}
\usepackage{graphicx}
\usepackage{amsmath,amssymb,amsthm}
\usepackage{xcolor}

\begin{document}
\title{Generalized Master Equation Approach to Time-Dependent Many-Body Transport}

\author{V. Moldoveanu}
\affiliation{National Institute of Materials Physics, Atomistilor 405A, Magurele 077125, Romania}

\author{A. Manolescu}
\affiliation{School of Science and Engineering, Reykjavik University, Menntavegur 1, IS-101 Reykjavik, Iceland}

\author{V. Gudmundsson}
\affiliation{Science Institute, University of Iceland, Dunhaga 3, IS-107 Reykjavik, Iceland}


\begin{abstract}
We  recall theoretical studies on transient transport through interacting mesoscopic systems.
It~is shown that a generalized master equation (GME) written and solved in terms of many-body states
provides the suitable formal framework to capture both the effects of the Coulomb interaction
and electron--photon coupling due to a surrounding single-mode cavity. We outline the derivation of this
equation within the Nakajima--Zwanzig formalism and point out technical problems related to its numerical
implementation for more realistic systems which can neither be described by non-interacting two-level
models nor by a steady-state Markov--Lindblad equation. We first solve the GME for a lattice model and
discuss the dynamics of many-body states in a two-dimensional nanowire, the dynamical onset of the current-current
correlations in electrostatically coupled parallel quantum dots and transient thermoelectric properties.
Secondly, we rely on a continuous model to get the Rabi oscillations of the photocurrent through a double-dot etched
in a nanowire and embedded in a quantum cavity. A~many-body Markovian version of the GME for cavity-coupled systems
is also presented.
\end{abstract}

\maketitle

\section{Introduction}

 Few-level open systems stand as everyday `lab rats' in corner stone experiments and future technologies in 
nanoelectronics \cite{di-ventra_2008} and quantum optics \cite{jahnke2016quantum}. Generically, they are electronic 
systems with a discrete spectrum 
(e.g., artificial atoms \cite{CHOW2013109}, nanowires or superconducting qubits \cite{GU20171}) 
connected to particle reservoirs or 
embedded in bosonic baths. Depending on the nature of the environment (i.e., fermionic or bosonic) to which the open systems are coupled, their theoretical 
investigation started with two toy-models, namely the single-level Hamiltonian of quantum transport 
and the Jaynes--Cummings (JC) Hamiltonian of a two-level system (TLS). 

Surprisingly or not, studying the sequential tunneling transport regime or the optical properties of quantum emitters eventually 
boils down to solve formally similar Markovian master equations (MEs) for the so called reduced density operator (RDO). The latter defines 
the non-unitary evolution of the small system in the presence of the infinite degrees of freedom of the reservoirs. 
Such MEs are derived by tracing out the reservoir's degrees of freedom and are known from the early days of condensed matter and 
quantum optics (see the seminal works of Bloch \cite{Bloch}, Wangsness \cite{Wangsness} and Redfield~\cite{Redfield}). 
The~master equation cleverly bypasses the fact that the Liouville--von-Neumann (LvN) equation of the coupled systems 
(i.e., the open system and the reservoirs) is impossible to solve and takes advantage of the fact that all observables associated 
to the small and open system can be calculated as statistical averages w.r.t. the RDO.  

Indeed, the RDO associated to the Jaynes--Cummings model has been a central object in quantum optics 
\cite{Scully,Carmichael} (e.g., in the study of lasing and for the calculation of photon correlation functions). 
In this context the master equation (ME) approach goes as follows: (i) one studies an atomic few-level 
system whose eigenvalues and eigenfunctions are supposed to be known; (ii) the dissipation in the system (e.g., cavity losses 
or various non-radiative recombination processes) is included through the so called `jump' operators; (iii) the occupation of 
atomic levels changes due to photon emission or absorption, but the particle number is conserved as the system is not coupled 
to particle reservoirs; (iv) under the Markov approximation the ME acquires a Lindblad form, usually solved in the steady-state regime. 

The above scenario changes when one aims to derive a quantum master equation describing transport phenomena. 
(i) The Coulomb interaction effects on the spectrum and eigenstates of the system cannot be always neglected, especially
for confined systems like quantum dots or nanowires; this requires a many-body derivation of the master equation;
(ii) The tunneling between source/drain probes prevents the charge conservation in the central system and the main 
quantity of interest is the electronic current; (iii) Finally, the steady-state regime does not cover the whole physics 
 and cannot even be guaranteed in general; moreover, the validity of the 
rotating-wave (RWA) and Markov approximations must be established more carefully \cite{Timm,Elenewski}. 
In fact it turns out that when applied to transport processes the master equation must rather be solved in 
its non-Markovian version. 

Such generalized master equations which take into account the memory effects have been mostly derived and 
implemented for time-dependent transport in non-interacting \cite{Moldoveanu09:073019,PhysRevB.81.085315} and interacting 
\cite{Moldoveanu10:155442} quantum dots, nanowires, and rings \cite{PhysRevB.71.235302}. It turns out that the generalized master equation (GME) method is 
a valuable tool for modeling and monitoring the dynamics of specific many-body states as well as for investigating time-dependent 
propagation  along a sample \cite{PhysRevB.85.245114} or capturing charge sensing effects \cite{PhysRevB.82.085311} 
and counting statistics in electrostatically parallel QDs \cite{PhysRevB.84.165114}. 
In particular, Harbola et al. \cite{Harbola} showed that a Lindblad form of the quantum master equation 
is still recovered in the high bias limit and by assuming the RWA. 

{Since then, a lot of theoretical work has been done to improve and refine the quantum master equation formalism.
A formally exact memory-kernel for the Anderson model was derived and calculated using real-time path
integral Monte Carlo methods \cite{PhysRevB.84.075150}. A hierarchical quantum master equation approach with good 
convergence at not too low  temperature was put forward by H\"{a}rtle~{et~al.}~\cite{PhysRevB.88.235426}. 
As for molecular transport calculations one can rely on the GME written in terms of the many-body states of the 
isolated molecule \cite{PhysRevB.79.205303,PhysRevB.78.125320}. A recent review on non-Markovian effects in open 
systems is also available \cite{RevModPhys.88.021002}.}

As we shall see below the implementation of GME approach to many-level systems with specific geometries poses considerable 
technical difficulties. These are related to the many-body structure of the central interacting system, to the accurate 
description of the contact regions and, more importantly, to the evaluation of the non-Markovian kernels which become 
complicated objects once we go beyond non-interacting single-level models.

A second useful extension of the ME method emerged in the context of cavity quantum electrodynamics. Here the system under study 
is a hybrid one, as the electronic system is still coupled to source/drain reservoirs (i.e., leads) but also interacts with a quantum 
cavity mode, the latter being subjected to dissipation into leaky modes described by a bosonic bath.  
Such systems are currently used in state-of-the-art measurements in cavity quantum electrodynamics 
\cite{Kulkarni,Viennot,Liu,Viennot408,PhysRevApplied.9.014030}. 
Again, the many-body nature of the problem is essential, as the electron-photon coupling leads to the formation of dressed states 
whose dynamics in the presence of both particle and dissipative bosonic reservoirs is far from being trivial. 
The relevant reduced density operator now acts in the many-body electron-photon Fock space and describes the dynamics of 
dressed-states. This fact brings new technical difficulties in the derivation \cite{PhysRevA.84.043832,PhysRevB.97.195442} 
and implementation of ME \cite{Gudmundsson12:1109.4728,doi:10.1021/acsphotonics.5b00115}. 
Let us also mention here recent studies on ground state electroluminescence \cite{PhysRevLett.116.113601,PhysRevLett.122.190403} 
and on cavity enhanced transport of charge \cite{Genes}.

In view of the abovementioned comments, the aim of this work is: (i) to briefly review the development of the generalized master equation
approach to time-dependent many-body transport in the presence of both fermionic and bosonic environments and (ii) to illustrate 
in a unified framework how the method really works, from formal technicalities to numerical implementation. In Section~\ref{S2} we shall 
therefore derive a non-Markovian master equation which describes the dynamics and the transport properties of rather general 
`hybrid' system consisting in an electronic component $S_1$ which is connected to particle reservoirs (i.e., leads) and a 
second subsystem $S_2$. The latter, although not coupled to particle reservoirs, interacts with system $S_1$ or with some leaky modes 
described as bosonic baths. Then we specialize this master equation to several systems of interest. More precisely, in~Section~\ref{S3} 
we recall GME results on transient charging of excited states and Coulomb-coupled quantum dots. Section~\ref{S4} deals with 
thermoelectric transport. Applications to transport in cavity quantum electrodynamics are collected 
in Sections \ref{S5} and \ref{S6}. We conclude in Section \ref{S7}.

\section{Formalism\label{S2}}
\subsection{Generalized Master Equation for Hybrid Systems}
Non-Markovian master equations for open systems have been derived in many recent textbooks or review papers via projection methods
(e.g., Nakajima--Zwanzig formalism or time-convolutionless approach \cite{Petruccione}). Nonetheless it is still instructive to 
outline here some theoretical and computational difficulties one encounters when {solving} transport master equation for 
interacting many-level systems. 


From the formal point of view the projection technique is quite general and the derivation of a master equation for the
RDO does not depend on a specific model (i.e., on the geometry and spectrum of the central system or on the correlation functions
 of fermionic/bosonic reservoirs). In general, as~long as one can write down a system-reservoir coupling Hamiltonian $H_{SR}$ 
a master equation can be~derived. 

For the sake of generality we shall consider a hybrid system $S$ made of an electronic structure $S_1$ which is 
coupled to $n_r$ particle reservoirs characterized by chemical potentials and temperatures $\{\mu_l,T_l\}$, $l=1,2,...,n_r$,
 and a second subsystem $S_2$ (i.e., a localized impurity, or an oscillator, or~a~single-mode quantum cavity). The 
subsystem $S_2$ can only be coupled to thermal or photonic baths which are described as a collection of oscillators with 
frequencies $\{\omega_k\}$. Let ${\cal F}_{S_1}$ and ${\cal F}_{S_2}$ be the Fock spaces associated to the two systems. 
Typically ${\cal F}_{S_1}$ is a set of interacting many-body configurations of the electronic system whereas 
${\cal F}_{S_2}$ is made by harmonic oscillator Fock states.

 The dynamics of the open system $S_1$ and of nearby `detector' system $S_2$ are intertwined by a coupling $V$. Under a voltage 
bias or a temperature difference the system $S_1$ carries an electronic or a heat current which need to be calculated in the presence
of the second subsystem. Conversely,~the~averaged observables of $S_2$ (e.q., mean photon number or the spin of a localized impurity) 
will also depend on the transport properties of $S_1$. The Hamiltonian of the hybrid structure is:
\begin{equation}\label{H-system}
 H_S=H_{S_1}+H_{S_2}+V.
\end{equation}

In this work $H_{S_1}$ will describe various Coulomb-interacting structures: a single quantum dot, a~2D wire or parallel 
quantum dots. We shall denote by $|\nu\rangle$ and $E_{\nu}$ the many-body configurations and eigenvalues of $H_{S_1}$, that is one has 
$H_{S_1}|\nu\rangle=E_{\nu}|\nu\rangle$. $H_{S_1}$ can be equally expressed in terms of creation 
and annihilation operators $\{ c^{\dagger}_{n\sigma}, c_{n\sigma}\}$ associated to a spin-dependent single-particle basis 
$\{\psi_{n\sigma}\}$ of a~single-particle Hamiltonian $h_{S_1}^{(0)}$ (see the next sections for specific models), 
such that: 
\begin{equation}\label{H_S1}
H_{S_1}=H_{S_1}^{(0)}+W,
\end{equation} 
where $H_{S_1}^{(0)}$ is the 2nd quantized form of $h_{S_1}^{(0)}$ and $W$ is the Coulomb interaction. Similarly, the eigenstates 
and eigenvalues of the second subsystem $S_2$ will be denoted by $|j\rangle$ and $e_j$ such that $H_{S_2}|j\rangle=e_j|j\rangle$. 
As for the coupling $V$ one can mention at least three examples: The exchange interaction between a quantum dot and a 
localized magnetic impurity with total spin $S$, the electron--photon coupling in a quantum-dot cavity and the electron--vibron 
coupling in nanoelectromechanical systems \cite{POOT2012273,doi:10.1002/pssb.201800443}. 

The total Hamiltonian of the system coupled to particle and/or bosonic reservoirs $R$ reads as:
\begin{equation}\label{total-H}
H(t)=H_S+H_R+H_{SR}(t):=H_0+H_{SR}(t),
\end{equation}
where the system-reservoir coupling $H_{SR}$ collects the coupling to the leads ($H_T$) and the coupling of a
bosonic mode to a thermal or leaky bosonic environments ($H_E$):
\begin{equation}\label{H_SR}
H_{SR}(t)=H_T(t)+H_E.
\end{equation}

Note that the interaction with the bosonic environment $H_E$ does not depend on time. The~lead-sample tunneling term $H_T$ 
carries a time-dependence that will be explained below. The~Hamiltonian of the reservoirs,
\begin{equation}\label{H_R}
H_R=H_{{\rm leads}}+H_{{\rm bath}}
\end{equation}
describes at least two semiinfinite leads (left-L and right-R) but could also contain a bosonic or a~thermal bath.


This general scheme allows one to recover several relevant settings. If $S_1$ describes an optically active structure and
$S_2$ defines a photonic mode then $V$ could become either the Rabi or the Jaynes--Cummings electron--photon coupling. 
The absence of the particle reservoirs simplifies $H_S$ to well known models in quantum optics, while by adding them one can 
study photon-assisted transport effects (e.g., Rabi oscillation of the photocurrents or electroluminescence). Also, by removing  
$S_2$, $V$~and the bosonic dissipation one finds the usual transport setting for a Coulomb interacting purely electronic structure. 

Let $\varepsilon^{l}(q)$ and $\psi^l_{q\sigma}$ be the single particle energies and wave functions of the $l$-th lead. 
For simplicity we assume that the states on the leads are spin-degenerate so their energy levels do not depend on the spin index. 
Using the creation/annihilation operators $c^{\dagger}_{ql\sigma}$/$c_{ql\sigma}$ associated to the single particle states, 
we can write:
\begin{equation}\label{Hleads}
H_{{\rm leads}}=\sum_l H_{l}=\int dq\sum_{\sigma}\varepsilon^{l}(q)c^{\dagger}_{ql\sigma}c_{ql\sigma} \,.
\end{equation}

As for the bosonic bath, it is described by a collection of harmonic oscillators with frequencies $\omega_k$ and by 
corresponding creation/annihilation operators $b_k^{\dagger}/b_k$:
\begin{equation}\label{Hbath}
H_{{\rm bath}}=\sum_k\hbar\omega_kb_k^{\dagger}b_k.
\end{equation}

The tunneling Hamiltonian has the usual form:
\begin{equation}\label{Htunnel}
H_T(t)=\sum_{l}\sum_{n\sigma}\int dq \chi_l(t)(T^l_{qn}c^{\dagger}_{ql\sigma}c_{n\sigma}+h.c),
\end{equation}
where we considered without loss of generality that the tunneling processes are spin conserving. 
For~the simplicity of writing the spin degree of freedom $\sigma$ will be henceforth tacitly merged with 
the single-particle index $n$ and restored if needed. 

The time-dependent switching functions $\chi_l(t)$ control the time-dependence of the contacts between the 
leads and the sample; these functions mimic the presence of a time dependent potential barrier. We~emphasize that in 
most studies based on ME method the coupling to the leads is suddenly switched at some initial instant $t_0$ 
such that for each lead $\chi_l(t)=\theta(t-t_0)$ where $\theta(x)$ is the Heaviside step function. This choice 
is very convenient if one imposes the Markov approximation in view of a time-local Master equation. Here we allow 
for more general switching functions: (i) a smooth coupling to the leads or (ii)~time-dependent signals applied at the 
contacts to the leads. In particular, if the potential barriers oscillate out of phase
 the system operates like a turnstile pump under a finite constant bias. 

The coupling $T^l_{qn}$ describes the tunneling strength between a state with momentum $q$ of the lead $l$ and the state $n$ 
of the isolated sample with wavefunctions $\psi_n$. In the next sections we shall show that these matrix elements have to 
be calculated for each specific model by taking into account the geometry of the system and of the leads.  

The associated density operator ${\cal W}$ of the open system obeys the Liouville--von Neumann~equation:
\begin{equation}\label{LvN}
i\hbar\frac{\partial{{\cal W}(t)}}{\partial t}={\cal L}(t){\cal W}(t),\quad {\cal W}(t_0)=\rho_S(t_0)\otimes\rho_R,
\end{equation}
where:
\begin{equation}
{\cal L}(t)={\cal L}_0+{\cal L}_{SR},\quad {\cal L}_0\cdot=[H_0,\cdot].
\end{equation}

We also introduce the notations:
\begin{equation}
 {\cal L}_S\cdot=[H_S,\cdot],\quad  {\cal L}_{SR}\cdot=[H_{SR},\cdot].
\end{equation}

The Nakajima--Zwanzig projection formalism leads to an equation of motion for the reduced density operator 
$\rho(t)={\rm Tr}_R\{{\cal W}\}$. The initial state ${\cal W}_0:={\cal W}(t_0)$ factorizes as: 
\begin{equation}\label{rho-initial}
{\cal W}_0=\rho_0\otimes\rho_{{\rm leads}}\otimes\rho_{{\rm bath}}:=\rho_0\otimes\rho_R, 
\end{equation}
where the equilibrium density operator of the leads reads: 
\begin{equation}\label{rho_L}
\rho_{{\rm leads}}=\prod_l\frac{e^{-\beta_l (H_l-\mu_l N_l)}}{{\rm Tr}_l \{e^{-\beta_l(H_l-\mu_l N_l)}\}},
\end{equation}
and $\beta_l=1/k_BT_l$, $\mu_l$ and $N_l$ denote the inverse temperature, chemical potential and the 
occupation number operator of the lead $l$. Similarly,
\begin{equation}\label{rho_B}
\rho_{{\rm bath}}=\prod_k e^{-\hbar\omega_kb_k^{\dagger}b_k/k_BT}(1-e^{-\hbar\omega_k/k_BT}).
\end{equation}

Finally, $\rho_0$ is simply a projection on one of the states of the hybrid system, and as such its calculation must 
take into account the effect of the hybrid coupling $V$ (see the discussion in Section~\ref{S2.2}).
We now define two projections:
\begin{equation}
P\cdot=\rho_R{\rm Tr}_R\{\cdot \},\quad Q=1-P.
\end{equation}

It is straightforward to check the following properties:
\begin{eqnarray}
P{\cal L}_S&=&{\cal L}_SP,\quad P{\cal L}_{SR}(t)P=0.
\end{eqnarray}

The Liouville Equation (\ref{LvN}) splits then into two equations:
\begin{eqnarray}\label{PW}
i\hbar P\frac{\partial W(t)}{\partial t}&=&P{\cal L}(t)PW(t)+P{\cal L}(t)QW(t)\\
i\hbar Q\frac{\partial W(t)}{\partial t}&=&Q{\cal L}(t)QW(t)+Q{\cal L}(t)PW(t),
\end{eqnarray}
and the second equation can be solved by iterations ($T$ being the time-ordering operator):
\begin{equation}\label{QW}
QW(t)=\frac{1}{i\hbar}\int_{t_0}^tdsT\exp \left\lbrace
-\frac{i}{\hbar}\int_s^tds'Q{\cal L}(s') \right\rbrace
Q{\cal L}(s)PW(s).
\end{equation}

Inserting Equation (\ref{QW}) in Equation (\ref{PW}) and using the properties of $P$ we get the Nakajima--Zwanzig equation:
\begin{eqnarray}\nonumber
i\hbar P\frac{\partial W(t)}{\partial t}&=&P{\cal L}_SW(t)\\\label{NZ-eq1}
&+&\frac{1}{i\hbar}P{\cal L}_{SR}(t)Q\int_{t_0}^tds
T\exp \left\lbrace
-\frac{i}{\hbar}\int_s^tds'Q{\cal L}(s')Q \right\rbrace Q{\cal L}_{SR}(s)PW(s).
\end{eqnarray}

In order to have an explicit perturbative expansion in powers of $H_{SR}(t)$ one has to factorize 
the time-ordered exponential as follows:
\begin{equation}
T\exp \left\lbrace
-\frac{i}{\hbar}\int_s^tds'Q{\cal L}(s')Q \right\rbrace =\exp \{Q{\cal L}_0Q\}
(1+{\cal R}),
\end{equation}
where the remainder ${\cal R}$ contains infinitely deep commutators with inconveniently embedded projection operators. 
Usually one considers a truncated version of the Nakajima--Zwanzig equation up to the second order contribution w.r.t. 
the system-reservoir $H_{SR}$:
\begin{equation}\label{NZ-eq2}
i\hbar {\dot\rho}(t)={\cal L}_S\rho(t)+\frac{1}{i\hbar}{\rm Tr}_R\left\lbrace {\cal L}_{SR}\int_{t_0}^t\,ds e^{-i(t-s){\cal L}_0}
{\cal L}_{SR}(s)\rho_R\rho(s)\right\rbrace.
\end{equation}

Now, by taking into account that for any operator $A$ acting on the Fock space of the hybrid system 
$e^{-i{t\cal L}_0}A=e^{-itH_0}Ae^{itH_0}$ and denoting by $U_0(t,s)=e^{-i(t-s)H_0}$ the unitary
evolution of the disconnected systems we arrive at the well known form of the GME:
\begin{equation}
\begin{array}{lll}
\nonumber
   i\hbar{\dot\rho}(t)&=&[H_S,\rho(t)]-\frac{i}{\hbar}U_0^{\dagger}(t,t_0){\rm Tr}_R\left\{\int_{t_0}^{t}\, ds
\left[\tilde{H}_{SR}(t),\left[\tilde{H}_{SR}(s),\tilde{\rho}(s)\rho_R\right]\right]\right\}U_0(t,t_0)\\\label{GME1}
&=&[H_S,\rho(t)]-\frac{i}{\hbar}U_S^{\dagger}(t,t_0){\rm Tr}_R\left\{\int_{t_0}^{t}\, ds 
\left[\tilde{H}_{SR}(t),\left[\tilde{H}_{SR}(s),\tilde{\rho}(s)\rho_R\right]\right]\right\}U_S(t,t_0),
\end{array}
\end{equation}

where in order to get to the last line we removed the evolution operators of the environment from both sides of the trace. 
At the next step one observes that when performing the trace over the reservoirs and environment degrees of freedom 
the mixed terms in the double commutator vanish because each of the coupling terms $H_T$ and $H_E$ carries only 
one creation or annihilation operator for the corresponding reservoir such that:
\begin{equation}
{\rm Tr}_R\{\tilde{c}^{\dagger}_{ql}(t)\tilde{b}_k(s)\rho_R\}={\rm Tr}_{{\rm leads}}\{\tilde{c}^{\dagger}_{ql}(t)\rho_{{\rm leads}}\}\cdot
{\rm Tr}_{{\rm bath}}\{\tilde{b}_k(s)\rho_{{\rm bath}}\}=0.
\end{equation} 

Moreover, the time evolution of each term can be simplified due to the commutation relations 
$[H_{{\rm bath}},H_T]=[H_{{\rm leads}},H_E]=0$:
\begin{eqnarray}\label{H_T-tilde}
{\tilde H_T}(t)&=&e^{\frac{i}{\hbar}t H_S}e^{\frac{i}{\hbar}t H_{{\rm leads}}}H_T
e^{-\frac{i}{\hbar}t H_S}e^{-\frac{i}{\hbar}t H_{{\rm leads}}},\\\label{H_E-tilde}
{\tilde H_E}(t)&=&e^{\frac{i}{\hbar}t H_S}e^{\frac{i}{\hbar}t H_{{\rm bath}}}H_E
e^{-\frac{i}{\hbar}t H_S}e^{-\frac{i}{\hbar}t H_{{\rm bath}}}.
\end{eqnarray} 

The GME then reads as:   
\begin{eqnarray}\nonumber
{\dot\rho}(t)&=&-\frac{i}{\hbar}[H_S,\rho(t)]-\frac{1}{\hbar^2}U_S^{\dagger}(t,t_0){\rm Tr}_{{\rm leads}}\left\{\int_{t_0}^{t}\, ds 
\left[\tilde{H}_T(t),\left[\tilde{H}_T(s),\tilde{\rho}(s)\rho_{{\rm leads}}\right]\right]\right\}U_S(t,t_0)\\\label{GME2}
&-&\frac{1}{\hbar^2}U_S^{\dagger}(t,t_0){\rm Tr}_{{\rm bath}}\left\{\int_{t_0}^{t}\, ds 
\left[\tilde{H}_E(t),\left[\tilde{H}_E(s),\tilde{\rho}(s)\rho_{{\rm bath}}\right]\right]\right\}U_S(t,t_0)\\\label{GME3}
&:=&-\frac{i}{\hbar}[H_S,\rho(t)]-{\cal D}_{{\rm leads}}[\rho,t]-{\cal D}_{{\rm bath}}[\rho,t].
\end{eqnarray}

Equation (\ref{GME3}) is the generalized master equation for our hybrid system. It provides the reduced density operator 
$\rho$ in the presence of particle and bosonic reservoirs and also takes into account the memory effects and
the non-trivial role of time-dependent signals applied at the contact regions through the switching functions $\chi_l$. 
The third term in Equation (\ref{GME3}) is needed only if $H_{S_2}$ describes a quantized optical or mechanical oscillation mode. 
{In our work on open QD-cavity systems we always assume that the coupling between the cavity photons and other 
leaky modes is much smaller that the electron-photon coupling $g_{EM}$ (see Section \ref{S5}). On the other hand, 
for our calculations $g_{EM}\ll \hbar\omega$, $\omega$ being the frequency of the cavity mode. Then the RWA holds and 
${\cal D}_{{\rm bath}}[\rho,t]$ can be cast in a Lindblad form. Let us stress that in the ultrastrong coupling regime on 
typically has $g_{EM}/\hbar\omega>0.2$ and the derivation of the dissipative term is more complicated and involves the 
dressed states of the QD-cavity system \cite{PhysRevA.84.043832}. In order to describe dissipative effects 
in the ultrastrong coupling regime beyond the RWA one needs more elaborate techniques \cite{PhysRevA.98.053834,PhysRevA.99.013807}.}

For further calculations one has to solve the GME as a system of coupled integro-differential equations for 
the matrix elements of the RDO with respect to a {suitable basis} in the Fock space ${\cal F}_S={\cal F}_{S_1}\otimes {\cal F}_{S_2}$. 
We discuss this issue in the next subsection.

\subsection{`Hybrid' States and Diagonalization Procedure\label{S2.2}}

The starting point in solving the GME is to write down the matrix elements of the system-environment operators 
$H_T$ and $H_E$ w.r.t. the `disjointed' basis formed by the eigenstates of $H_{S_1}$ and $H_{S_2}$, that is 
$|\nu ,j\rangle :=|\nu\rangle\otimes |j\rangle$. However this strategy does not help much when evaluating the time 
evolution (see Equations (\ref{H_T-tilde}) and (\ref{H_E-tilde})) as $H_S$ is not diagonal w.r.t. to $|\nu ,j\rangle$ such 
that one cannot easily write down the matrix elements of the unitary evolution $U_S(t,t_0)$.  
In fact we are forced to solve the GME by using the eigenstates $|\varphi_p)$ and eigenvalues ${\cal E}_p$ 
of the Hamiltonian $H_S$. The former are written as:
\begin{equation}\label{phi_p}
|\varphi_p)=\sum_{\nu,j}{\cal V}^{(p)}_{\nu j}|\nu,j\rangle.
\end{equation}

Here the round bracket notation $|\varphi_p)$ is meant to underline that the state $\varphi_p$ describes the 
interacting system $S$, in the sense that both Coulomb interactions and the coupling to the bosonic modes 
were taken into account when diagonalizing $H_S$. This notation also prevents any confusion if the `free' states 
$|\nu,j\rangle $ were also labeled by a single index $p'$. In that case the above equation is 
conveniently rewritten as $|\varphi_p)=\sum_{p'}{\cal V}^{(p)}_{p'}|p'\rangle$. Note that $p$ is usually a multiindex carrying 
information on relevant quantum numbers. In most cases of interest the coupling $V$ between the two systems leads to a strong mixing of 
the unperturbed basis elements $|\nu,j\rangle$ and is not necessarily small. Therefore we shall not follow a perturbative approach but 
rather calculate ${\cal E}_p$  and the weights ${\cal V}^{(p)}_{\nu j}$ by numerically diagonalizing $H_S$ on a relevant subspace of 
`disjointed' states. 

Prior to any model specific calculations or numerical implementations it is useful to comment a bit on the two dissipative contributions
in Equation (\ref{GME3}). It is clear that the evolution operator $U_S$ describes the joint systems $S_1$ and $S_2$ and therefore the 
hybrid interaction cannot be simply neglected neither in ${\cal D}_{{\rm leads}}$ nor in ${\cal D}_{{\rm bath}}$; in fact 
one can easily check that $V$ does not commute with $H_E$ or $H_T$. Moreover, as has been clearly pointed out by Beaudoin {et al.} 
\cite{PhysRevA.84.043832}, by disregarding the qubit-resonator interaction when calculating ${\cal D}_{{\rm bath}}$ one ends up with 
unphysical results. In what concerns the role of $V$ in the leads' contribution, a recent work emphasized that for QD-cavity systems 
the corresponding master equation must be derived in the basis of dressed-states \cite{PhysRevB.97.195442}.

The diagonalization of $H_S$ poses serious technical problems because both spaces ${\cal F}_{S_1}$ and 
${\cal F}_{S_2}$ are in principle infinite dimensional. Besides that, the Coulomb interaction 
in $H_{S_1}$ prevents one to derive the interacting many-body configurations $\{|\nu\rangle\}$ analytically. 
We now propose a step-by-step diagonalization procedure leading to a relevant set of 
interacting states of the full Hamiltonian. The~procedure requires several `intermediate' diagonalization operations: 

($D_1$) Analytical or numerical calculation of the single-particle states of the Hamiltonian ${\hat h}_{S_1}^{(0)}$ which 
describes the non-interacting electronic system $S_1$. As we shall see in the next sections, this step may not be trivial 
if the geometry of the sample is taken seriously into account. Let us select a subset of $N_{{\rm ses}}$ single-particle states 
$\{\psi_1, \psi_2,...,\psi_{N_{{\rm ses}}}\}$ (if needed this set of states includes the spin degree of freedom). 
 Typically we choose the lowest energy single-particle states
but in some cases \cite{Moldoveanu09:073019} it is more appropriate to select the subset of states which effectively contribute to the
transport (i.e., states located within the bias window).

($D_2$) The construction of a second set of $N_{{\rm mes}}$ non-interacting many-body configurations (NMBS) 
$\{|\lambda\rangle\}_{\lambda =1,..,N_{{\rm mes}}}$ from the $N_{{\rm ses}}$ single-particle states introduced above. 
Note that for computational reasons we have to keep $N_{{\rm mes}}<2^{N_{{\rm ses}}}$ for larger $N_{{\rm ses}}$. 
Then, if $i_n^{\lambda}$ is the occupation number of the single-particle state $\psi_n$, a non-interacting many-body 
configuration $|\lambda\rangle$ reads as:
\begin{equation}\label{N-MBS}
|\lambda\rangle=|i_1^{\lambda}, i_1^{\lambda},...,i_{N_{{\rm ses}}}^{\lambda}\rangle .
\end{equation}
 
($D_3$) Diagonalization of the Coulomb-interacting electronic Hamiltonian $H_{S_1}=H_{S_1}^{(0)}+W$ on the subspace of 
non-interacting many-body states from ${\cal F}_{S_1}$. As a result one gets $N_{{\rm mes}}$ interacting many-body states (IMBS) 
$|\nu\rangle$ and the associated energy levels $E_{\nu}$ introduced in Section \ref{S2}.1.
We also introduce the `free' energies of $H_{S_1}+H_{S_2}$, that is ${\cal E}_{\nu, j}^{(0)}=E_{\nu}+e_j$.
Note that in view of diagonalization the interaction $V$ between the two subsystems must be also written w.r.t.\ the 
`free' states $\{|\nu,j\rangle \}$. If the second subsystem is not needed then the GME must be solved w.r.t.\ the set $\{|\nu\rangle\}$
and the diagonalization procedure stops here. {It is worth pointing out here that even in the absence of bosonic fields 
and electron-photon coupling, the master equation for Coulomb interacting systems cannot be written in terms of single particle
states. In spite of the fact that the unitary evolution $U_S$ is diagonal w.r.t. the many-body basis $\{|\nu\rangle\}$,
the matrix elements of the fermionic operators in the interaction picture $\langle\nu|\tilde{c}_{n\sigma}(t)|\nu'\rangle$
depend on energy differences $E_{\nu}-E_{\nu'}$ which, due to the Coulomb effects, cannot~be reduced to the single-particle
energy $\varepsilon_n$.}

($D_4$) Diagonalization of the fully interacting Hamiltonian $H_S$ on a subspace of ${\cal F}$ made by the lowest energy $N_{{\rm mesT}}$ 
interacting MBS of $H_{S_1}$ and $j_{{\rm max}}$ eigenstates of $H_{S_2}$. Remark that after the 1st truncation w.r.t. 
NMBSs we perform here a 2nd double truncation both w.r.t. IMBS (as $N_{{\rm mesT}}<N_{{\rm mes}}$) and w.r.t. the states in ${\cal F}_{S_2}$. 

Once this procedure is performed, one can express the system-environment couplings $H_T$ and $H_B$ in the fully interacting basis 
and use the eigenvalues ${\cal E}_p$ to replace the unitary evolution $U_S$ by the corresponding diagonal matrix 
$e^{-it{\cal E}_p}\delta_{pp'}$. Finally, the GME is to be solved w.r.t. the fully interacting basis (see subsection 2.3).  

Now let us enumerate and explain the advantages of this stepwise procedure when compared to a single and direct diagonalization of 
$H_S$. 

(a) {Numerical efficiency and accuracy.} Both diagonalization methods (stepwise and direct) require a truncation of both bases 
and are not free of numerical errors which in principle should diminish as the size of the bases increase. It is clear that in the 
stepwise procedure the $N_{{\rm mes}T}$ interacting MBSs are derived from a {larger} set of non-interacting states 
$\{|\lambda\rangle\}_{\lambda =1,..,N_{{\rm mes}}}$. Then the calculated \mbox{$N_p:=N_{{\rm mes}T}\times j_{{\rm max}}$} fully interacting states 
are more accurate than the ones obtained by diagonalizing once a $N_p\times N_p$ matrix. On the other hand, enlarging 
the full space to $N_{{\rm mes}}\times j_{{\rm max}}$ elements could be challenging in terms of CPU times.
 Convergence calculations relevant to circuit quantum electrodynamics have been presented in \cite{PhysRevE.86.046701}. In particular
it was shown that the inclusion of the (usually neglected) diamagnetic term in the electron-photon coupling improves the convergence of the
diagonalization procedure.

{The size of various effective bases used in the numerical calculation is decided both by physical considerations and
convergence tests. Typically, out of the $N_{{\rm ses}}$ single-electron states we construct the set of non-interacting
MBSs containing up to $N_e$ electrons, the size of this set being, of course, ${N_{\rm{ses}}}\choose{N_e}$. 
The~accuracy of numerical diagonalization which leads to the interacting many-body configurations  with up to $N_e$ electrons 
is essentially assessed by comparing the spectra associated obtained for different $N_{\rm{ses}}$. In particular, if we discretize 
our open system in a small number of lattice sites we can use all single-electron states as a basis, and we can calculate all many-body
electron states (like in the discrete case presented in \cite{Moldoveanu09:073019}. Obviously, this is no longer possible for a more 
complex geometry, and~then we need to evaluate the convergence of the results when the basis is truncated.}

{For the continuous model an extensive discussion on the convergence of the numerical diagonalization 
w.r.t. the various truncated bases  is given in a previous publication \cite{PhysRevE.86.046701}. 
Let us stress here that 
once the geometry of the system and the spatially-dependent interactions are accounted for there 
is no simple way to count ahead how many states one needs to get stable transport simulations,
and only extensive tests can be performed to resolve this issue.}

(b) {Physical interpretation.} It is obvious that the Coulomb interaction $W$ mixes only the non-interacting many-body
configurations $|\lambda\rangle$ while the hybrid coupling $V$ mixes both $\lambda$ and $j$ states. For this reason the 
weights of a non-interacting state $|\lambda,j\rangle$ in a fully interacting state $|\varphi_p)$ (as provided by a 
single diagonalization) cannot be easily associated with one of the interacting terms. In view of physical discussion 
it is more intuitive and natural to analyze the dynamics of the Coulomb-interacting system $S_1$ in the presence of the 
second subsystem $S_2$. One such example is  a self-assembled quantum dot embedded in a single-mode quantum cavity 
\cite{PhysRevB.97.195442}.
In this system the optical transitions couple electron-hole pairs which are genuine interacting many-body states. A second example
is a double quantum dot patterned in a 2D quantum wire which is itself placed in a cavity. There the interdot Coulomb 
interaction affects the optical transitions as well.  

On the other hand, the above procedure will not be appropriate if one is interested in including a~time-dependent driving 
term in $H_S$. This would be the case for a pumping potential or for a~modulating optical signal.   

Finally we shall comment on the typical sizes of many-body Fock spaces used to model steady-state or transient
transport in various systems. One of the simplest yet promising system for solid-state quantum computation is a 
double quantum dot accommodating at most two electrons on each dot such that the relevant Fock space already comprises 
256 interacting many-body states (counting~the spin). In~this case transport simulations can be obtained even without 
truncating the basis, especially if the spectral gaps allows one to disregard the contribution of higher energy configurations. 
However,~when~studying transport on edge states due to a strong perpendicular magnetic field in 2D systems (e.g., graphene or 
phosforene) one is forced to consider the low energy bulk-states as well. In~our previous numerical studies we find that 
one has to take into account at least 10 single-particle states; obviously, performing time-dependent simulations 
for $2^{10}$ MBSs is quite inconvenient so a truncation is needed. More~importantly, a realistic description of 
complex systems like rings or double dots etched in a 2DEG cannot be obtained with only few single-particle states. 
Note that the optical selection rules and matrix elements of the electron--photon interaction depend on the these states as well.

{In~the calculation of one, rather deep, QD embedded in a short quantum wire we are using 52
single-electron states, asking for 52 one-electron states, 1326 two-electron states and 560 three electron states.
Of these we take the lowest in energy 512 and tensor multiply by 17 photon states to obtain a basis
of 8704 MBS to calculate the dressed MBS. Then for the transport, we select the lowest in energy 128 dressed
states and construct the 16,384 dimensional Liouville space. All this choice is tailored for a rather
narrow section of a parameter space, if we consider the wire length, the confinement energy and the shape
of the QD and the range of the magnetic field.}

{Markovian or non-Markovian master equation method have been also developed for transport simulations in molecular junctions;
here a truncation is required w.r.t. to the basis states describing the molecular vibrations. In particular, Schinabeck {et al.}
\cite{PhysRevB.94.201407} proposed a hierarchical polaron master equation which was successfully implemented numerically
for two molecular orbitals and several tens of vibrational states.}

\subsection{Numerical Implementation and Observables}

The last step before numerical implementation requires the calculation of the system-environment couplings $H_T$ and $H_E$
w.r.t.\ the full basis $|\varphi_p)$. Clearly, to this end we shall use the unitary transformations 
$|\lambda\rangle\leftrightarrow |\nu\rangle$ and $|\nu,j\rangle\leftrightarrow |\varphi_p)$ which are already at hand due to the 
stepwise diagonalization procedure introduced in the previous section. Then let us introduce some generalized `jump' operators 
collecting all transitions, between pairs of fully interacting states, generated by tunneling of an electron with momentum $q$ from the 
$l$-th lead to the single-particle levels of the electronic system $S_1$:
\begin{equation}\label{Toperator}
      {\cal T}_l(q)=\sum_{p,p'}{\cal T}_{pp'}^l(q)
      |\varphi_p)(\varphi_{p'}| \ , \quad
      {(\cal T}_l(q))_{pp'}=\sum_nT^l_{qn}(\varphi_p|c_n^{\dagger}|\varphi_{p'}) \ .
\end{equation}

Then the dissipation operator associated to the particle reservoirs reads:
\begin{equation}\label{D_leads}
{\cal D}_{{\rm leads}}[\rho,t]=-\frac{1}{\hbar^2}\sum_{l=\mathrm{L,R}}\int dq\:\chi_l(t)
      ([{\cal T}_l,\Omega_{ql}(t)]+h.c.)\ ,
\end{equation}
with the following notation:
\begin{eqnarray}
      &&\Omega_{ql}(t)=U_\mathrm{S} (t,t_0) \int_{t_0}^tds\:\chi_l(s)
      \Pi_{ql}(s)e^{i((s-t)/\hbar )\varepsilon_l(q)}U_\mathrm{S}^\dagger(t,t_0),\\
      &&\Pi_{ql}(s)=U_\mathrm{S}^\dagger(s,t_0)\left ({\cal T}_l^{\dagger}
      \rho(s)(1-f_l)-\rho(s){\cal T}_l^{\dagger}f_l\right )U_\mathrm{S}(s,t_0),
\end{eqnarray}
and where for simplicity we omit to write the energy dependence of the Fermi function $f_l$.
Similarly,~the~bosonic operators have to written down w.r.t.\ the full basis which then leads to the calculation of
${\cal D}_{{\rm bath}}$. Under the Markov approximation w.r.t. the correlation function of the bosonic reservoir the latter
becomes local in time.

{The GME is solved numerically by time discretization using the Crank--Nicholson method which allows us to compute the 
reduced density operator for discrete time steps $\rho(t_{n})$, starting with an~initial condition corresponding to a
given state of the isolated central system, i.e., before the onset of the coupling with the leads. We take advantage of the fact 
that, by discretizing the time domain, the~operator $\Omega_{ql}(t_{n+1})$ obeys a recursive formula generated by the incremental
integration between $t_n$ and $t_{n+1}$, that is:
\begin{equation}
\Omega_{ql}(t_{n+1})=U_\mathrm{S}(t_{n+1},t_n)\Omega_{ql}(t_n)U_\mathrm{S}^\dagger(t_{n+1},t_n)+{\cal A}_{ql}(t_{n+1},t_n;\rho(t_{n+1}),\rho(t_n)),
\end{equation}
where the second term of the right-hand side depends on the yet unknown $\rho(t_{n+1})$. For any pair of time steps $\{t_n,t_{n+1}\}$
we initially approximate $\rho(t_{n+1})$ in ${\cal A}_{ql}$ by the already calculated $\rho(t_n)$, and~perform iterations to recalculate
$\rho(t_{n+1})$ via the GME, each time updating $\rho(t_{n+1})$ in ${\cal A}_{ql}$, until a convergence test for $\rho(t_{n+1})$ is 
fulfilled.}
 At any step of the iteration we also calculate and include ${\cal D}_{{\rm bath}}[\rho,t_m]$
into the iterative procedure; its calculation is much simpler as the Markov approximation w.r.t. the bath degrees of freedom takes care
of the time integral so this dissipative term becomes local in time. Finally,~we~check {numerically} the conservation
of probability and the positivity of the diagonal elements of $\rho(t_m)$, {i.e.,} the populations of fully
interacting states $|\varphi_p)$ at the corresponding time step {and for each iteration}.

 There are several reasons to extend the GME method beyond single-level models. (1) The electronic transport
at finite bias collects contributions from all the levels within the bias window. This feature leads to the well known
stepwise structure of the current-voltage characteristics; (2) In the presence of Coulomb interaction the GME must
be derived in the language of many-body states which allows us to perform exact diagonalization on appropriate Fock
subspaces; (3) The minimal model which describes the effect of the field-matter coupling in optical cavities with embedded
quantum dots requires at least two single-particle levels.

Both the GME and non-equilibrium Green's function formalism (NEGF) rely on the partitioning approach and allow for many-body
interaction in the central system, while the leads are assumed to be non-interacting (this assumption leads in particular
to the Fermi distribution of the particle reservoirs). There is however a crucial difference between the two methods.
The perturbative expansion of the dissipative kernel forces restricts the master equation approach to weak lead-sample
tunnelings while the interaction effects are accounted for exactly. In contrast, the Keldysh formalism is not limited to small
system-reservoir couplings but the Coulomb effects have to be calculated from appropriate interaction self-energies. Which method
fits better is simply decided by the particular problem at hand.

As stated in the Introduction, the advantage of the RDO stems from the fact that it can be used to calculate 
statistical averages of various observables ${\cal O}$ of the hybrid system: 
\begin{equation}\label{observables}
\langle{\cal O}\rangle={{\rm Tr}}_{{\cal F}}\{\rho(t){\cal O}\} \ .
\end{equation}

Useful examples are averages of the photon number operator ${\cal N}^{ph}=a^{\dagger}a$ and of the charge operator 
${\cal Q}=\sum_n c^{\dagger}_nc_n$. Also, the average currents in a two-lead geometry (i.e., $l=L,R$ can be identified from the 
continuity equation:
\begin{equation}\label{continuity}
\langle{\dot{\cal Q}}\rangle={\rm Tr}_{\cal F}\{{\cal Q}{\dot\rho(t)}\}=J_L(t)-J_R(t).
\end{equation}

\subsection{Coupling between Leads and Central System}
The modeling of the central systems and the reservoirs can be performed either 
by using continuous confining potentials or a spatial grid. Examples are a 
short parabolic 
wire \cite{Gudmundsson09:113007,Gudmundsson:2013.305}, ring~\cite{PhysRevB.87.035314,2040-8986-17-1-015201},
parallel wires with a window coupler \cite{Abdullah10:195325}, and wire with embedded 
dot \cite{Abdullah2014254,Gudmundsson09:113007} or dots~\cite{Gudmundsson16:AdP}.   
The coupling between the leads and the central system with length $L_x$ is described by
Equation~(\ref{Htunnel}), and in order to reproduce scattering effects seen in 
a Lippmann--Schwinger formalism~\cite{Vargiamidis03:597,PhysRevB.70.245308,PhysRevB.71.235302} the 
coupling tensor is defined as  
\begin{equation}
T^l_{qn} = \int_{\Omega_S^l\times \Omega_l} d{\bf r}d{\bf r}'
\left(\Psi^l_q ({\bf r}') \right)^*\Psi^S_n({\bf r})
g^l_{qn} ({\bf r},{\bf r'})+h.c.,
\label{T_aq}
\end{equation}
for states with wavefunction $\Psi^l_q$ in lead $l$, and $\Psi^S_n$ in the central
system. The domains for the integration of the wavefunctions in the leads are chosen to be
\begin{equation}
\begin{array}{ccc}
\Omega_\mathrm{L}&=&\left\{ (x,y)|\left[-\frac{L_x}{2}-2a_w,
-\frac{L_x}{2}\right]\times [-3a_w,+3a_w]\right\}, \\
\Omega_\mathrm{R}&=&\left\{ (x,y)|\left[+\frac{L_x}{2},
+\frac{L_x}{2}+2a_w\right]\times [-3a_w,+3a_w]\right\},
\end{array}
\end{equation}
and for the system as
\begin{equation}
\begin{array}{ccc}
\Omega_\mathrm{S}^\mathrm{L}&=&\left\{ (x,y)|\left[-\frac{L_x}{2},
-\frac{L_x}{2}+2a_w\right]\times [-3a_w,+3a_w]\right\},\\
\Omega_\mathrm{S}^\mathrm{R}&=&\left\{ (x,y)|\left[+\frac{L_x}{2}-2a_w,
+\frac{L_x}{2}\right]\times [-3a_w,+3a_w]\right\}.
\end{array}
\end{equation}

\begin{figure}[t]
		\includegraphics[width=0.64\textwidth,angle=0]{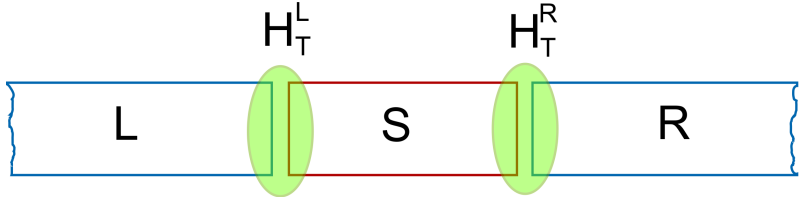}
	\caption{A schematic of the coupling of the system
		to the leads. The transparent green areas correspond to the contact regions
		defined by the nonlocal overlap function $g^{\mathrm{L,R}}_{qn}$ in $H_T(t)$.}
	\label{System-vg}
\end{figure}

The function
\begin{equation}
g^l_{qn} ({\bf r},{\bf r'}) =
g_0^l\exp{\left[-\delta_1^l(x-x')^2-\delta_2^l(y-y')^2\right]}
\exp{\left(\frac{-|E_n-\epsilon^l(q)|}{\Delta_E^l}\right)}.
\label{gl}
\end{equation}
with ${\bf r}\in\Omega_\mathrm{S}^l$ and ${\bf r}'\in\Omega_l$
determines the coupling of any two single-electron states by the ``nonlocal overlap''
of their wave functions in the contact region of the leads and the system, and their energy affinity. 
A schematic view of the coupling is seen in Figure  \ref{System-vg}. The parameters
$\delta_1^l$ and $\delta_1^l$ define the spatial range of the coupling within the 
domains $\Omega_\mathrm{S}^l\times\Omega_l$ \cite{Gudmundsson09:113007}.

The short quantum wire is considered to have hard wall confinement in the transport direction,
the $x$-direction, at $x=\pm L_x/2$, and parabolic confinement in the $y$-direction with characteristic 
energy $\hbar\Omega$. Possibly, the leads and the central system are considered to be placed in a
perpendicular homogeneous external magnetic field $\mathbf{B}=B\hat{\mathbf{z}}$. Together they lead to a
natural length scale, the effective magnetic length $a_w$ with $a_w^2\Omega_w=\hbar /m$, with 
$\Omega_w^2=[(\Omega_0)^2+(\omega_s)^2]^{1/2}$, and the cyclotron frequency $\omega_c=(eB/m)$. 
For GaAs with effective mass $m=0.067m_e$ relative dielectric constant $\kappa_c=12.3$ and
confinement energy $\hbar\Omega_0=2.0$ meV, $a_w=23.8$ nm. The magnetic field $B=0.1$ T.

The energy spectrum of the quasi 1D semi-infinite lead $l$ is represented by $\epsilon^l(q)$, 
with $q$ standing for the momentum quantum number of a standing wave state, and the subband
index $n^l$. The spectrum in the absence of spin orbit interactions can be evaluated 
exactly analytically \cite{ThorstenPhD}. The coupling of the leads and the central systems
in the continuous representation conserves parity of the electron states across the 
tunneling barrier.

The full strength of the continuous approach emerges as it is applied to describe the
transport of interacting electrons through 3D photon cavities in the transient time regime
or the long time regime ending in a steady state of the system. This will be reported below (see Sections 5 and 6).  
The numerical calculations can sometimes be simplified by describing the leads and 
the central system on a discrete spatial lattice, where the geometric details of the 
central system are usually implemented by hard walls and Dirichlet boundary 
conditions. The spatial integral of the coupling tensor (\ref{T_aq}) are then 
reduced to a set of contact points between the leads and the central system  
\cite{Moldoveanu09:073019,PhysRevB.82.085311}.
 
\section{Many-body Effects in the Transient Regime\label{S3}}

In this section we review some results on the transient transport in interacting systems described by a lattice model
\cite{Moldoveanu09:073019,Moldoveanu10:155442}. For the sake of generality we extend the GME method by including as well 
the spin degree or freedom which was previously neglected. The lattice model matches naturally to the partitioning 
transport setting, facilitates the geometrical description of the central sample (e.g.,~a~parallel quantum dot) and 
captures the dependence of the tunneling coefficients on the localization of the single-particle wavefunctions 
at the contact regions. A more realistic description is provided by the continuous model (see the previous section) 
which requires however a very careful tailoring of the confining potentials.

 The results presented in this section are also meant to illustrate the usefulness of the GME approach in 
describing the transient regime in terms of the dynamical occupations of the interacting many-body configurations. 
Such a description cannot be recovered within the non-equilibrium Greens' function formalism. 

{Developing the GME method  in the language of interacting many-body states was equally motivated by experimental works.
Recording the charging of excited states of QDs in the Coulomb blockade regime constitutes the core of transient current
spectroscopy and pump-and-probe techniques~\cite{Fujisawa_2003}. Also, transient currents through split-gate quantum point 
contacts (QPSs) and Ge quantum dots have been measured some time ago by Nasser {et al.} \cite{Naser} and by Lai {et al.} 
\cite{LAI2009886}. Another~relevant class of transport phenomena which can be modeled and understood within the GME method is 
the electron pumping through QDs with tunable-barriers (see e.g., the recent review \cite{Kaestner_2015}). 
In this context we investigated the transient response of a quantum dot submitted to a sequence of rectangular pulses 
applied at the contact to the input \cite{PhysRevB.80.205325} and the turnstile protocol for single-molecule magnets 
\cite{Moldoveanu_2015}.}

\subsection{Transient Charging of Excited States}

We consider a two-dimensional system of length $L_x$ and width $L_y$ described by a lattice with $N_x$ sites on the $x$ 
axis and $N_y$ sites on the $y$ axis. The total number of sites is denoted by $N_{xy}=N_xN_y$. By setting the two lattice 
constants $a_x$ and $a_y$ one has $L_x=a_xN_x$ and  $L_y=a_yN_y$. Once we know the single-particle eigenstates of the 
electronic subsystem $S_1$ we can write down its Hamiltonian $H_{S1}:=H_{S_1}^{(0)}+W$ in a second quantized form w.r.t. this basis, that is:
\begin{equation}
\begin{array}{lll}
 H_{S_1}^{(0)}&=& \sum_{n,n'=1}^{N_{xy}}\langle\psi_n|{\hat h}_{S_1}^{(0)}|\psi_{n'}\rangle c_n^\dagger c_{n'} 
= \sum_{n} \epsilon_n c_n^\dagger c_n \\  \label{eq:VCoul}
W&=&\frac{1}{2}\sum_{n,m,n',m'}V_{nmn'm'}c^{\dagger}_nc^{\dagger}_mc_{m'}c_{n'}
\end{array}
\end{equation}
where the Coulomb matrix elements are given by (${\bf r,r'}$ are sites of the 2D lattice):
\begin{equation}\label{V_C}
V_{nmn'm'}=\sum_{{\bf r,r'}}\psi_n^*({\bf r})\psi_m^*({\bf r'})V_C({\bf r}-{\bf r'})\psi_{n'}({\bf r})\psi_{m'}({\bf r'}).
\end{equation}

The Coulomb potential itself is given by
\begin{equation}\label{eq:CoulInt}
 V_C(\mathbf r,\mathbf r') = \frac{e^2}{4\pi\epsilon(|\mathbf r -\mathbf r'|+\eta)} \ ,
\end{equation}
where $\eta$ is a small positive regularization parameter. 

Like in the continuous model, the tunneling coefficients $T_{qn}^l$ are associated to a pair of states $\{{\psi}^l_q,\psi_n\}$
from the lead $l$ and the sample $S_1$. However the lattice version is much simpler:
\begin{equation}\label{Tqn}
T^{l}_{qn}=V_l{\psi}^{l*}_q(0_l)\psi_n(i_l),
\end{equation}
where $0_l$ is the site of the lead $l$ which couples to the contact site $i_l$ in the sample.
The wavefunctions of the semi-infinite lead are known analytically:
\begin{equation}\label{psiq}
\psi^l_q(j)=\frac{\sin(q(j+1))}{\sqrt{2\tau\sin q}},\quad \varepsilon_q=2\tau\cos q .
\end{equation}

In the above equation $\tau$ is the hopping energy of the leads. 
The integral over $q$ in the tunneling Hamiltonian 
(see Equation (\ref{Htunnel}) from Section \ref{S2}) counts the momenta of the 
incident electrons such that $\varepsilon^l(q)$ scans the continuous spectrum of the semi-infinite leads 
$\sigma_l\in [-2\tau+\Delta,2\tau+\Delta]$ where $\Delta$ is a shift which is chosen such that $\sigma_l$ covers the 
lowest-energy many-body spectrum of the central system. The~construction of the coupling coefficients $T^{l}_{qn}$ shows that a 
single-particle state which vanishes at the contact sites does not contribute to the currents. This is the case for states 
which are mostly localized at the center of the sample, while in the presence of a strong magnetic field the currents will 
be carried by edge states.

In \cite{Moldoveanu09:073019} we implemented GME for a non-interacting lattice Hamiltonian, whereas the Coulomb interaction 
effects were introduced in \cite{Moldoveanu10:155442}. In what concerns the geometrical effects we essentially 
showed that the transient currents depend on the location of the contacts (through the value of the single-particle wavefunctions 
of the sample at those points) but also on the initial state and on the switching functions $\chi_l(t)$ of the leads. 
It turns out that the stationary current does not depend on the last two parameters, in agreement with rigorous results 
\cite{PhysRevB.84.075464,Cornean2009}. We also identified a delay of the output currents which was attributed to the electronic 
propagation time along the edge states of the Hofstadter spectrum. 

The presence of Coulomb interaction brings in specific steady-state features known from previous calculations 
like the Coulomb blockade and the step-like structure of the current--voltage characteristics. On the other hand the GME
method naturally allows a detailed analysis of the time-dependent currents associated to each many-body configuration  
as well as of the relevant populations.    

Since the Hamiltonian $H_{S_1}$ of the interacting system commutes with the total number operator $Q=\sum_n c_n^{\dagger} c_n$,
its eigenstates $|\nu\rangle$ can still be labeled by the occupation $N_{\nu}$ of the non-interacting MBSs from which the state is
built. Then the single index $\nu$ can be replaced by two indices, the particle number $N_{\nu}$ and an index $i_{\nu}=0,1,2,...$ 
for the ground ($i_{\nu}=0$) and excited states ($i_{\nu}>0$, where $i_{\nu}$ also counts the spin degeneracy). The notation 
for the interacting many-body energies is changed accordingly ${\cal E}_{\nu}\to{\cal E}^{(i_{\nu})}_{N_{\nu}}$. 

We now define some useful quantities for our time-dependent analysis. The charge accumulated on $N$-electrons states
is calculated by collecting the associated populations:
\begin{equation}\label{N-charging}
q_N(t)=eN\sum_{\nu,n_{\nu}=N}\langle\nu|\rho(t)|\nu\rangle,
\end{equation}
where the sum counts all states whose total occupation $n_{\nu}=N$. Similarly one can identify the transient currents 
$J_{l,N}$ carried by $N$-particle states. These currents can be traced back form the right hand side of the GME:
\begin{equation}\label{partialC}
\langle{\dot Q}\rangle=\sum_N(J_{L,N}(t)-J_{R,N}(t))=\sum_N {\dot q}_N(t).
\end{equation}

Throughout this work we shall adopt the following sign convention for the currents associated to each lead: $J_L>0$
if the electrons flow from the left lead towards the sample and $J_R>0$ if they flow from
the sample towards the right lead. 

The sequential tunneling processes change the many-body configurations of the electronic system. The energy required 
to bring the system to the $i$-th MBS with $N$ particles is measured w.r.t.\ the ground state with $N-1$
electrons ($i=0,1,..$). We introduce two classes of chemical potentials of the sample:
\begin{equation}\label{mu_gN}
\mu^{(i)}_{g,N}={\cal E}^{(i)}_N-{\cal E}^{(0)}_{N-1},
\end{equation}
\begin{equation}\label{mu_xN}
\mu^{(i)}_{x,N}={\cal E}^{(i)}_N-{\cal E}^{(1)}_{N-1},
\end{equation}
where $\mu^{(i)}_{g,N}$ characterizes transitions from the ground state $(N-1)$-particle configuration to various 
$N$-particle configurations.  
In particular $\mu^{(0)}_{g,N}$ describes addition processes involving ground-states with $N-1$ and $N$ electrons while 
$\mu^{(i>0)}_{g,N}$ refers to transitions from $(N-1)$-particle ground state to excited $N$-particle configurations. 
The chemical potentials $\mu^{(i)}_{x,N}$ describe transitions from the 1st $(N-1)$-particle excited states to 
configurations with $N$ particles. In a transition of this type an electron tunnels on the lowest single particle state
to the central system which already contains one electron on the excited single-particle state $|\sigma_2\rangle$. As a result some 
of the triplet states are being populated. We shall see that these transitions play a role especially in the transient regime. 

For numerical calculations we considered a 2D quantum wire of length $L_x=75$ nm and width $L_y=10$ nm. The lowest two spin-degenerate single-particle levels are $\varepsilon_1=0.375$ meV and \mbox{$\varepsilon_2=3.37$ meV}. The non-interacting MBSs are described by the 
spins of the occupied single-particle levels, e.g., $|\sigma_1\overline{\sigma}_2\rangle$ is a two-particle configuration with a spin 
$\sigma$ associated to the lowest single-particle state and a second electron with opposite spin orientation on the energy level 
$\varepsilon_2$. Besides the usual singlet ($S$) and triplet ($T$) states we find that the Coulomb interaction induces the 
configuration mixing of the antiparallel configurations $|\uparrow_1\downarrow_1\rangle$ and $|\uparrow_2\downarrow_2\rangle$. 
More precisely, we get an interacting ground two-particle state mostly made of $|\uparrow_1\downarrow_1\rangle$ 
(whose weight is 0.86) and with a small component (weight~0.14) of state $|\uparrow_2\downarrow_2\rangle$. Conversely, 
$|\uparrow_1\downarrow_1\rangle$ is also found in the highest energy two-particle state. We stress here that the configuration 
mixing decreases and eventually vanishes if the gap $E_{\uparrow_2\downarrow_2}-E_{\uparrow_1\downarrow_1}$ between two non-interacting 
energies is much larger that the corresponding matrix element.

\begin{figure}[t]
        \includegraphics[width=0.45\textwidth]{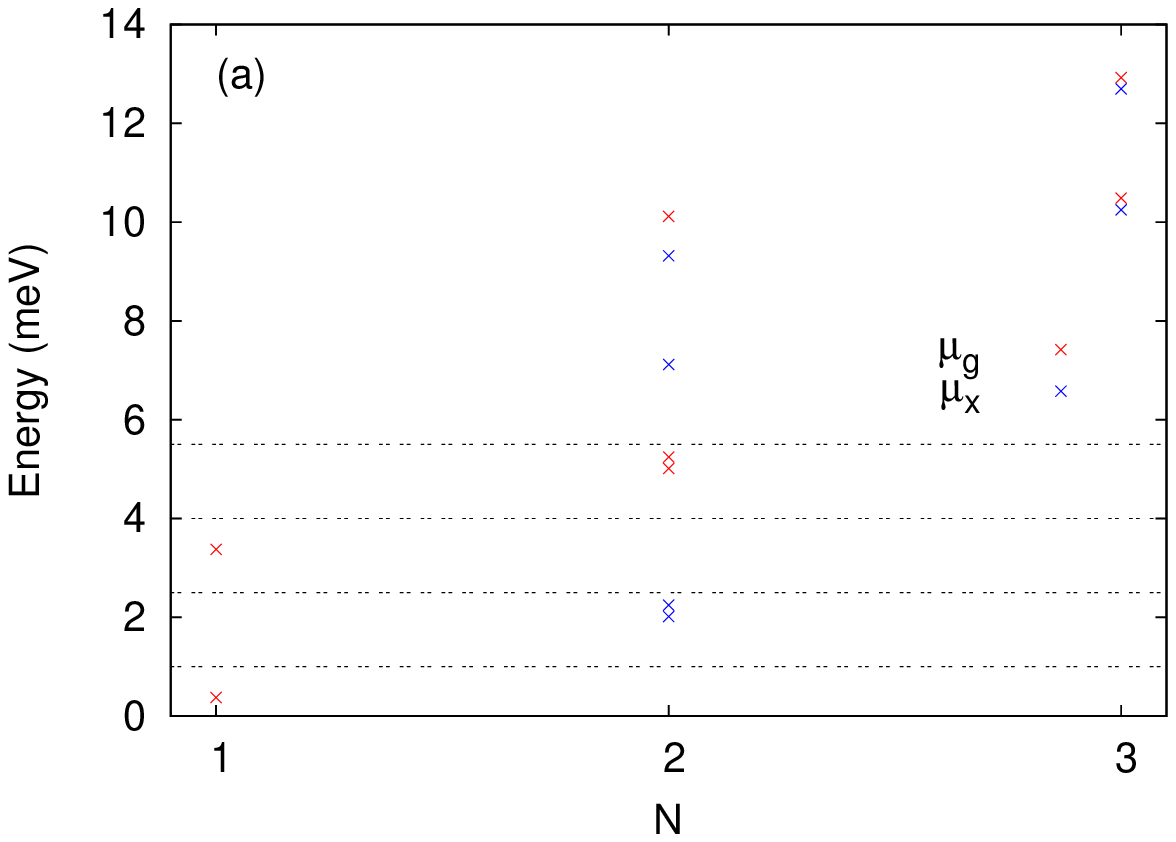}
        \includegraphics[width=0.45\textwidth]{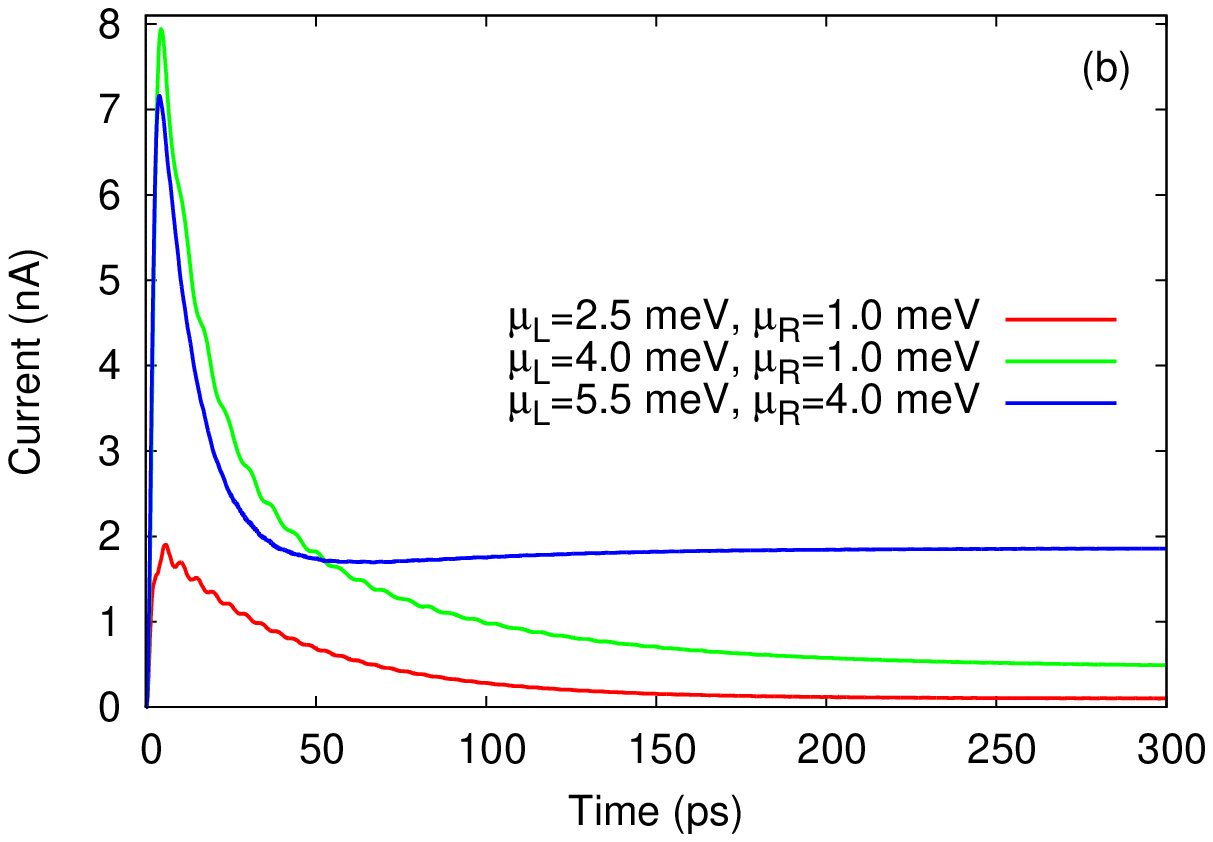}
        \caption{(\textbf{a}) The chemical potentials $\mu^{(i)}_{g,N}$ (red crosses) and $\mu^{(i)}_{x,N}$ (blue crosses) for $N$-particle
configurations, $N=1,2,3$. For a given particle number $N$ the chemical potentials are ordered vertically according to the index 
$i=0,1,..$. The horizontal lines correspond to specific values of the chemical potentials in the leads (see the discussion in the text); (\textbf{b}) 
The time-dependent currents in the left lead at different values of the bias window $\mu_L-\mu_R$.}
        \label{Fig-vm01}
\end{figure}

In Figure \ref{Fig-vm01}a we show the chemical potentials corresponding to interacting MB configurations with up to three electrons.
As long as the chemical potential $\mu^{(i)}_{g,N}$ lies within the bias window the corresponding state will contribute 
both to the transient and steady-state currents. We shall see that if $\mu^{(i)}_{g,N}<\mu_R$ the state $|i,N\rangle$
contributes only to the transient currents. Finally, when $\mu^{(i)}_{g,N}\gg\mu_L$ the state $|i,N\rangle$ is poorly 
populated and will not contribute to transport. Let us stress here a rather unusual transition from $|\sigma_2\rangle$ 
to the ground two-particle state which is mostly made of $|\sigma_1{\overline\sigma}_1\rangle$. The corresponding addition 
energy $\mu_{x,2}^{(0)}=2$ meV is smaller than the energy required for the usual transition 
$|\sigma_1\rangle\to|\sigma_1{\overline\sigma}_1\rangle$. This happens because of the Coulomb mixing between 
$|\uparrow_1\downarrow_1\rangle$ and $|\uparrow_2\downarrow_2\rangle$ which makes possible the transition from the 
excited single-particle state to the mixed interacting two-particle groundstate.
The~chemical potential $\mu_{x,2}^{(2)}$ describes the transition from the excited single-particle state 
$|\sigma_2\rangle$ to the triplet states. 

Already by analyzing Figure \ref{Fig-vm01} one can anticipate to some extend how the transport takes place in terms of allowed 
sequential tunneling processes. Suppose that the chemical potentials of the leads are selected such that 
$\mu_{g,1}^{(0)}<\mu_R<\mu_{g,1}^{(1)}<\mu_L<\mu_{g,2}^{(0)}$ (as an example we set $\mu_R=1$ meV and \mbox{$\mu_L=4$ meV}). 
Then both single-particle levels are available for tunneling but one expects that the 
double occupancy is excluded because $\mu_L<\mu_{g,2}^{(0)}\sim$ 5 meV. According to this scenario, more charge will accumulate 
on $\varepsilon_1$, the excited states $|\sigma_2\rangle$ will eventually deplete and the steady-state current vanishes in 
the steady-state. This is the well known Coulomb blockade effect. However, we see in Figure \ref{Fig-vm01}b that the steady-state 
current vanish only when $\mu_L<\mu_{g,1}^{(1)}$ as well, which suggest that the presence of the excited single-particle states within the bias window leads to 
a partial lifting of the Coulomb blockade. We stress that such an effect cannot be predicted within a single-site model with onsite 
Coulomb interaction. A third 
curve shows the current for $\mu_L=5.5$\, meV and $\mu_L=4$\,meV.

\begin{figure}[t]
\includegraphics[width=0.45\textwidth]{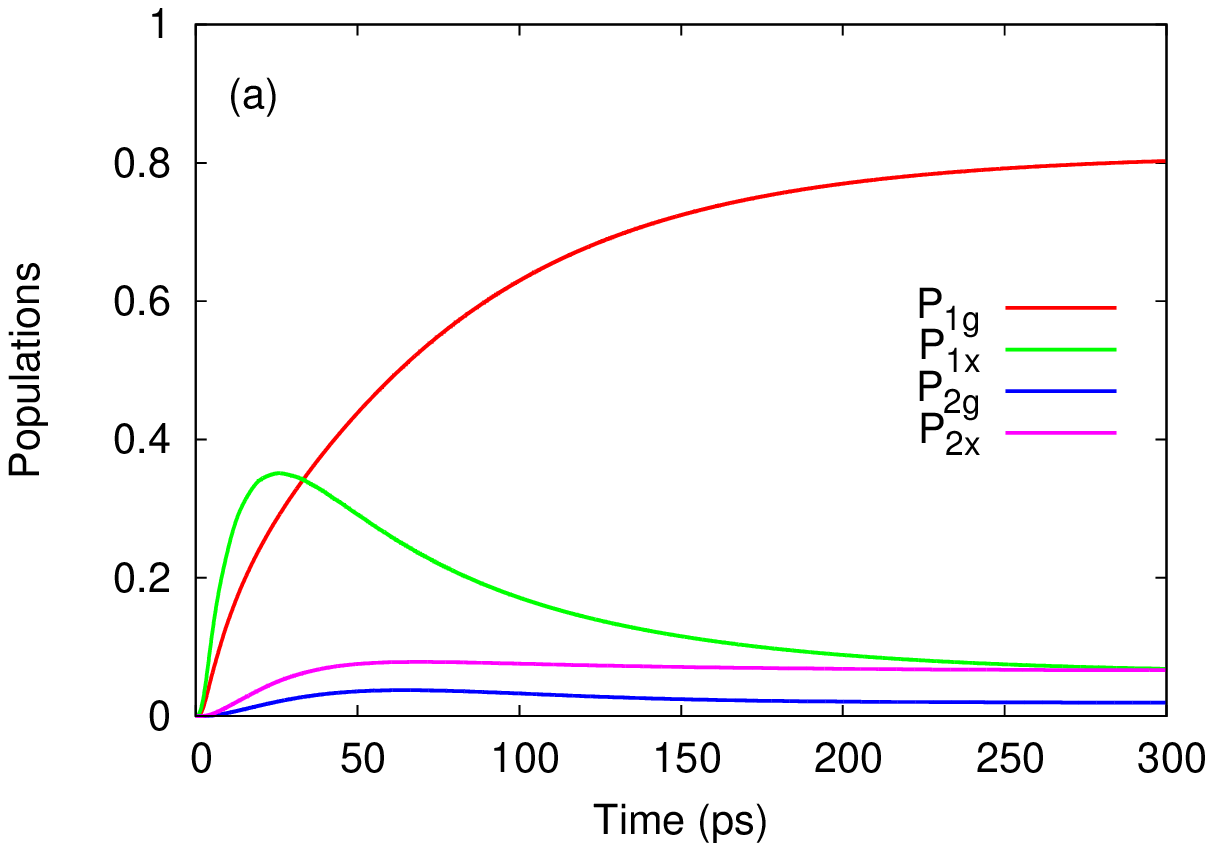}
\includegraphics[width=0.45\textwidth]{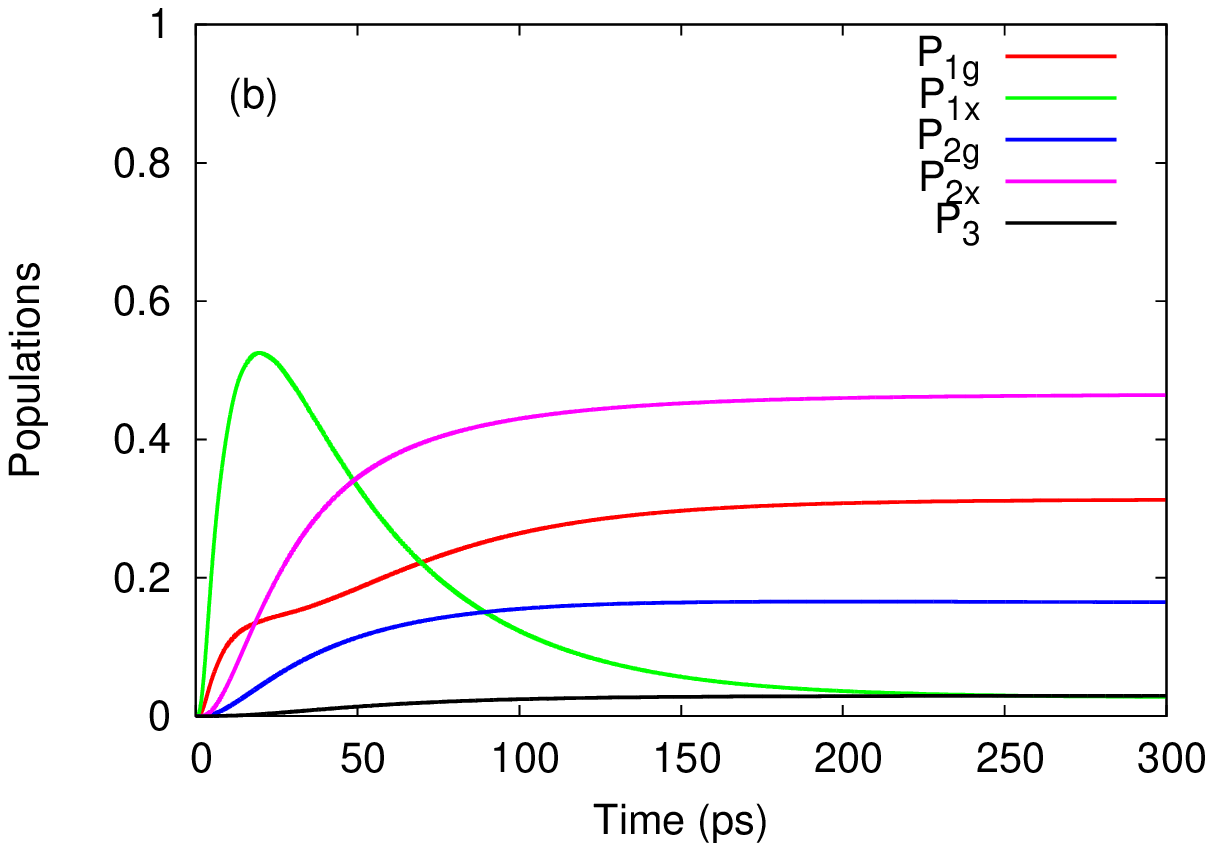}
        \caption{The populations of ground (g) and excited (x) $N$-particle states ($N=1,2,3$) for two bias windows;
(\textbf{a}) $\mu_L=4$\,meV, $\mu_R=2$\,meV; (\textbf{b}) $\mu_L=5.5$\,meV, $\mu_R=4$\,meV. In panel (\textbf{a}) $P_{3}$ is negligible and was omitted.}
        \label{Fig-vm02}
\end{figure}

 Figure \ref{Fig-vm02} presents the evolution of the relevant populations at two values of the bias window. In~Figure \ref{Fig-vm02}a
the population $P_{1g}=P_{\uparrow_1}+P_{\downarrow_1}$ of the ground single-particle states dominates in the steady state. This is
expected, as the corresponding chemical potential lies below the bias window so this state will be substantially populated. The other 
configurations contributing to the steady-state are just the ones which can be populated by tunnelings from the left lead, that is
the excited single-particle states and all two-particle states except for the single configuration which cannot be accessed. 
By looking at  Figure \ref{Fig-vm01}a one infers that the two-particle states are being populated when one more electron is added from 
the left lead on the initial excited single-particle state $|\sigma_2\rangle$. In particular, the ground two-particle state 
is populated only due to the Coulomb-induced configuration mixing.    
 
A completely different behavior is noticed in Figure \ref{Fig-vm02}b. As the bias window is pushed upwards such that 
$\mu_{g,1}^{(1)}<\mu_R<\mu_{g,2}^{(1)}<\mu_L$ the transitions from the lowest states $|\sigma_1\rangle$ to two-particle states are also 
activated. Consequently, the population $P_{2x}$ of the excited two-particle configurations exceeds $P_{1g}$ and dominate in the 
steady-state regime. Note that $P_{2x}>P_{2g}$ because it collects the population of the degenerate triplet states. In the transient 
regime the excited single particle states are populated much faster than the ground states. This happens because of the different 
localizations of the single-particle wavefunctions on the contact regions. We find that the wavefunction associated to the 2nd 
single-particle state has a larger value at the endpoints of the leads. A drop of $P_{1x}$ follows as the ground one-electron states 
and the other two-particle configurations become active (a similar feature is noticed in Figure \ref{Fig-vm02}a). A small 
populations of the three particle states can be also observed. The steady-state current increases considerably 
(see Figure \ref{Fig-vm01}b) and is due to the two-particle states.   

We end this section with a discussion on the partial currents $J_{L,N}$ and $J_{R,N}$ associated to $N$-particle states. 
Although they cannot be individually measured, these currents provide further insight into the transport processes, in particular on the 
way in which the steady-state regime is achieved.

\begin{figure}[t]
\includegraphics[width=0.3\textwidth,angle=-90]{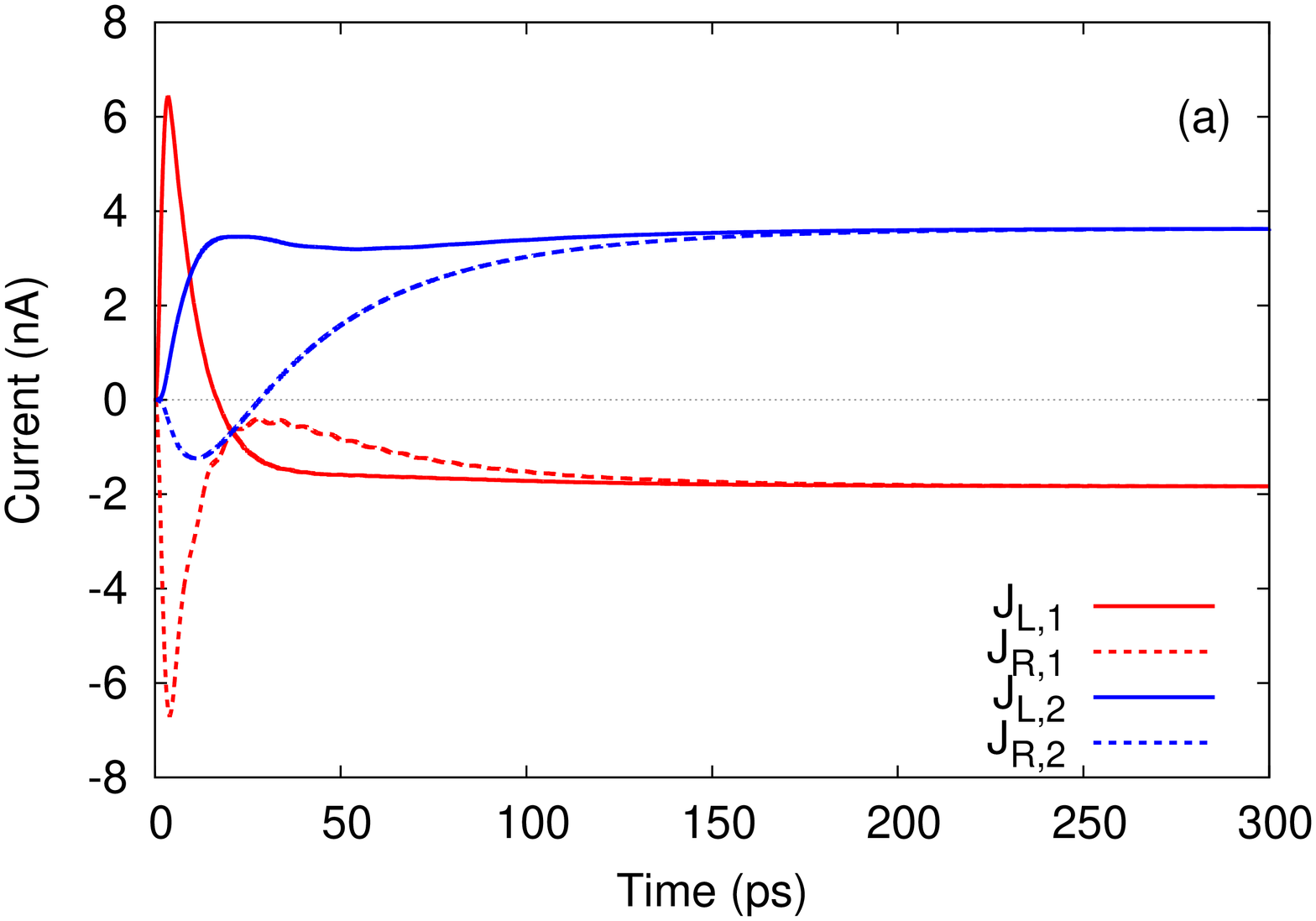}
\includegraphics[width=0.3\textwidth,angle=-90]{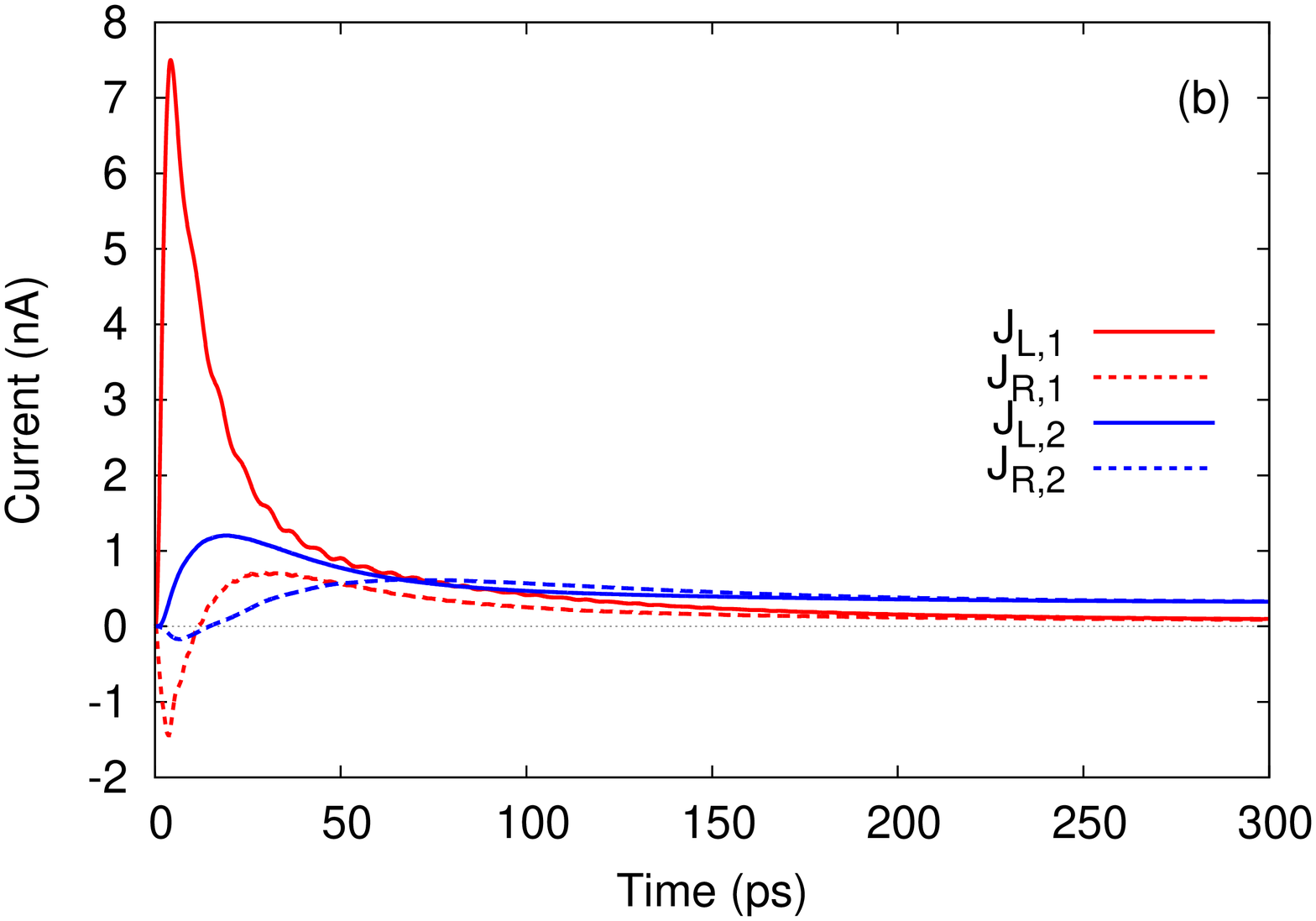}
\includegraphics[width=0.3\textwidth,angle=-90]{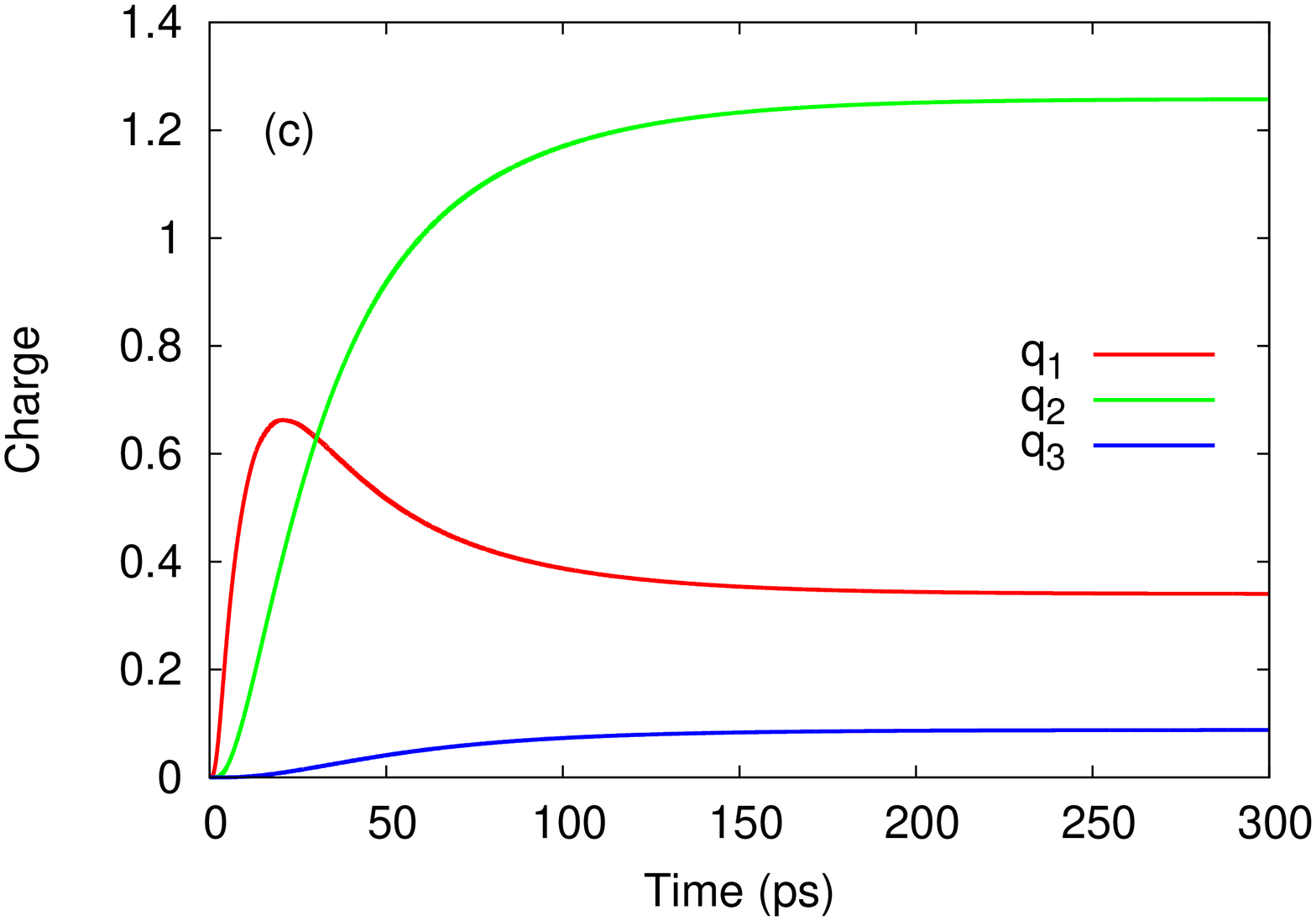}
\includegraphics[width=0.3\textwidth,angle=-90]{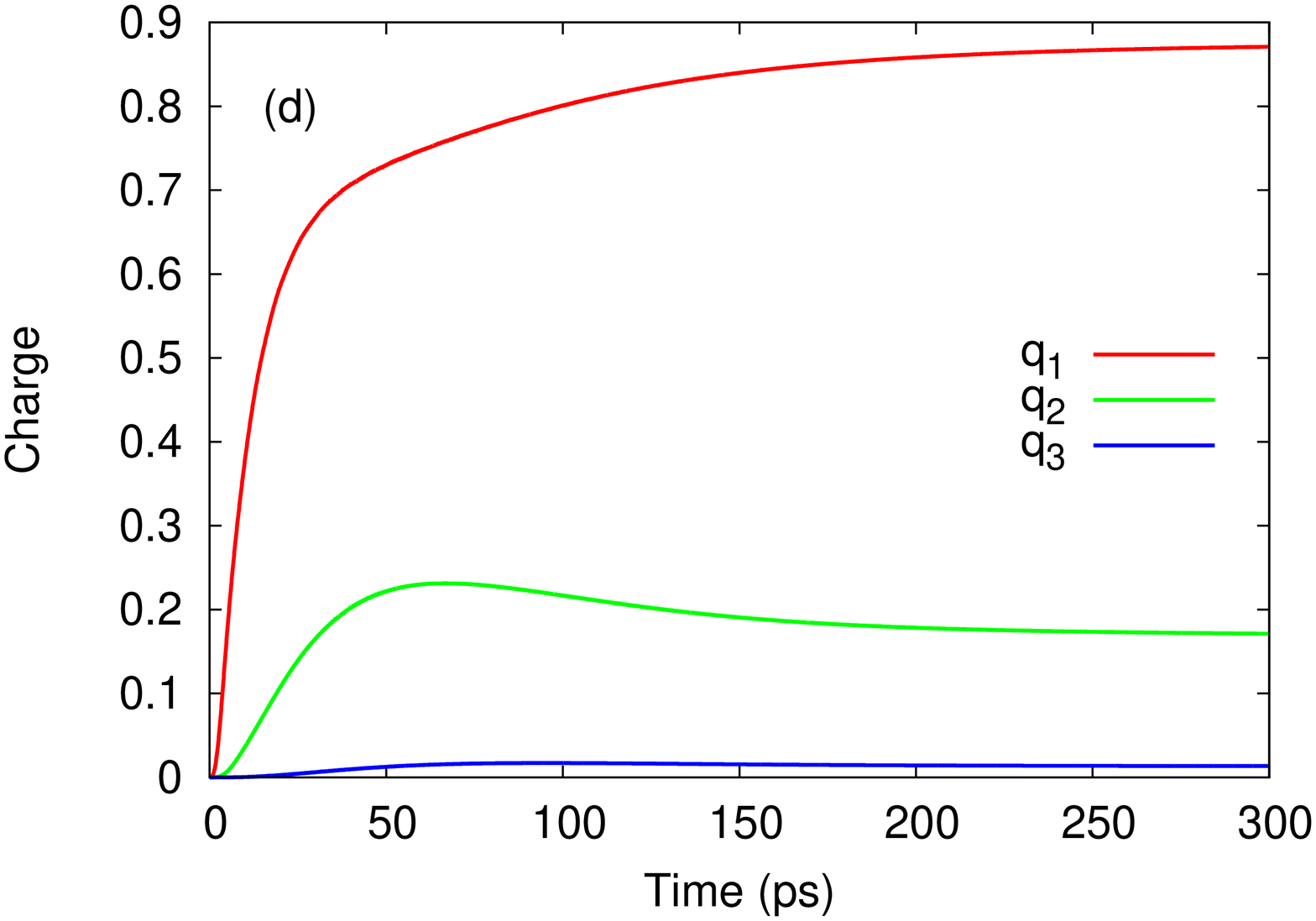}
        \caption{ (\textbf{a}) The transient currents $J_{L,N}$ and $J_{R,N}$ associated to one and two-particle configurations for 
$\mu_L=5.5$\,meV, $\mu_R=4$\,meV; (\textbf{b}) The same currents for a bias window $\mu_L=4$\,meV, $\mu_R=2$\,meV; 
(\textbf{c})~The charge $q_N$ accumulated on $N$-particle states at $\mu_L=5.5$\,meV, $\mu_R=4$\,meV; (\textbf{d}) $q_N$ for $\mu_L=4$\,meV, 
$\mu_R=2$\,meV, and $q_N$  are given in units of electron charge $e$.}
\label{Fig-vm03}
\end{figure}

Figure \ref{Fig-vm03}a shows that in the steady-state regime the currents carried by the one-particle states $J_{L,1}$ and $J_{R,1}$ 
achieve a negative value when $\mu_{g,1}^{(0)}<\mu_{g,1}^{(1)}<\mu_R<\mu_{g,2}^{(0)}<\mu_L$, whereas the two-particle currents evolve to a 
larger positive value such that the total current $J_L$ 
will be positive as already shown in Figure \ref{Fig-vm01}b. When the bias window is shifted down to $\mu_L=4$\,meV and $\mu_R=2$\,meV
all transients are mostly positive (see Figure \ref{Fig-vm03}b). One observes that the single-particle configurations 
are responsible for the spikes of the 
total current $J_L$ and that the two-particle currents display a smooth behavior. These~features can be explained by looking 
at the charge occupations $q_N$ shown in Figures \ref{Fig-vm03}c,d. 
As~$\mu_{g,1}^{(0)}$ and $\mu_{g,1}^{(1)}$ are both below $\mu_R=4$\,meV while $\mu_{g,2}^{(0)}$ is well within the 
bias window, the population of the single-particle states increases rapidly in the transient regime but then also drops 
in favour of $P_2$, the~total occupation of two-particle states. Such a redistribution of charge among configurations with different 
particle numbers is less pronounced in Figure \ref{Fig-vm03}d, because in this case the smaller contribution of the two-particle states 
is only due to transitions allowed by $\mu_{x,2}^{(0)}$ and $\mu_{x,2}^{(1)}$ which are now located within the bias window. 
The slope of $q_2$ also changes sign in the transient regime and one can check from Figure \ref{Fig-vm03}b that on the corresponding
time range $J_{R,2}$ slightly exceeds $J_{L,2}$.  

The occupation of the three-particle configurations is negligible so $q_3$ is also small and was included here only for completeness 
while the associated currents were omitted.

\subsection{Coulomb Switching of Transport in Parallel Quantum Dots}

After using the GME formalism to describe transient transport via excited states in a single interacting nanowire we now 
extend its applications to capacitively coupled quantum systems. Besides~Coulomb blockade, the electron-electron interaction
cause momentum-exchange which leads to the well known Coulomb drag effect in double-layer structures  \cite{drag-RMP} and double 
quantum dots~\cite{PhysRevLett.104.076801,APJauho-drag,Lim_2018} or wires \cite{PhysRevB.99.035423}. Also, theoretical calculations
on thermal drag between Coulomb-coupled systems were recently presented \cite{Sanchez_2017,Th-drag}. 
   
\begin{figure}[t]
\includegraphics[width=0.25\textwidth,angle=0]{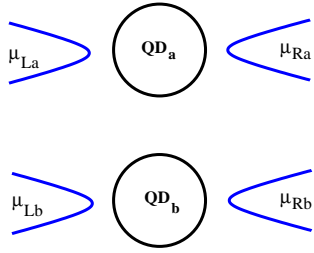}
        \caption{A sketch of the parallel double-dot system. Each QD is coupled to source-drain particle 
reservoirs described by chemical potentials $\mu_{Ls}$ and $\mu_{Rs}$, $s=a,b$. There is no interdot 
electron tunneling but the systems are correlated via Coulomb interaction.}
\label{Fig-vm04}
\end{figure}

Here we consider a very simple model for two parallel quantum dots \cite{PhysRevB.82.085311} (a sketch of the system is given 
in Fig.\,\ref{Fig-vm04}). Each system is described by a 1D four-sites
chain and for simplicity we neglect the spin degree of freedom which will only complicate the discussion of the effects.
The diagonalization procedure provides all 256 many-body configurations emerging from the 8 single-particle states. Let us point out that
the interdot and intradot interactions are treated on equal footing beyond the single-capacitance model. 
The hopping energy within the dots is $t_D=1$ meV and the time unit is expressed in units of $\hbar/t_D$.
Then the currents are calculated in units of $et_D/\hbar$. The tunneling rates to the four leads are all equal 
$V_{La}=V_{Ra}=V_{Lb}=V_{Rb}$. 

We shall use the GME method to study the {onset} of the interdot Coulomb interaction. In order to distinguish the  
transient features due to mutual capacitive coupling we consider a transport setting in which each dot is connected to 
the leads at different times. More precisely, one system, say $QD_a$ is open at the initial instant $t_a=0$ and then 
reaches a stationary state ($J_{La}=J_{Ra}$) at some later time $T_a$. The coupling of the nearby system to its leads is 
switched on at $t_b>T_a$ such that the changes in the current $J_a$ can only be due to mutual Coulomb interaction.
Note that the usual Markov--Lindblad version of the master equation simulate the transport when the four leads are coupled 
suddenly and simultaneously to the double-dot structure.  

As before, the interacting many-body configurations can be labeled by to the occupations of each dot according to 
the correspondence ${\cal E}_{\nu}\to{\cal E}^{(i_{\nu})}_{N_{a,\nu},N_{b,\nu}}$. Here $N_{s,\nu}$ is the number of 
electrons in the system $s$ associated to a many-body configuration $\nu$. If the two systems are identical the lowest 
chemical potentials are introduced as:
\begin{equation}\label{mu_gMN}
\mu^{(0)}_{g}(N_a,N_b)={\cal E}^{(0)}_{N_a,N_b}-{\cal E}^{(0)}_{N_a-1,N_b}={\cal E}^{(0)}_{N_a,N_b}-{\cal E}^{(0)}_{N_a,N_b-1},
\end{equation}
because of the degeneracy w.r.t. to the total occupation number ${\cal E}^{(0)}_{N_a,N_b}={\cal E}^{(0)}_{N_b,N_a}$
For the parameters chosen here one finds: $\mu^{(0)}_{g}(1,1)=3$ meV, $\mu^{(0)}_{g}(2,0)=4$ meV and $\mu^{(0)}_{g}(2,1)=4.5$ meV.
The location of the several chemical potentials w.r.t. the two bias windows already suggests the possible interdot 
correlation effects. The main point is that the transport channels through one dot also depend on the occupation of the 
nearby dot. One therefore expects that the currents $J_{La}$ and $J_{Ra}$ also depend on the bias applied on the nearby system. 

\begin{figure}[t]
\includegraphics[width=0.4\textwidth,angle=0]{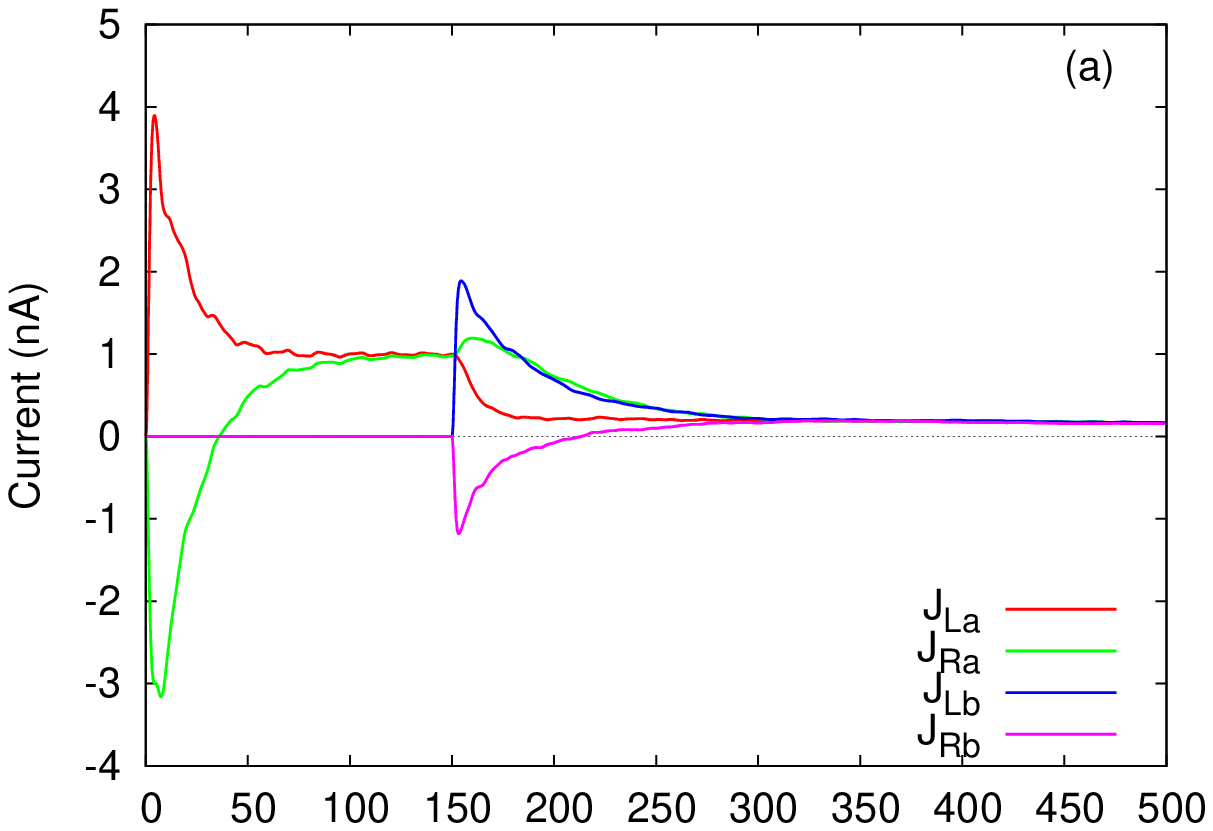}
\includegraphics[width=0.4\textwidth,angle=0]{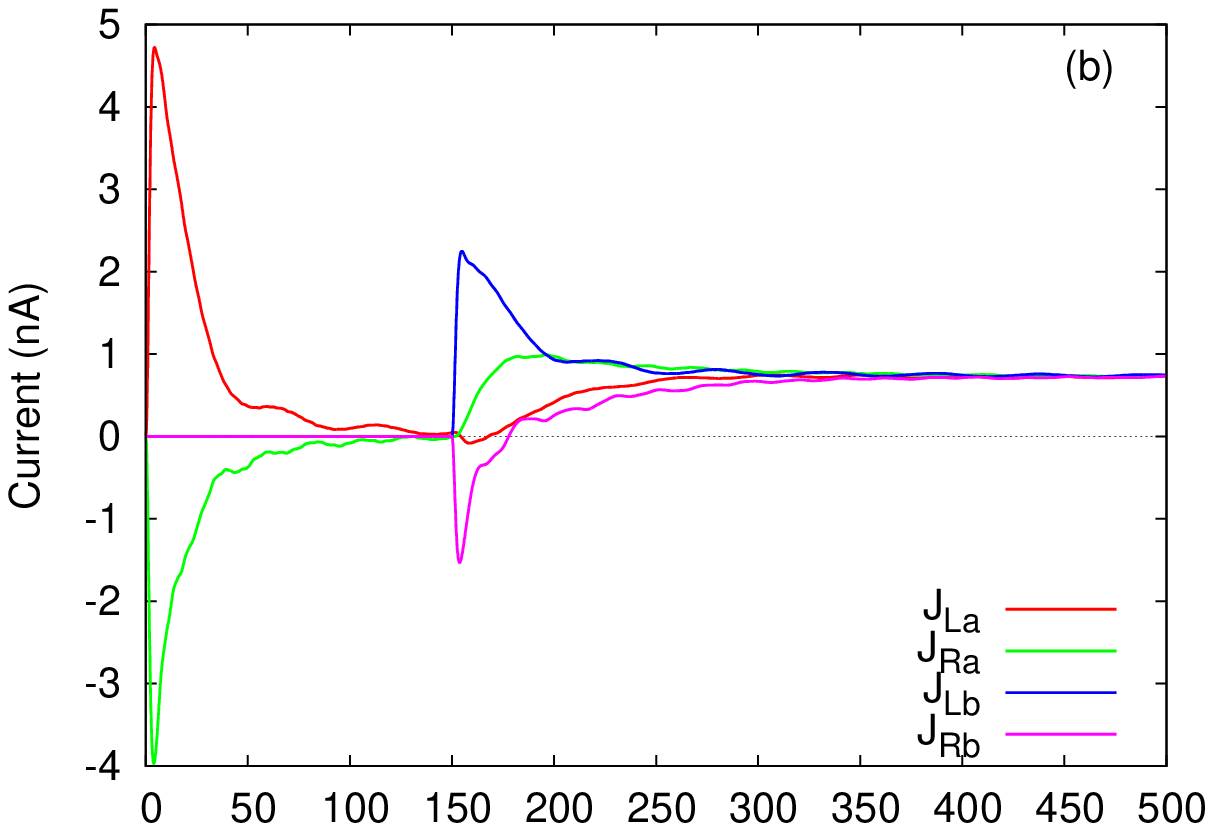}
\includegraphics[width=0.4\textwidth,angle=0]{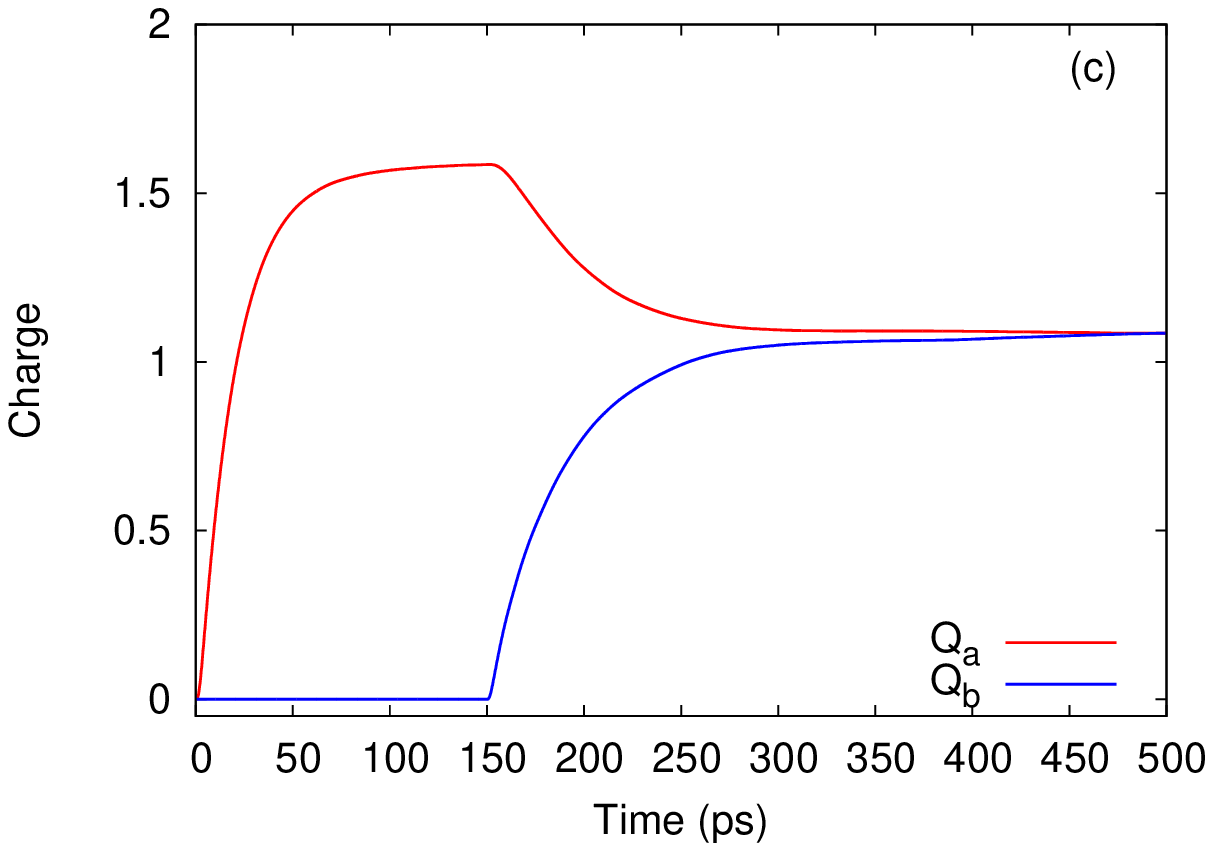}
\includegraphics[width=0.4\textwidth,angle=0]{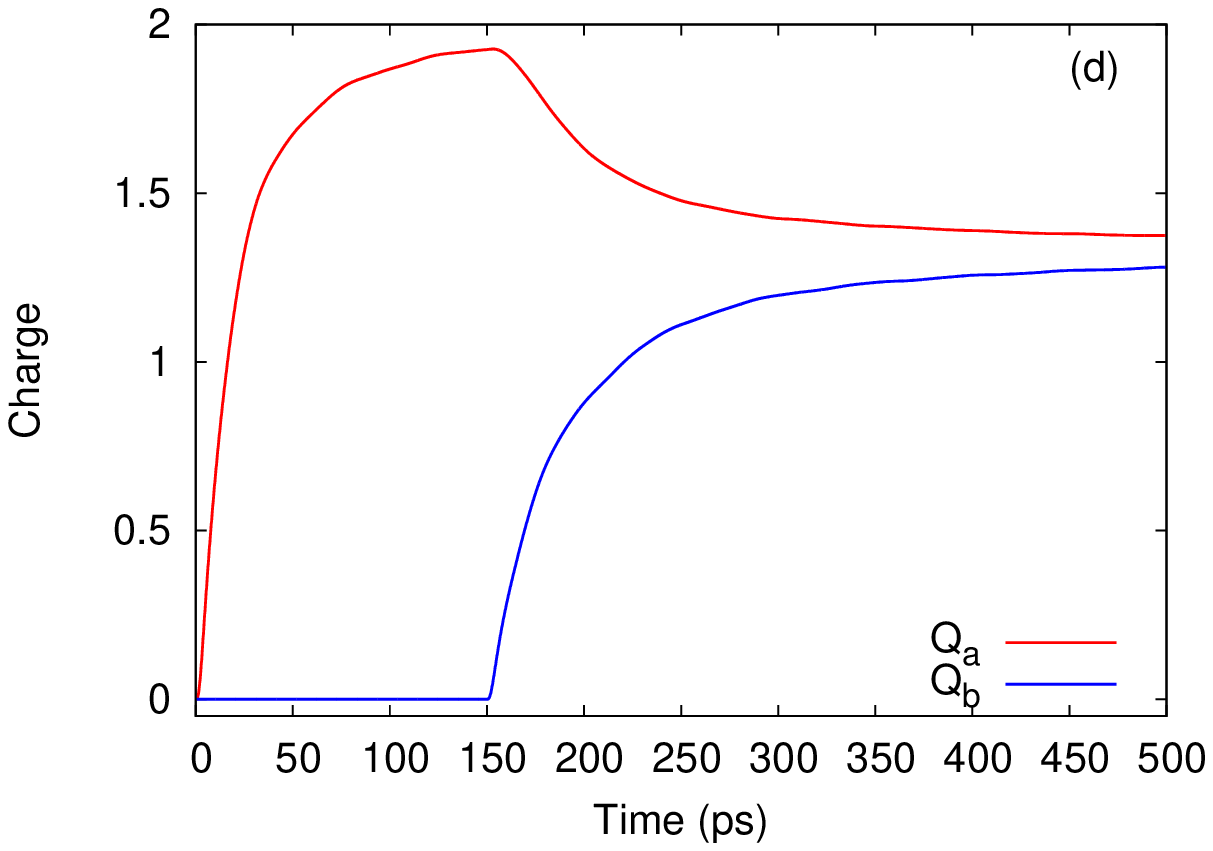}
        \caption{ The transient currents in the two systems for different chemical potentials 
of the leads: (\textbf{a})~$\mu_{La}=\mu_{Lb}=4.25$ meV, $\mu_{Ra}=\mu_{Rb}=3.75$ meV;
(\textbf{b})  $\mu_{La}=\mu_{Lb}=4.75$ meV, $\mu_{Ra}=\mu_{Rb}=4.35$ meV; (\textbf{c,d})~The charge occupations 
of the two systems associated to the currents in Figures  \ref{Fig-vm05}a,b. The charges $Q_{a,b}$ are given 
in units of electron charge $e$.}
\label{Fig-vm05}
\vspace{-5mm}
\end{figure}

To discuss this effect we performed transport simulations for two arrangements (A and B) of the bias window $\mu_{Ls}-\mu_{Rs}$.
In the A-setup we select the four chemical potentials such that the chemical potentials associated to the many-body configurations 
relevant for transport obey the inequalities $\mu^{(0)}_{g}(1,1)<\mu_{Rs}<\mu^{(0)}_{g}(2,0)<\mu_{Ls}<\mu^{(0)}_{g}(2,1)$. 
The scenario is easy to grasp: As $QD_a$ is coupled to the leads and the nearby dot is disconnected and empty, it will accumulate 
charge and evolve to a steady state where the current is essentially given by tunneling assisted transitions between 
${\cal E}^{(0)}_{2,0}\leftrightarrow {\cal E}^{(0)}_{1,0}$. This~behavior is observed in  Figure \ref{Fig-vm05}a up 
to $t_b=150$ ps when $QD_b$ is also coupled to its leads. Note~also that the charge occupation of $QD_a$ almost saturates 
at $Q_a=1.6$. As expected, for $t>t_b$ a transient current develops in $QD_b$, but a simultaneous drop the $J_{La}$ and $J_{Ra}$
shows the dynamical onset of the charge sensing effect between the two systems.
In the final steady-state the two currents nearly vanish, thus proves their negative correlation due to the mutual Coulomb 
interaction. The~charges $Q_{a,b}$ reach the same value and suggest that in the long time limit the double system contains
one electron on each dot. Remark that in the final steady-state the dominant population corresponds to the many-body 
energy ${\cal E}^{(0)}_{1,1}$ which is not favorable for transport through any of the dots as long as 
$\mu^{(0)}_{g}(2,1)={\cal E}^{(0)}_{2,1}-{\cal E}^{(0)}_{1,1}$ is outside the bias window.    

Figures \ref{Fig-vm05}b,d present the currents and the charge occupations for the second setup B which is defined 
by the inequalities  $\mu^{(0)}_{g}(2,0)<\mu_{Rs}<\mu^{(0)}_{g}(2,1)<\mu_{Ls}$. Following the same reasoning as before one
infers that now $QD_a$ will enter the Coulomb blockade regime before $t=t_b$ because there are no transport channels 
within the bias window. However, the blockade is removed due to the second dot whose charging activates tunneling through 
$\mu^{(0)}_{g}(2,1)=\mu^{(0)}_{g}(1,2)$. This is an example of positive correlations between the two systems. Further discussions
can be found in a previous publication \cite{PhysRevB.82.085311}. 

\section{Thermoelectric Transport\label{S4}}

Until now we showed results for the charge transport driven by an electric 
bias of the leads due to different chemical potentials. 
The GME formalism allows also, in a 
straightforward way, the presence of a temperature bias.  Instead of different 
chemical potentials in the left and right leads, $\mu_{L,R}$, one can 
easily consider different temperatures, $T_{L,R}$, and calculate the resulting 
currents after switching on the contacts between the leads on the central system.  
Notice that, like in the case of an electric bias, there~is no requirement that the
temperature bias is small, such that the nonlinear thermoelectric regime is 
directly accessible \cite{Sanchez16}.  In addition, since the Coulomb interaction 
between electrons in the central system is already incorporated via the Fock 
space, the GME allows the inclusion of Coulomb blocking and other
electron correlation effects in the thermoelectric transport
\cite{Sierra14,Torfason13}.

The thermoelectric transport at nanoscale is a reach and active topic
within the context of the modern quantum thermodynamics, partly motivated
by novel ideas on the conversion of wasted heat into electricity, and
partly by the characterization of nanoscale system by methods complementary to pure
electric transport \cite{NJP2014}.  For example, an effect 
specific to nanosystems is the sign change of the thermoelectric current
or voltage when the electronic energy spectrum consists of discrete levels.
This~effect was predicted in the early 90' \cite{Beenakker92} and detected
experimentally for quantum dots \cite{Staring93,Dzurak93,Svensson12} and 
molecules \cite{Reddy1568}. This means that thermoelectric current in
a nanoelectronic system may flow from the hotter contact to the 
colder one, but also from the colder to the hotter, although the second
possibility might appear counter-intuitive.  

A simple explanation of this sign change of the current is that in
a nanoscale system with discrete resonances the current can be seen
as having two components, one carried by populated states above the
Fermi energy, and another one carried by depopulated states below it.
By analogy with a semiconductor, the former states correspond to electrons
in the conduction band and the later states to holes in the valence band.
Whereas an electric bias drives the electric currents due to particles
and holes in the same direction, such that they always add up, a thermal
bias drives them in opposite directions, such that the net current is
their difference, which can be positive, negative, or zero. 

We can describe this effect with the GME, first assuming a simple model with 
unidimensional and discretized leads, and just a single site in between them as  
central system.  By using the Markov approximation one can show analytically that
the current in the leads, in the steady state, are obtained as \cite{Torfason13}
\begin{equation}
J_{L,R}=\frac{1}{\tau^2}\frac{V_L^2V_R^2}{V_L^2+V_R^2}\left[f_L(E)-f_R(E)\right]\ ,
\label{JLR}
\end{equation}
where$V_{L,R}$ are the coupling parameters of the leads with the central site, 
$\tau$ is the hopping energy on the leads, and $E$ is the energy of the 
central site.  We see that the sign of the current depends on the difference 
between the Fermi energies in the leads at the resonance energy,
\begin{equation}
f_l(E)=\frac{1}{e^{(E-\mu_l)/k_BT_l}+1}\ , \ \ l=L,R \ .
\end{equation}

Thus, in the presence of a thermal bias, say $T_L>T_R$, but in the absence
of an electric bias, i.e., $\mu_L=\mu_R$, the current is zero and changes
sign around $\mu_l=E$.  In addition, the current may also vanish if the
chemical potential in the leads is sufficiently far from the resonance
such that the two Fermi functions are both close to zero or one. Which
means that if the central system has more resonant energies
the current may also change sign when $\mu_l$ is somewhere between two 
of them.

\begin{figure}[t] 
\includegraphics[width=0.45\textwidth]{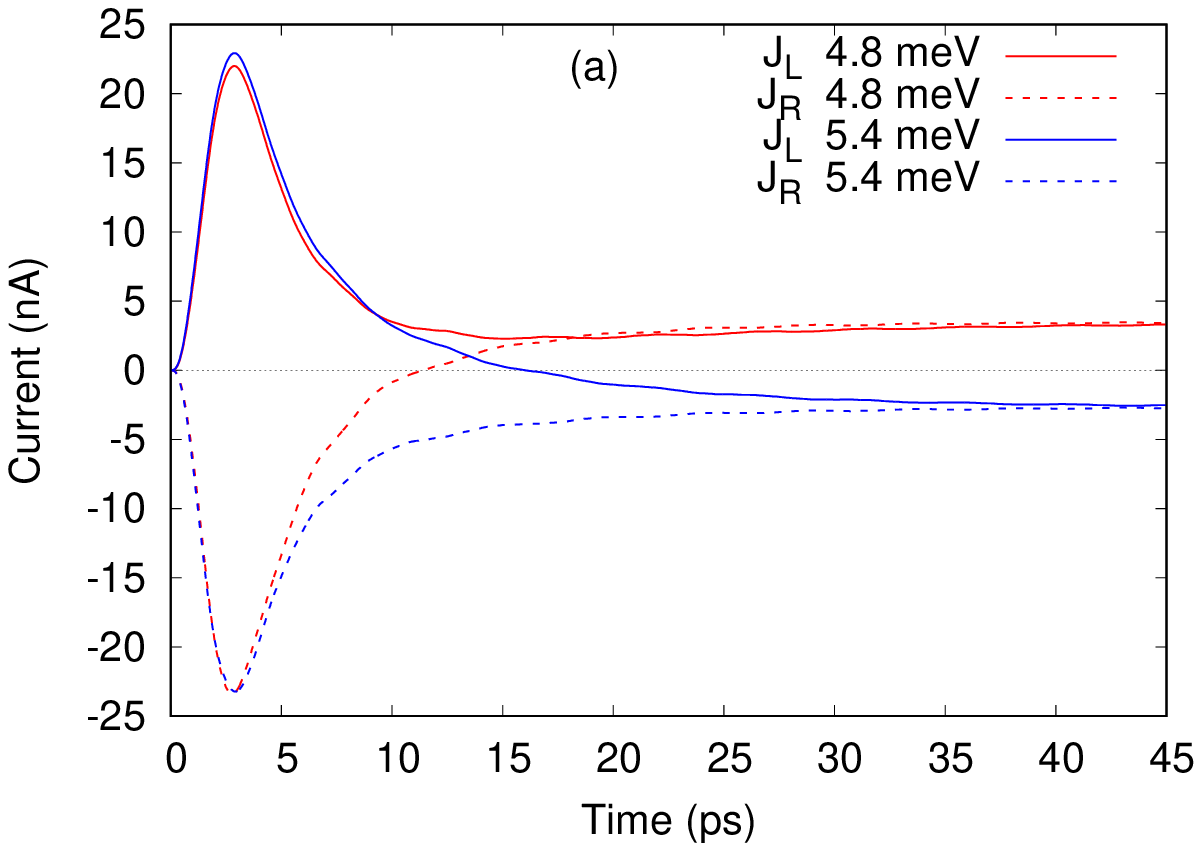}
\includegraphics[width=0.45\textwidth]{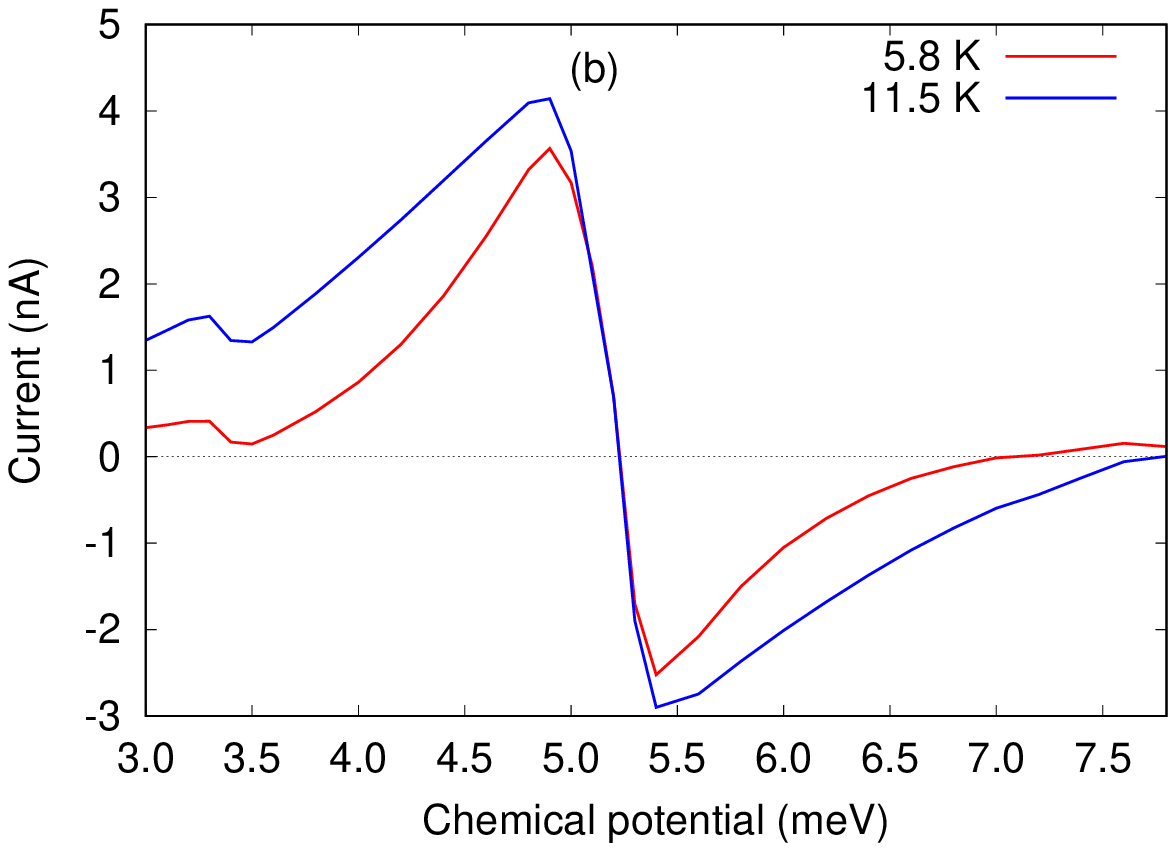}
\caption{ (\textbf{a}) The time evolution of the currents in the left
and right leads, $J_{L,R}$, driven by a temperature bias where $T_L=5.8$
K and $T_R=0.58$ K. With red color the results for the chemical potential
$\mu_L=\mu_R=48$ meV, and with blue color for $\mu_L=\mu_R=54$ meV.
In the steady state the currents have opposite sign;  (\textbf{b}) The current
in the steady state for two different temperatures of the left lead,
$T_L=5.8$ K (red) and $T_L=11.5$ K (blue), for variable chemical
potentials $\mu_L=\mu_R$.  } 
\label{Fig-thermo} 
\end{figure}

In Figure \ref{Fig-thermo} we show an example of thermoelectric currents 
calculated with the GME, using the same model as in Section \ref{S3}.  
The lowest single-particle levels having energies $\varepsilon_1=0.375$ meV and 
$\varepsilon_2=3.37$ meV are followed by the two-particle singlet 
state with $E_s=5.39$ meV and triplet with $E_t=5.62$ meV, and then 
by another excited two-body state with zero spin with energy $E_x=10.5$ meV.  
We consider temperatures $k_BT_L=0.5$ meV and $k_BT_R=0.05$ meV in the
left and right lead, respectively (or $T_L=5.8$ K and $T_R=0.58$ K),
and equal chemical potentials. In Figure \ref{Fig-thermo}a one can see
the time dependence of the currents in the leads after they are coupled
to the central system, for two values of the chemical potentials, 4.8
meV and 5.4 meV, selected on each side of the singlet state.  Compared to
the results shown in Section \ref{S3} here we increased the coupling
parameters between the leads and the central system 1.4 times, such that
the steady state is reached sooner.

As predicted by Equation (\ref{JLR}), the currents in the steady state
have opposite sign.  But in fact, as~shown by the red curve of Figure 
\ref{Fig-thermo}b, here we do not resolve the energy interval between
the singlet and triplet states with $k_BT_L > E_t-E_s=0.23$ meV, such
that we obtain one single (common) sign change for these two levels (or
``resonances''). Next, by increasing the chemical potential within the larger gap
between $E_s$ and $E_x$ the current in the steady state approaches zero 
and changes sign again, for $\mu_l \approx 7.0$ meV, and
for  $\mu_l \approx 7.8$ meV when the temperature of the hot lead is 
doubled, $T_L=11.5$ K.

By varying the chemical potential below the singlet energy $E_s$ we
obtain a similar decreasing trend of the current, except that now there
is no sign change close to the energy $\varepsilon_2=3.37$ meV, but~only
a succession of minima and maxima. The reason is the level broadening due
to the coupling of the central system with the leads \cite{Torfason13}.
Still, from such data one can observe experimentally the charging energy,
as the interval between consecutive maxima, or minima, or mid points
between them \cite{Reddy1568}.

In the present review we show only the thermoelectric current, which corresponds
to a short-circuit experimental setup, i.e., a circuit without a load.  To obtain
a voltage with the GME method one has to simulate a load by considering also a 
chemical potential bias. Thus, one can obtain the open-circuit voltage, which
corresponds to that electric bias $\mu_R-\mu_L$ which totally suppresses the 
thermoelectric current, or the complete I-V characteristic of the 
``thermoelectric device''.  Interestingly, the sign change of the thermoelectric
current or voltage can also be obtained by increasing the temperature of the 
hot lead, while keeping the other lead as cold as possible 
\cite{Svensson13,Sierra14,Zimbovskaya15,Stanciu15}.  

A novel example of sign reversal of the thermoelectric current has
been recently predicted in tubular nanowires, either with a core-shell
structure or made of a topological insulator material, in~the presence of
a transversal magnetic field \cite{Erlingsson17}.  In this case the energy spectra are
continuous, but~organized in subbands which are nonmonotonic functions
of the wavevector along the nanowire, yielding a transmission function
nonmonotonic with the energy, and the reversal of the thermoelectric
current, even in the presence of moderate perturbations \cite{Thorgilsson17,Erlingsson18}.

\section{Electron Transport Through Photon Cavities\label{S5}}
\unskip
\subsection{The Electron-Photon Coupling}
From the beginning our effort to model electron transport through a nano scale system placed in a 
photon cavity has been geared towards systems based on a two-dimensional electron
gas in GaAs or similar heterostructures. We have emphasized intersubband transitions in the 
conduction band, active~in the terahertz range, in anticipation of experiments in this promising
system \cite{Zhang1005:2016}.

Here, subsystem $S_1$ is a two-dimensional electronic nanostructure placed in a static (classical) external magnetic field.
The~leads are subjected to the same homogeneous external field.
The~electronic nanostructure, via split-gate configuration, is parabolically confined in the $y$-direction with
a characteristic frequency $\Omega_0$. The ends of the  nanostructure in the $x$-direction at $x = \pm L_x / 2$ are etched, 
forming a hard-wall confinement of length $L_x$. The external classical magnetic field is given by $\mathbf B=B\mathbf{\hat{z}}$
with a vector potential $\mathbf A=(-By,0,0)$. The single-particle Hamiltonian reads:
\begin{eqnarray}\nonumber
      {\hat h}_{S_1}^{(0)}&=& \frac{1}{2m}\Big(\mathbf p + q\mathbf A\Big)^2 + \frac12m\Omega_0^2y^2\\ \label{eq:shr}
      &=& \frac{1}{2m} p_x^2 + \frac{1}{2m} p_y^2 + \frac12 m\Omega_w^2 y^2 + i\omega_c y p_x\ ,
\end{eqnarray}
where $m$ is the effective mass of an electron, $-q$ its charge, $\mathbf p$ the canonical momentum operator, 
$\omega_c = qB/m$ is the cyclotron frequency and $\Omega_w = \sqrt{\omega_c^2+\Omega_0^2}$ is the modified parabolic
confinement. The spin degree of freedom is included with either a Zeeman term added to the 
Hamiltonian \cite{2016arXiv161109453G}, or~with Rashba and Dresselhaus spin orbit interactions, 
additionally \cite{ARNOLD2014170}.

$H_{S_2}$ is simply the free field photon term for one cavity mode and by ignoring the zero point energy can be written as
$H_{S_2}=\hbar\omega_p a^\dagger a$ where $\hbar \omega_p$ is the single photon energy
and $a$ ($a^\dagger$) is the bosonic annihilation (creation) operator. The electron-photon interaction
term $V_{{\rm el-ph}}$ can be split into two terms $V_{{\rm el-ph}}=V_{{\rm el-ph}}^{(1)} + V_{{\rm el-ph}}^{(2)}$
where
\begin{eqnarray}\label{eq:H1}
      V_{{\rm el-ph}}^{(1)} &:=& \sum_{ij}\sum_{i,j}\left\langle\psi_i\left|\frac{q}{2m}\Big(\boldsymbol\pi\cdot 
      \mathbf A_{{\rm EM}}+\mathbf A_{{\rm EM}}\cdot\boldsymbol\pi\Big)\right|\psi_j\right\rangle c_i^\dagger c_j  \\\label{eq:H2} 
      V_{{\rm el-ph}}^{(2)} &:=& \sum_{ij}\sum_{i,j}\left\langle\psi_i\left|\frac{q^2}{2m}\mathbf A_{{\rm EM}}^2\right|\psi_j\right\rangle c_i^\dagger c_j,
\end{eqnarray}
with $\boldsymbol \pi\equiv \mathbf p + q\mathbf A$ the mechanical momentum. The term in Equation (\ref{eq:H1})
is the paramagnetic interaction, whereas the diamagnetic term is defined by Equation (\ref{eq:H2}). 
By assuming that the photon wavelength is much larger than characteristic length scales of the system one can 
approximate the vector potential amplitude to be constant over the electronic system. Let us stress here that, in contrast to
the usual dipole approximation, we will not omit the diamagnetic electron-photon interaction term. Then the vector potential is written 
as:
\begin{equation}\label{eq:AEM}
      \mathbf A_{{\rm EM}} \simeq \mathbf{\hat e} A_{{\rm EM}}\left(a+a^\dagger\right) = \mathbf{\hat e} \frac{\mathcal E_c}{q\Omega_wa_w}\left(a+a^\dagger\right) \ ,
\end{equation}
where $\mathbf{\hat e}$ is the unit polarization vector and $\mathcal E_c \equiv qA_{{\rm EM}}\Omega_wa_w$ is the electron-photon 
coupling strength. For a 3D rectangular Fabry Perot cavity we have $A_{{\rm EM}}= \sqrt{\hbar/(2\omega_pV\epsilon_0)}$ where
$V$ is the cavity volume. Linear polarization in the $x$-direction is achieved for a $\mathbf{TE}_\mathrm{011}$ mode, 
and in the $y$-direction with a $\mathbf{TE}_\mathrm{101}$~mode.

Using the approximation in Equation (\ref{eq:AEM}), the expressions for the electron-photon interaction
in Equations (\ref{eq:H1}) and (\ref{eq:H2}) are greatly simplified by pulling $\mathbf A_{{\rm EM}}$ in front of the integrals.
For the paramagnetic term, we get
\begin{equation}\label{eq:Hint1}
      V_{{\rm el-ph}}^{(1)} \simeq \mathcal E_c\left(a+a^\dagger\right)\sum_{ij} g_{ij} c_i^\dagger c_j \ .
\end{equation}
where we introduced the dimensionless coupling between the electrons and the cavity mode 
\begin{equation}\label{eq:gij}
      g_{ij} = \frac{a_w}{2\hbar} \mathbf{\hat e}\cdot \int d{\mathbf r} \left[ \psi_{i}^*({\mathbf
      r}) \left\{ \boldsymbol{\pi}\psi_{j}({\mathbf r}) \right\} 
      +  \left\{ \boldsymbol{\pi} \psi_{i}^*({\mathbf r})
      \right\} \psi_{j}({\mathbf r})
      \right] \ .
\end{equation}

As for the diamagnetic term, we get
\begin{equation}
      V_{{\rm el-ph}}^{(2)} \simeq \frac{\mathcal E_c^2}{\hbar\Omega_w} \left [ \left (a^\dagger a + \frac{1}{2}\right )+
      \frac{1}{2} \left (a^\dagger a^\dagger + aa\right )\right ]\mathcal N^e \ ,
\end{equation}
where $\mathcal N^e$ is the number operator in the electron Fock space. Note that $V_{{\rm el-ph}}^{(2)}$ does not depend
on the photon polarization or geometry of the system in this approximation. We do not use the rotating wave 
approximation as in our multilevel systems even though a particular electron transition could be in resonance
with the photon field we want to include the contribution form others not in resonance.

For the numerical diagonalization of $H_S$ we shall use the lowest $N_{{\rm mesT}}\ll N_{{\rm mes}}$ IMBS of $H_{S_1}$ 
and photon states containing up to $N_{{\rm EM}}$ photons, resulting in a total of $N_{{\rm mesT}}\times(N_{{\rm EM}}+1)$ 
states in the 'free' basis $\{|\nu,j\rangle\}$. 

\subsection{Results}
Groups modeling the near resonance interaction of one cavity mode with a two level electronic
system have expressed the importance of using a large enough, or the correct type, of a photon
basis in the strongly interacting regime \cite{Feranchuk96:4035,Li09:044212}. In many level systems
where wavefunction and geometric effects are accounted for our experience is that convergence in
numerical diagonalization is more sensitive to proper truncation of the electronic sector of the
Fock many-body space. This reflects the polarizability of the electric charge by a cavity field
in the construction of the photon-dressed electronic states. At the same time the inclusion of the 
diamagnetic interaction curbs the need for states with a very high photon 
number \cite{Jonasson2011:01,PhysRevE.86.046701,Gudmundsson16:AdP}.

\begin{figure}[t]
	\centerline{\includegraphics[width=0.82\textwidth]{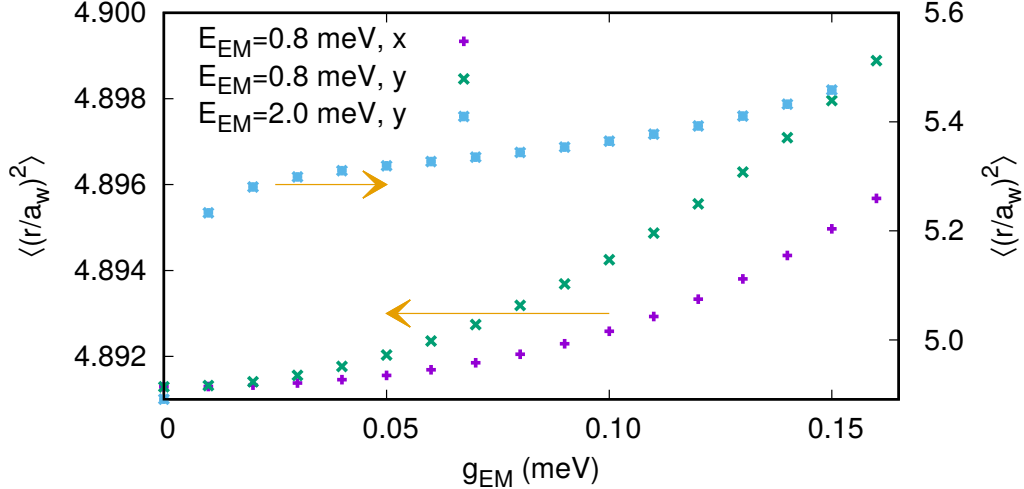}}
	\caption{The expectation value $\langle (r/a_w)^2 \rangle$ for the first photon
    replica of the two-electron ground state in the
    closed system at $t=0$ for $x$- and $y$-polarization of the photon field.
    $\hbar\omega =2.0$ meV, $B=0.1$ T. Two parallel quantum dots are embedded in 
    the central system.}
	\label{Fig-vg07}
\end{figure}

The polarizability of the first photon replica of the two-electron ground state is displayed in
Figure~\ref{Fig-vg07} as a function of $g_\mathbf{EM}$, the photon energy $\hbar\omega$ 
and its polarization \cite{Gudmundsson16:AdP}.
The polarizability is nonlinear, anisotropic,  and largest for the cavity photon close to a resonance with the
confinement energy in the $y$-direction.

\begin{figure}[t]
	\centerline{\includegraphics[width=0.78\textwidth]{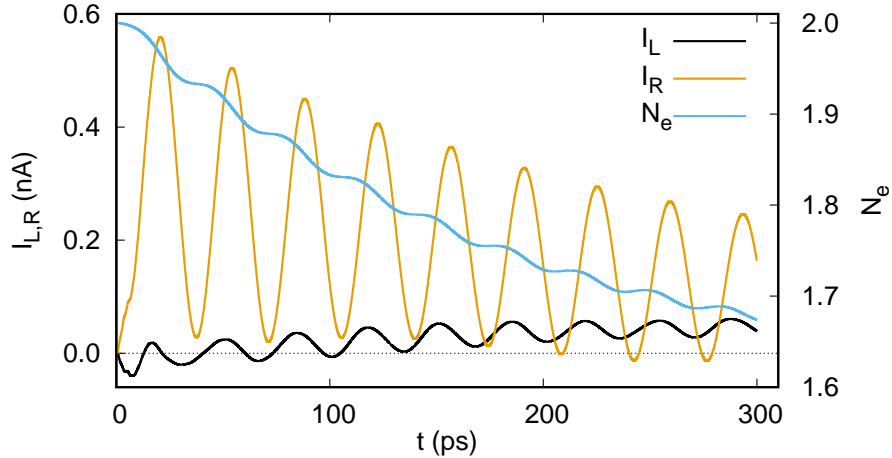}}
	\caption{The left (black) and right (gold) currents and the mean
	electron number (blue) for initially fully entangled Rabi-split singlet two-electron states 
    as the interacting system discharges in the transient regime. $\hbar\omega =2.0$ meV,
    $B=0.1$ T. Two parallel quantum dots are embedded in the central system.}
	\label{Fig-vg06}
\end{figure}

A Rabi oscillation of two electrons in the double quantum dot system embedded in the short quantum wire
leads to oscillating charge with time in the system. The oscillating probability of charge presence in 
the contact areas of the short wire thus lead to oscillations in the current leaving the system through
the left and right leads \cite{doi:10.1021/acsphotonics.5b00115}, see Figure \ref{Fig-vg06}.
Alternatively, one may view this as the consequence of the Rabi resonance entangling two states
with different tunneling probability to the leads. 

In the transient or the late transient regime we have used the non-Markovian GME to investigate
several results: Thorsten Arnold et al.\ used a time-convolution-less (TCL) version of the GME to study
the effects of magnetic field and photons \cite{PhysRevB.87.035314} on the transport of interacting electrons through 
a quantum ring with spin-orbit interactions in a photon cavity with circular \cite{ARNOLD2014170}
and linear polarization~\cite{Arnold2014}. Aharonov-Bohm oscillations were established in the time-dependent
transport through a ring structure with additional vortexes in the contact region of the quantum wire.
$x$-polarized photons with energy 0.3 meV attenuate the Aharonov-Bohm oscillations over a broad range
of magnetic field, but $y$-polarized photons influence the transport in a more complex fashion. 
The oscillations are generally attenuated, but one oscillation peak is split and the charge current is 
enhanced at a magnetic field corresponding to a half-integer flux quantum \cite{PhysRevB.87.035314}.
With the spin-orbit interactions the spin polarization and the spin photo currents of the quantum ring are 
largest for circularly polarized photon field and a destructive Aharonov--Casher (AC) phase interference. 
The dip in the charge current caused by the destructive AC phase becomes threefold under the circularly 
polarized photon field as the interaction of the angular momentum of the electron and the spin angular 
momentum of the light create a many-body level splitting \cite{ARNOLD2014170}.
The detailed balance between the para- and the diamagnetic electron-photon interactions has been
studied for an electron in the quantum ring structure when excited by a short classical dipole
pulse \cite{2040-8986-17-1-015201}.   

Nzar Rauf Abdullah et al.\ have used the GME formalism to investigate photon assisted transport~\cite{0953-8984-25-46-465302}, 
photon mediated switching in nanostructures \cite{Abdullah10:195325,Abdullah2014254,ABDULLAH2016280}, the 
balancing of magnetic and forces caused by cavity photons \cite{0953-8984-28-37-375301},
cavity-photon affected thermal transport \cite{ABDULLAH2018199,doi:10.1021/acsphotonics.5b00532}, 
and the influence of cavity photons on thermal spin currents in a system with spin orbit interactions
\cite{ABDULLAH2018102,Abdullah_2018}.   

\section{Steady-State\label{S6}}

The investigation of the time dependent transport of electrons through a photon cavity 
soon made it clear that for the continuous model the inherent time scales can lead to 
relaxation times far beyond what is accessible with simple integration of the GME
\cite{0953-8984-25-46-465302,Abdullah2014254,Gudmundsson:2013.305,PhysRevB.87.035314,Gudmundsson12:1109.4728}.
The underlying cause for the diverse relaxation times is on one hand 
electron tunneling rates affected by the shape or geometry of the system
and the condition of weak coupling. Different many-body states can have a high or low probability for
electrons to be found in the contact areas of the central system. 
On the other hand are slow rates of FIR or terahertz active transitions, that are furthermore affected by
the geometry of the wavefunctions of the corresponding final and initial states. 
In addition, the cavity decay, or coupling to the environment, affects relaxation times
as we address below \cite{Nzar-2019-Rabi}. To avoid confusion it is important to remember that
we calculate the eigenstates of the closed central system, the interacting electron and photon system,
and the opening up of the system to the leads or the external photon reservoir is always a necessary
triggering mechanism for all transitions later in time, photon active or not.

\subsection{The Steady-State Limit}
In order to investigate the long-time evolution and the steady state of the central system under the influences
of the reservoirs we resort to a Markovian version of the GME, whereby we assume memory effects in the kernel 
of the GME (\ref{GME2}) to vanish, relinquishing the reduced density operator local in time enabling the approximation
\cite{2016arXiv161003223J}
\begin{equation}
      \int^\infty_0 ds\; e^{is(E_\beta-E_\alpha -\epsilon)} \approx \pi\delta (E_\beta-E_\alpha -\epsilon ),
\end{equation}
where a small imaginary principle part is ignored. We have furthermore assumed instant lead-system coupling at
$t=0$ with $\chi_l(t)=\theta (t)$, the Heaviside unit step function, in Equation (\ref{Htunnel}) for $H_T$.
In~order to transform the resulting Markovian equation into a simpler form
we use the vectorization operation~\cite{IMM2012-03274}, that stacks the columns of a matrix into a vector,
and its property
\begin{equation}
      \mathrm{vec}(A\rho B) = (B^T\otimes A)\mathrm{vec}(\rho )
\end{equation}
through which the reduced density matrix can always be moved to the right side
of the corresponding term, and a Kronecker product has been introduced with the
property $B\otimes A=\{B_{\alpha,\beta}A\}$. The Kronecker product of two $N_\mathrm{mes}\times N_\mathrm{mes}$
matrices results in a $N_\mathrm{mes}^2\times N_\mathrm{mes}^2$ matrix, and effectively the vectorization
has brought forth that the natural space for the Liouville-von Neumann equation is not the standard
Fock space of many-body states, but the larger Liouville space of transitions 
\cite{Weidlich71:325,Nakano2010,Petrosky01032010}. 

No further approximations are used to attain the Markovian master equation and due to the 
complex structure of the non-Markovian GME we have devised a general recipe published 
elsewhere~\cite{2016arXiv161003223J} to facilitate the analytical construction and the numerical
implementation. The~Markovian master equation has the form
\begin{equation}
       \partial_t\mathrm{vec}(\rho )={\cal L}\mathrm{vec}(\rho ),
       \label{markovianGEM}
\end{equation}     
and as the non-Hermitian Liouville operator ${\cal L}$ is independent of time
the analytical solution of Equation (\ref{markovianGEM}) can be written as
\begin{equation}
      \mathrm{vec}(\rho(t)) = \{{\cal U}[\exp{({\cal L}_\mathrm{diag}t)}] {\cal V} \}\mathrm{vec}(\rho(0)),
      \label{SolMarkovianGME}
\end{equation}
in terms of the matrices of the column stacked left ${\cal U}$, and the right ${\cal V}$ eigenvectors of 
${\cal L}$ 
\begin{equation}
      {\cal L}{\cal V} = {\cal V}{\cal L_\mathrm{diag}}, \quad\mbox{and}\quad 
      {\cal U}{\cal L} = {\cal L_\mathrm{diag}}{\cal U} ,
\end{equation}
obeying
\begin{equation}
{\cal U}{\cal V} = I, \quad\mbox{and}\quad 
{\cal V}{\cal U} = I.
\end{equation}

Special care has to be taken in the numerical implementation of this solution
procedure as many software packages use another normalization for ${\cal U}$
and ${\cal V}$. Calculations in the Liouville space using~(\ref{SolMarkovianGME}) are
memory (RAM) intensive, but bring several benefits: No time integration combined with
iterations is needed, thus time points can be selected with other criteria in mind.
The solution is thus convenient for long-time evolution, that is not easily accessible
with numerical integration. The complicated structure of the left and right eigenvector matrices for
a complex system with nontrivial geometry makes Equation  (\ref{SolMarkovianGME}) the best choice to
find the properties of the steady state by monitoring the limit of it as time gets very large.
The complex eigenvalue spectrum of the Liouville operator ${\cal L}$ reveals information about
the mean lifetime and energy of all active transitions in the open system, and the zero eigenvalue
defines the steady state. 

In the steady state the properties of the system do not change with time, but underneath the
{\lq\lq}quiet surface{\rq\rq} many transitions can be active to maintain it. The best experimental
probes to gauge the underlying processes are measurements of noise spectra for a particular
physical variable. They~are available through the two-time correlation functions of the respective 
measurable quantities.  For~a~Markovian central system weakly coupled to reservoirs the two-time
correlation functions can be calculated applying the Quantum Regression Theorem 
(QRT) \cite{0305-4470-14-10-013,Wallis-QO} stating that the the equation of motion for a two-time
correlation function has the same form as the Markovian master equation (\ref{markovianGEM}) for
the operator \cite{doi:10.1063/1.3570581}
\begin{equation}
      \chi (\tau ) = \mathrm{Tr_R}\left\{ e^{-iH\tau /\hbar} X\rho_\mathrm{T} (0) e^{+iH\tau /\hbar} \right\} , 
      \label{QRT-chi}     
\end{equation}
where $H$ is the total Hamiltonian of the system, $\rho_\mathrm{T}$ its density operator, 
and the trace is taken with respect of all reservoirs.
For photon correlations $X=a+a^\dagger$ as in \cite{Gudmundsson16:AdP}, or $X=Q\Lambda^l$ for
current correlations as in \cite{GUDMUNDSSON20181672}, where $Q=\sum_ic_i^\dagger c_i$ is the fermionic 
charge operator and $\Lambda^l$ is the Liouville dissipation operator for lead $l$. The structure of $\chi$
(\ref{QRT-chi}) indicates that the two-time correlation function is then
\begin{equation}
      \langle X(\tau )X(0)\rangle = \mathrm{Tr_S}\left\{X(0)\chi (\tau ) \right\} ,
      \label{Correlation-Heisenberg}
\end{equation}
with
\begin{equation}
      \mathrm{vec}(\chi (\tau)) = \{{\cal U}[\exp{({\cal L}_\mathrm{diag}t)}] {\cal V} \}\mathrm{vec}(\chi (0)) .
\end{equation}   
The left side of Equation (\ref{Correlation-Heisenberg}), the two-time correlation function, is written in the 
Heisenberg picture, in~contrast to the Schr\"odinger picture used elsewhere in the article. 
The Fourier spectral density for the photon two-time correlation function is denoted by $S(E)$,
and for the current-current correlation the corresponding Fourier spectral density denoted by
$D_{ll'}(E)$, where $l$ and $l'$ refer to $L$ and $R$, the Left and Right leads.

\subsection{Results}
To date we have used the Markovian version of the master equation to investigate properties
of the steady state, and how the system with electrons being transported through a photon
cavity reaches it. We assume GaAs parameters with effective mass $m=0.067m_e$, effective relative 
dielectric constant $\epsilon_r=12.3$, and effective Land{\'e} $g$-factor $g=-0.44$.
The characteristic energy of the parabolic confinement of the semi-infinite leads and the
central system in the $y$-direction is $\hbar\Omega_0=2.0$ meV. The~length of the short 
quantum wire is $L_x$, and the overall coupling coefficient for the leads to the system 
is $g_{LR}a_w^{3/2}=0.124$ meV.

\begin{figure}[t]
	\centerline{\includegraphics[width=0.98\textwidth]{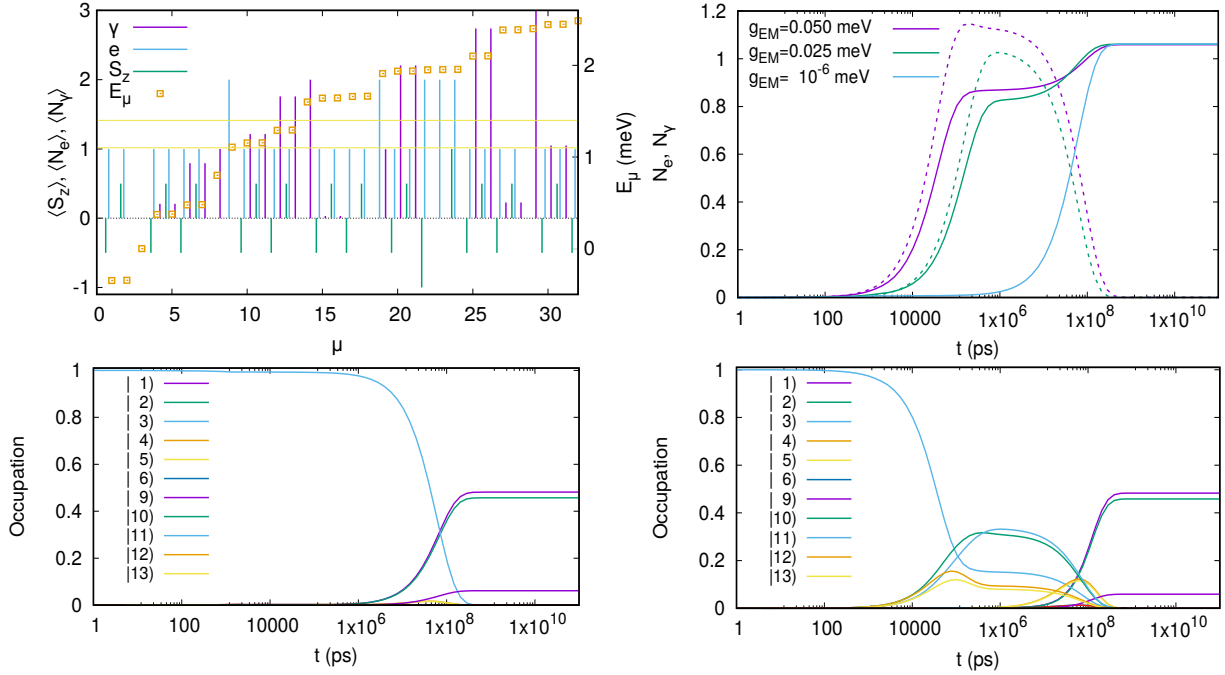}}
	\caption{(\textbf{upper left}) For the closed system as functions of the number of the eigenstate $\mu$, 
		     the~many-body energy (squares), the mean photon ($\gamma$) and electron content ($e$), 
		     and the mean spin $z$-component ($S_z$). The horizontal yellow lines represent the 
		     chemical potentials of the left ($\mu_L$) and right leads ($\mu_R$) when the system
		     will be coupled to them. (\textbf{upper right}) The mean electron (solid) and photon number (dashed) 
		     in the central system as a function of time. The mean occupation of the many-body eigenstates 
		     of the system for $g_\mathrm{EM}=1\times 10^{-6}$ meV (\textbf{lower~left}), and \mbox{$g_\mathrm{EM}=0.05$ meV}
		     (\textbf{lower~right}). $V_g=-1.6$ mV, $\hbar\omega = 0.8$ meV, $x$-polarization, $\kappa = 1\times 10^{-5}$ meV, 
		     $L_x=150$ nm, and~$B=0.1$ T. No quantum dots in the short wire.}
	\label{Fig-vg01}
\end{figure}

We start with a central system made of a finite parabolic quantum wire without
any embedded quantum dots. Figure \ref{Fig-vg01} demonstrates that the approach to build 
and solve the Markovian master Equations (\ref{markovianGEM})--(\ref{SolMarkovianGME}) works for an 
interacting system with 120 many-body states participating in the transport \cite{Gudmundsson16:AdP_10}. 
The upper right panel displays the properties of the lowest 32 many-body states at the plunger gate
voltage $V_g=-1.6$ mV. With $\mu_L=1.4$ meV and $\mu_L=1.1$ meV there are 8 states below the
bias window and five states within it. In the bias window is one spin singlet two-electron state 
(the~two-electron ground state) and two spin components of two one-electron states 
with a non-integer mean photon content indicating a Rabi splitting.  
The upper left panel of Figure \ref{Fig-vg01} show the mean electron and photon 
numbers in the central system when it is initially empty. With a very low coupling, $g_\mathrm{EM}=1\times 10^{-6}$ meV,
between the electrons and photons, the charging is very slow with the probability approaching unity around
$t\approx 10^8$ s. With increasing $g_\mathrm{EM}$ the charging becomes faster, and during the phase 
the mean photon number in the system rises. The lower panels of Figure \ref{Fig-vg01} reveal what
is happening. With the low photon coupling (lower left panel) electrons tunnel non-resonantly into
the two spin components of the ground state, $|1)$ and $|2)$ as the vacuum state $|3)$ looses occupation,
and~to a small fraction the two-electron state $|9)$ gets occupied.  When the coupling of the electrons and
the photons is not vanishingly small (lower right panel) the charging of the system takes a different
rout. The finite $g_\mathrm{EM}$ allows the incoming electron to enter the Rabi-split one-electron states
in the bias window as these are a linear combination of electron states with a different photon number.
This~explains the growing mean number of photons in the system for intermediate times. These states are eigenstates 
of the central system, but not of the open system, so at a later time they decay into the the one- and 
two-electron ground states as before bringing the system into the same Coulomb blocked steady state as before.
We thus observe electromagnetically active transitions in the system in an intermediate time 
regime \cite{Gudmundsson16:AdP_10}.  

\begin{figure}[t]
	\centerline{\includegraphics[width=0.98\textwidth]{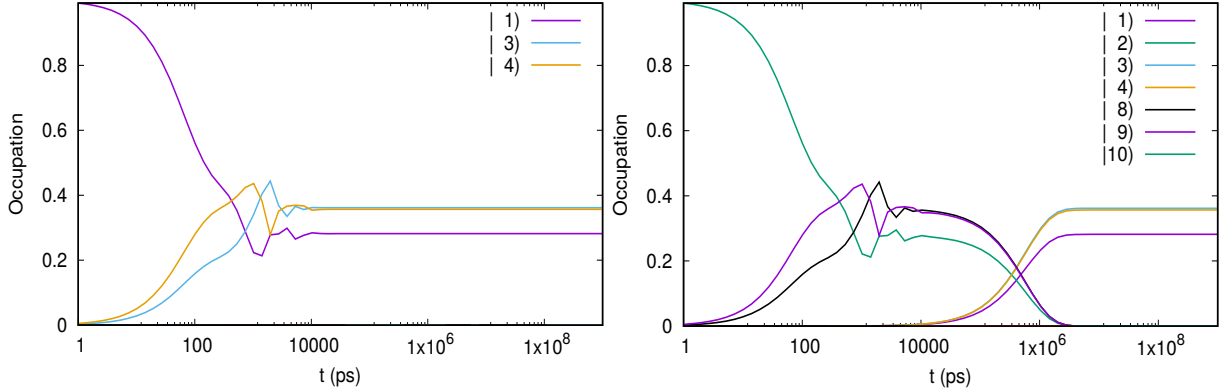}}
	\caption{The mean occupation of the many-body eigenstates of the system when the initial
	         state is the ground state $|1)$ (\textbf{left}), or the first photon replica of the ground
             state $|2)$ (\textbf{right}). $g_\mathrm{EM}=0.05$ meV. $V_g=-2.0$ mV, $\hbar\omega = 0.8$ meV, 
             $x$-polarization, $L_x=150$ nm, and $B=0.1$ T. Two parallel quantum dots embedded in 
             the short wire, but no photon reservoir.}
	\label{Fig-vg03}
\end{figure}

The on-set of the steady state regime is difficult to judge only from the shape of the 
charge being accumulated in the system or the current through or into it as a function
of time \cite{2016arXiv161109453G}. For a system of two parallel quantum dots embedded in
a short quantum wire ($L_x=150$) nm the charging and the current as functions of time
look the same (see Figures 4 and 5 in ref.\ \cite{2016arXiv161109453G}), but when the occupation 
of the eigenstates of the closed system is analyzed,
see Figure \ref{Fig-vg03},
a clear difference is seen for the approach to the steady state depending on whether the initial state
contains only one or no photon \cite{2016arXiv161109453G}. In the case of neither photon nor an electron in the cavity
initially an electron tunnels into the system into the two spin components of the one-electron ground state,
which happens to be in the bias window for $V_g=-2.0$ mV. Thus, the steady state is a combination of the 
empty state and these two one-electron states. In the case of one photon and no electron initially in the system
an electron tunnels non-resonantly into the 1-electron states $|8)$ and $|9)$ with energy slightly below 2 meV,
and thus well above the bias window. The mean photon content of these states is close to unity and at a later
time the electron ends up in the two spin components of the one-electron ground state via a radiative 
transition \cite{2016arXiv161109453G}). 

Note that the {\lq\lq}irregularly{\rq\rq} looking structure around $t\approx 2000$ ps will be addressed below. 
Please~note that the numbering of interacting many-body state depends on the structure of the system, and the 
plunger gate voltage $V_g$.

\begin{figure}[t]
	\centerline{\includegraphics[width=0.98\textwidth]{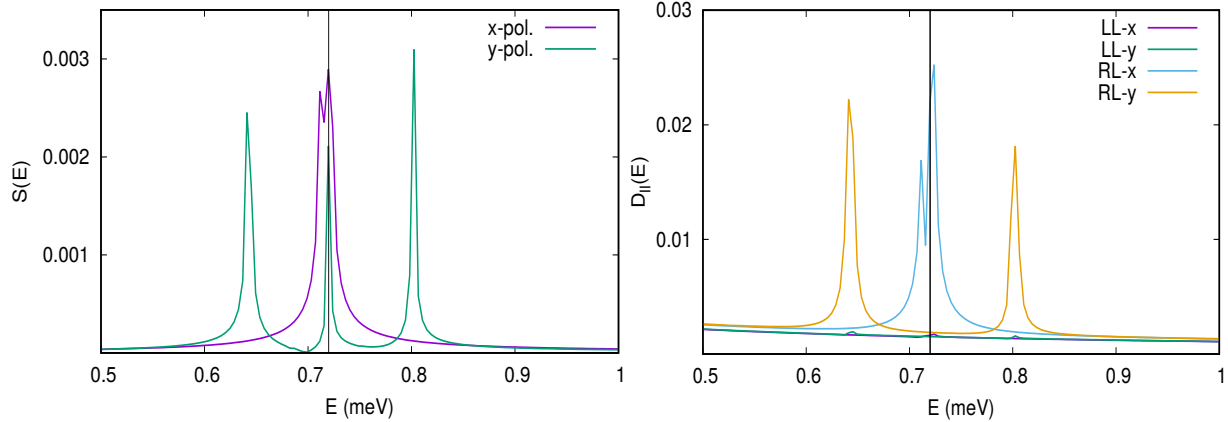}}
	\caption{The spectral density $S(E)$ of the emitted cavity radiation for the central system
		     in a steady state (\textbf{left}), and the spectral densities for the current--current correlations  
		     $D_{ll'}(E)$ (\textbf{right}). \mbox{$g_\mathrm{EM}=0.1$ meV}, $V_g=-2.0$ mV, $\hbar\omega = 0.72$ meV,
             $\kappa = 1\times 10^{-3}$ meV, and $L_x=150$ nm. 
             Two parallel quantum dots embedded in the short wire.}
	\label{Fig-vg02}
\end{figure}

In the steady state all the mean values of the open system have reached a constant value.
In order to query about the active underlying processes it is necessary to calculate the 
spectral densities of the photon or current correlations. We present these for the central 
system consisting of a short quantum wire ($L_x=150$) nm with two embedded quantum dots in
Figure \ref{Fig-vg02} (see refs.\ \cite{2017arXiv170603483G} and \cite{GUDMUNDSSON20181672}).
Importantly we show in Ref.\ \cite{Gudmundsson16:AdP} that both the paramagnetic and the diamagnetic
electron-photon interactions can lead to a Rabi resonance. The resonance for the diamagnetic interactions
is much smaller, but the symmetry of the two parallel quantum dots leads to selection rules where for
$x$-polarized cavity photon field the paramagnetic interaction is blocked, but both are present for the
$y$-polarized field. Here, the active states are the one-electron ground state and the first excited one-electron
state, with which the first photon replica of the ground state interacts for $\hbar\omega = 0.72$ meV. 
The spectral density of the photon-photon two-time correlation function, $S(E)$ seen in the left panel of
Figure \ref{Fig-vg02} shows one peak at the energy of the cavity mode $\hbar\omega = 0.72$ meV, and~two side 
peaks for the $y$-polarization. The central peak is the ground state state electroluminescence and the side
peaks are caused by the Rabi-split states \cite{PhysRevLett.116.113601,PhysRevA.80.053810,2017arXiv170603483G}. 
Here, we observe the ground state electroluminescence even though the electron-photon coupling is not in 
the ultra strong regime, as~we diagonalize the Hamiltonian in a large many-body Fock space instead 
of applying conventional perturbative calculations.  

For the $x$-polarized cavity field we find a much weaker ground state electroluminescence caused by the
diamagnetic electron-photon interaction \cite{2017arXiv170603483G}. In addition, we identify these effects
for the fully interacting two-electron ground state, where they are partially masked by many concurrently 
active transitions. The spectral density for the current--current correlation functions $D_{ll'}(E)$ displayed
in the right panel of Figure \ref{Fig-vg02} show only peaks at the Rabi-satellites, as could be 
expected \cite{GUDMUNDSSON20181672}. An~inspection of $D_{ll'}(E)$ over a larger range of energy reveals
more transitions active in maintaining the steady state, both radiative transitions and non 
radiative \cite{GUDMUNDSSON20181672}. Moreover, we notice that when the steady state is not in a
Coulomb blocking regime the spectral density of the current-current correlations always shows
a background to the peaks with a structure reminiscent of a $1/f$ behavior, that is known in
multiscale systems.

\begin{figure}[t]
	\centerline{\includegraphics[width=0.64\textwidth]{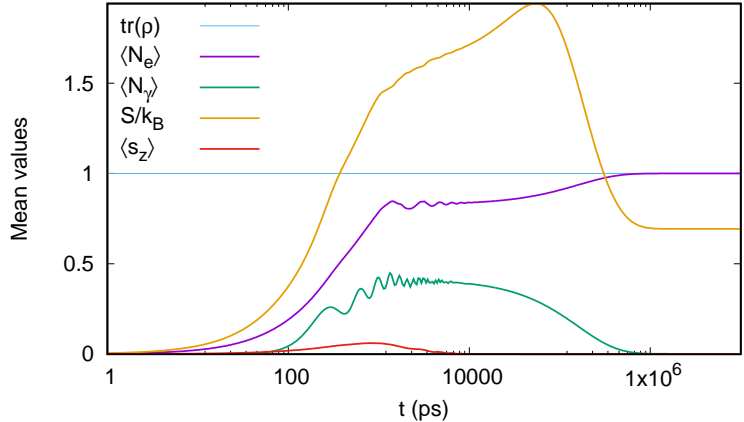}}
	\caption{The mean electron ($e$), photon ($\gamma$), $z$-component of the spin ($S_z$), trace of the
	         reduced density matrix, and the R{\'e}niy-2 entropy ($S$) as functions of time.
             $\hbar\omega = 0.373$ meV,   $x$-polarization,  $\kappa = 1\times 10^{-5}$~meV, 
             $g_\mathrm{EM}=0.05$ meV, and $L_x=180$ nm.
             Two asymmetrically embedded quantum dots in the short~wire.}
	\label{Fig-vg04}
\end{figure}

An {\lq\lq}irregularly{\rq\rq} looking structure in the mean occupation, the current current, and the mean number
of electrons and photons. This is a general structure seen in all types of central system we have investigated
in the continuous model. In Figure \ref{Fig-vg04} we analyze it in a short parabolically confined quantum wire
of length $L_x=180$ nm with two asymmetrically embedded quantum dots \cite{2018arXiv180906930G}. 
An increased number of time points on the logarithmic scale shows regular oscillations. 
A careful analysis reveals two independent oscillations: A spatial charge oscillation 
between the quantum dots with the Rabi frequency in the system, and a still slower
nonequilibrium oscillation of the spin populations residing as the system is brought to a
steady state \cite{2018arXiv180906930G}. 

The steady-state Markovian formalism has been used to investigate oscillations
in the transport current as the photon energy or the electron--photon coupling
strength are varied with or without flow of photons from the external 
reservoir \cite{2019arXiv190303655A,2019arXiv190404888A}. Moreover, the formalism
has been used to establish the signs of the Purcell effect \cite{PhysRev.69.681} in the transport 
current \cite{Nzar-2019-Rabi}. 

In light of the experimental interest of using a two-dimensional electron gas in a GaAs
heterostructure \cite{Zhang1005:2016} we have calculated the exact matrix elements for 
the electron-photon interaction taking into account the spatial variation of the vector field 
$\mathbf{A}$ of the electronic system. This is a small correction in most cases but may be important
when studying high order transitions or nonperturbational effects caused by the photon field.
This has led us to discover a very slow high order transition between the ground states of
two slightly dissimilar quantum dots \cite{2019arXiv190510883G}.

The fist steps have been taken to investigate thermoelectric effects 
in the central system coupled to cavity photons, in the steady state.
In a short quantum wire with one embedded quantum dot in the resonant regime, an inversion of 
thermoelectric current is found caused by the Rabi-splitting. The~photon field can change both
the magnitude and the sign of the thermoelectric current induced by the temperature
gradient in the absence of a voltage bias between the leads \cite{2018arXiv181205665A}.

\section{Summary\label{S7}}

It goes without saying that as transport experiments at nanoscale become more involved the formal tools must be 
suitably extended or adapted. In particular, the unavoidable charging and correlation effects at finite bias pushed 
the theoretical calculations from the very convenient single-particle (or at most mean-field) Landauer--B\"{u}ttiker 
picture to the complicated many-body perturbation theory of the non-equilibrium Keldysh--Green's functions 
\cite{StefanucciBook:2013}.

Here we summarized some results on time-dependent transport in open interacting systems which argue for the similar
idea that if one looks for transient effects and dynamics of excited states the simple rate equation approach must 
be extended to the non-Markovian generalized master equation.     

 The GME we used in all examples is constructed and solved w.r.t\ the exact many-body states of the central 
open system and can be therefore implemented numerically without major changes to study both Coulomb-interacting   
and hybrid systems where the fermion-boson interaction is crucial, like QD-cavity systems or nano-electromechanical 
systems. A consistent derivation of the GME the full knowledge of the eigenvalues and eigenfunctions of complicated 
interacting Hamiltonian (e.g.,~cavity-coupled systems must be described by `dressed' states). With very few 
exceptions coming from quantum optics (i.e., the Jaynes-Cummings model for two-level or $\Lambda$ and $V$ three-level
 systems) such a task can only be achieved via numerically exact diagonalization of large matrices, especially~for
electron-photon systems. To bypass this difficulty we proposed and successfully used a stepwise diagonalization procedure.

The dynamics of excited states in a quantum wire, the onset of current-current correlations for a pair of electrostatically 
coupled quantum dots and thermoelectric effects were presented within a simple lattice model which however captures the 
relevant physics.  

 When turning to QED-cavity system we developed the GME within a continuous model which accounts for the geometrical
details of the sample and of the contact regions. Moreover, the calculations were performed by taking into account both the 
paramagnetic and diamagnetic contributions to the electron-photon coupling and without relying on the rotating-wave approximation.
 This is an important step beyond the Jaynes--Cummings model. Also, the number of many-body stated needed in the calculations 
increased considerably. Thus, the accuracy of the stepwise numerical diagonalization had to be carefully discussed. 
Finally, for systems with long relaxation time a Markovian version of GME was proposed and implemented via a clever 
vectorization procedure.

{We end this review by pointing out possible improvements of the GME method and some of its future applications.
At the formal level, perhaps the most challenging upgrade is the inclusion of time-dependent potentials describing
laser pulses or microwave driving signals. Provided this is successfully achieved, one could study transport through
driven nano-electromechanical systems (NEMS) or the physics of Floquet states emerging in strongly driven systems 
\cite{PhysRevLett.115.133601,PhysRevLett.121.043603}. Let us mention here that at least for closed systems (i.e., not 
connected to particle reservoirs) studies based on Floquet master equations for two-level system are already available 
\cite{PhysRevA.96.041802,PhysRevA.92.022349}. As for more immediate applications we aim at the theoretical modeling of 
transport in Tavis--Cummings systems, motivated~by the recent observation of state readout in a system of distant coupled 
quantum dots individually connected to a pair of leads and interacting via cavity photons \cite{PhysRevApplied.9.014030}.}

\begin{acknowledgments}
This work was partially supported by the Research Fund of the University of Iceland, the~Icelandic Research Fund,
grant no. 163082-051, the Icelandic Instruments Fund, and Reykjavik University, grant no. 815051. Some of the computations were
performed on resources provided by the Icelandic High Performance Computing Centre at the University of Iceland. V.M. 
also acknowledge financial support from CNCS-UEFISCDI grant PN-III-P4-ID-PCE-2016-0084 and from the Romanian Core 
Program PN19-03 (contract no. 21 N/08.02.2019).
\end{acknowledgments}


\bibliography{mod_qd-VM}

\begin{thebibliography}{121}%
\makeatletter
\providecommand \@ifxundefined [1]{%
 \@ifx{#1\undefined}
}%
\providecommand \@ifnum [1]{%
 \ifnum #1\expandafter \@firstoftwo
 \else \expandafter \@secondoftwo
 \fi
}%
\providecommand \@ifx [1]{%
 \ifx #1\expandafter \@firstoftwo
 \else \expandafter \@secondoftwo
 \fi
}%
\providecommand \natexlab [1]{#1}%
\providecommand \enquote  [1]{``#1''}%
\providecommand \bibnamefont  [1]{#1}%
\providecommand \bibfnamefont [1]{#1}%
\providecommand \citenamefont [1]{#1}%
\providecommand \href@noop [0]{\@secondoftwo}%
\providecommand \href [0]{\begingroup \@sanitize@url \@href}%
\providecommand \@href[1]{\@@startlink{#1}\@@href}%
\providecommand \@@href[1]{\endgroup#1\@@endlink}%
\providecommand \@sanitize@url [0]{\catcode `\\12\catcode `\$12\catcode
  `\&12\catcode `\#12\catcode `\^12\catcode `\_12\catcode `\%12\relax}%
\providecommand \@@startlink[1]{}%
\providecommand \@@endlink[0]{}%
\providecommand \url  [0]{\begingroup\@sanitize@url \@url }%
\providecommand \@url [1]{\endgroup\@href {#1}{\urlprefix }}%
\providecommand \urlprefix  [0]{URL }%
\providecommand \Eprint [0]{\href }%
\providecommand \doibase [0]{http://dx.doi.org/}%
\providecommand \selectlanguage [0]{\@gobble}%
\providecommand \bibinfo  [0]{\@secondoftwo}%
\providecommand \bibfield  [0]{\@secondoftwo}%
\providecommand \translation [1]{[#1]}%
\providecommand \BibitemOpen [0]{}%
\providecommand \bibitemStop [0]{}%
\providecommand \bibitemNoStop [0]{.\EOS\space}%
\providecommand \EOS [0]{\spacefactor3000\relax}%
\providecommand \BibitemShut  [1]{\csname bibitem#1\endcsname}%
\let\auto@bib@innerbib\@empty
\bibitem [{\citenamefont {Di~Ventra}(2008)}]{di-ventra_2008}%
  \BibitemOpen
  \bibfield  {author} {\bibinfo {author} {\bibfnamefont {M.}~\bibnamefont
  {Di~Ventra}},\ }\href {\doibase 10.1017/CBO9780511755606} {\emph {\bibinfo
  {title} {Electrical Transport in Nanoscale Systems}}}\ (\bibinfo  {publisher}
  {Cambridge University Press},\ \bibinfo {year} {2008})\BibitemShut {NoStop}%
\bibitem [{\citenamefont {Jahnke}(2016)}]{jahnke2016quantum}%
  \BibitemOpen
  \bibfield  {author} {\bibinfo {author} {\bibfnamefont {F.}~\bibnamefont
  {Jahnke}},\ }\href {https://books.google.ro/books?id=8YCnDAEACAAJ} {\emph
  {\bibinfo {title} {Quantum Optics with Semiconductor Nanostructures}}},\
  Woodhead Publishing Series in Electronic and Optical Materials\ (\bibinfo
  {publisher} {Elsevier Science},\ \bibinfo {year} {2016})\BibitemShut
  {NoStop}%
\bibitem [{\citenamefont {Chow}\ and\ \citenamefont
  {Jahnke}(2013)}]{CHOW2013109}%
  \BibitemOpen
  \bibfield  {author} {\bibinfo {author} {\bibfnamefont {W.~W.}\ \bibnamefont
  {Chow}}\ and\ \bibinfo {author} {\bibfnamefont {F.}~\bibnamefont {Jahnke}},\
  }\href {\doibase https://doi.org/10.1016/j.pquantelec.2013.04.001} {\bibfield
   {journal} {\bibinfo  {journal} {Progress in Quantum Electronics}\ }\textbf
  {\bibinfo {volume} {37}},\ \bibinfo {pages} {109 } (\bibinfo {year}
  {2013})}\BibitemShut {NoStop}%
\bibitem [{\citenamefont {Gu}\ \emph {et~al.}(2017)\citenamefont {Gu},
  \citenamefont {Kockum}, \citenamefont {Miranowicz}, \citenamefont {xi~Liu},\
  and\ \citenamefont {Nori}}]{GU20171}%
  \BibitemOpen
  \bibfield  {author} {\bibinfo {author} {\bibfnamefont {X.}~\bibnamefont
  {Gu}}, \bibinfo {author} {\bibfnamefont {A.~F.}\ \bibnamefont {Kockum}},
  \bibinfo {author} {\bibfnamefont {A.}~\bibnamefont {Miranowicz}}, \bibinfo
  {author} {\bibfnamefont {Y.}~\bibnamefont {xi~Liu}}, \ and\ \bibinfo {author}
  {\bibfnamefont {F.}~\bibnamefont {Nori}},\ }\href {\doibase
  https://doi.org/10.1016/j.physrep.2017.10.002} {\bibfield  {journal}
  {\bibinfo  {journal} {Physics Reports}\ }\textbf {\bibinfo {volume}
  {718-719}},\ \bibinfo {pages} {1 } (\bibinfo {year} {2017})},\ \bibinfo
  {note} {microwave photonics with superconducting quantum
  circuits}\BibitemShut {NoStop}%
\bibitem [{\citenamefont {Bloch}(1957)}]{Bloch}%
  \BibitemOpen
  \bibfield  {author} {\bibinfo {author} {\bibfnamefont {F.}~\bibnamefont
  {Bloch}},\ }\href {\doibase 10.1103/PhysRev.105.1206} {\bibfield  {journal}
  {\bibinfo  {journal} {Phys. Rev.}\ }\textbf {\bibinfo {volume} {105}},\
  \bibinfo {pages} {1206} (\bibinfo {year} {1957})}\BibitemShut {NoStop}%
\bibitem [{\citenamefont {Wangsness}\ and\ \citenamefont
  {Bloch}(1953)}]{Wangsness}%
  \BibitemOpen
  \bibfield  {author} {\bibinfo {author} {\bibfnamefont {R.~K.}\ \bibnamefont
  {Wangsness}}\ and\ \bibinfo {author} {\bibfnamefont {F.}~\bibnamefont
  {Bloch}},\ }\href {\doibase 10.1103/PhysRev.89.728} {\bibfield  {journal}
  {\bibinfo  {journal} {Phys. Rev.}\ }\textbf {\bibinfo {volume} {89}},\
  \bibinfo {pages} {728} (\bibinfo {year} {1953})}\BibitemShut {NoStop}%
\bibitem [{\citenamefont {Redfield}(1965)}]{Redfield}%
  \BibitemOpen
  \bibfield  {author} {\bibinfo {author} {\bibfnamefont {A.~G.}\ \bibnamefont
  {Redfield}},\ }\href {https://doi.org/10.1016/B978-1-4832-3114-3.50007-6}
  {\bibfield  {journal} {\bibinfo  {journal} {Adv. Magn. Reson}\ }\textbf
  {\bibinfo {volume} {1}},\ \bibinfo {pages} {1} (\bibinfo {year}
  {1965})}\BibitemShut {NoStop}%
\bibitem [{\citenamefont {Scully}\ and\ \citenamefont
  {Zubairy}(1997)}]{Scully}%
  \BibitemOpen
  \bibfield  {author} {\bibinfo {author} {\bibfnamefont {M.~O.}\ \bibnamefont
  {Scully}}\ and\ \bibinfo {author} {\bibfnamefont {M.~S.}\ \bibnamefont
  {Zubairy}},\ }\href@noop {} {\emph {\bibinfo {title} {{Optics}}}}\ (\bibinfo
  {publisher} {Cambridge University Press},\ \bibinfo {year}
  {1997})\BibitemShut {NoStop}%
\bibitem [{\citenamefont {Carmichael}(2003)}]{Carmichael}%
  \BibitemOpen
  \bibfield  {author} {\bibinfo {author} {\bibfnamefont {H.~J.}\ \bibnamefont
  {Carmichael}},\ }\href@noop {} {\emph {\bibinfo {title} {{Statistical Methods
  in Quantum Optics 1: Master Equations and Fokker-Planck Equations}}}}\
  (\bibinfo  {publisher} {Springer},\ \bibinfo {year} {2003})\BibitemShut
  {NoStop}%
\bibitem [{\citenamefont {Timm}(2008)}]{Timm}%
  \BibitemOpen
  \bibfield  {author} {\bibinfo {author} {\bibfnamefont {C.}~\bibnamefont
  {Timm}},\ }\href {\doibase 10.1103/PhysRevB.77.195416} {\bibfield  {journal}
  {\bibinfo  {journal} {Phys. Rev. B}\ }\textbf {\bibinfo {volume} {77}},\
  \bibinfo {pages} {195416} (\bibinfo {year} {2008})}\BibitemShut {NoStop}%
\bibitem [{\citenamefont {Elenewski}\ \emph {et~al.}(2017)\citenamefont
  {Elenewski}, \citenamefont {Gruss},\ and\ \citenamefont
  {Zwolak}}]{Elenewski}%
  \BibitemOpen
  \bibfield  {author} {\bibinfo {author} {\bibfnamefont {J.~E.}\ \bibnamefont
  {Elenewski}}, \bibinfo {author} {\bibfnamefont {D.}~\bibnamefont {Gruss}}, \
  and\ \bibinfo {author} {\bibfnamefont {M.}~\bibnamefont {Zwolak}},\ }\href
  {\doibase 10.1063/1.5000747} {\bibfield  {journal} {\bibinfo  {journal} {The
  Journal of Chemical Physics}\ }\textbf {\bibinfo {volume} {147}},\ \bibinfo
  {pages} {151101} (\bibinfo {year} {2017})},\ \Eprint
  {http://arxiv.org/abs/https://doi.org/10.1063/1.5000747}
  {https://doi.org/10.1063/1.5000747} \BibitemShut {NoStop}%
\bibitem [{\citenamefont {Moldoveanu}\ \emph
  {et~al.}(2009{\natexlab{a}})\citenamefont {Moldoveanu}, \citenamefont
  {Manolescu},\ and\ \citenamefont {Gudmundsson}}]{Moldoveanu09:073019}%
  \BibitemOpen
  \bibfield  {author} {\bibinfo {author} {\bibfnamefont {V.}~\bibnamefont
  {Moldoveanu}}, \bibinfo {author} {\bibfnamefont {A.}~\bibnamefont
  {Manolescu}}, \ and\ \bibinfo {author} {\bibfnamefont {V.}~\bibnamefont
  {Gudmundsson}},\ }\href {http://stacks.iop.org/1367-2630/11/073019}
  {\bibfield  {journal} {\bibinfo  {journal} {New Journal of Physics}\ }\textbf
  {\bibinfo {volume} {11}},\ \bibinfo {pages} {073019} (\bibinfo {year}
  {2009}{\natexlab{a}})}\BibitemShut {NoStop}%
\bibitem [{\citenamefont {Vaz}\ and\ \citenamefont
  {Kyriakidis}(2010)}]{PhysRevB.81.085315}%
  \BibitemOpen
  \bibfield  {author} {\bibinfo {author} {\bibfnamefont {E.}~\bibnamefont
  {Vaz}}\ and\ \bibinfo {author} {\bibfnamefont {J.}~\bibnamefont
  {Kyriakidis}},\ }\href {\doibase 10.1103/PhysRevB.81.085315} {\bibfield
  {journal} {\bibinfo  {journal} {Phys. Rev. B}\ }\textbf {\bibinfo {volume}
  {81}},\ \bibinfo {pages} {085315} (\bibinfo {year} {2010})}\BibitemShut
  {NoStop}%
\bibitem [{\citenamefont {Moldoveanu}\ \emph
  {et~al.}(2010{\natexlab{a}})\citenamefont {Moldoveanu}, \citenamefont
  {Manolescu}, \citenamefont {Tang},\ and\ \citenamefont
  {Gudmundsson}}]{Moldoveanu10:155442}%
  \BibitemOpen
  \bibfield  {author} {\bibinfo {author} {\bibfnamefont {V.}~\bibnamefont
  {Moldoveanu}}, \bibinfo {author} {\bibfnamefont {A.}~\bibnamefont
  {Manolescu}}, \bibinfo {author} {\bibfnamefont {C.-S.}\ \bibnamefont {Tang}},
  \ and\ \bibinfo {author} {\bibfnamefont {V.}~\bibnamefont {Gudmundsson}},\
  }\href {http://link.aps.org/doi/10.1103/PhysRevB.81.155442} {\bibfield
  {journal} {\bibinfo  {journal} {Phys. Rev. B}\ }\textbf {\bibinfo {volume}
  {81}},\ \bibinfo {pages} {155442} (\bibinfo {year}
  {2010}{\natexlab{a}})}\BibitemShut {NoStop}%
\bibitem [{\citenamefont {Gudmundsson}\ \emph {et~al.}(2005)\citenamefont
  {Gudmundsson}, \citenamefont {Lin}, \citenamefont {Tang}, \citenamefont
  {Moldoveanu}, \citenamefont {Bardarson},\ and\ \citenamefont
  {Manolescu}}]{PhysRevB.71.235302}%
  \BibitemOpen
  \bibfield  {author} {\bibinfo {author} {\bibfnamefont {V.}~\bibnamefont
  {Gudmundsson}}, \bibinfo {author} {\bibfnamefont {Y.-Y.}\ \bibnamefont
  {Lin}}, \bibinfo {author} {\bibfnamefont {C.-S.}\ \bibnamefont {Tang}},
  \bibinfo {author} {\bibfnamefont {V.}~\bibnamefont {Moldoveanu}}, \bibinfo
  {author} {\bibfnamefont {J.~H.}\ \bibnamefont {Bardarson}}, \ and\ \bibinfo
  {author} {\bibfnamefont {A.}~\bibnamefont {Manolescu}},\ }\href {\doibase
  10.1103/PhysRevB.71.235302} {\bibfield  {journal} {\bibinfo  {journal} {Phys.
  Rev. B}\ }\textbf {\bibinfo {volume} {71}},\ \bibinfo {pages} {235302}
  (\bibinfo {year} {2005})}\BibitemShut {NoStop}%
\bibitem [{\citenamefont {Torfason}\ \emph {et~al.}(2012)\citenamefont
  {Torfason}, \citenamefont {Manolescu}, \citenamefont {Molodoveanu},\ and\
  \citenamefont {Gudmundsson}}]{PhysRevB.85.245114}%
  \BibitemOpen
  \bibfield  {author} {\bibinfo {author} {\bibfnamefont {K.}~\bibnamefont
  {Torfason}}, \bibinfo {author} {\bibfnamefont {A.}~\bibnamefont {Manolescu}},
  \bibinfo {author} {\bibfnamefont {V.}~\bibnamefont {Molodoveanu}}, \ and\
  \bibinfo {author} {\bibfnamefont {V.}~\bibnamefont {Gudmundsson}},\ }\href
  {\doibase 10.1103/PhysRevB.85.245114} {\bibfield  {journal} {\bibinfo
  {journal} {Phys. Rev. B}\ }\textbf {\bibinfo {volume} {85}},\ \bibinfo
  {pages} {245114} (\bibinfo {year} {2012})}\BibitemShut {NoStop}%
\bibitem [{\citenamefont {Moldoveanu}\ \emph
  {et~al.}(2010{\natexlab{b}})\citenamefont {Moldoveanu}, \citenamefont
  {Manolescu},\ and\ \citenamefont {Gudmundsson}}]{PhysRevB.82.085311}%
  \BibitemOpen
  \bibfield  {author} {\bibinfo {author} {\bibfnamefont {V.}~\bibnamefont
  {Moldoveanu}}, \bibinfo {author} {\bibfnamefont {A.}~\bibnamefont
  {Manolescu}}, \ and\ \bibinfo {author} {\bibfnamefont {V.}~\bibnamefont
  {Gudmundsson}},\ }\href {\doibase 10.1103/PhysRevB.82.085311} {\bibfield
  {journal} {\bibinfo  {journal} {Phys. Rev. B}\ }\textbf {\bibinfo {volume}
  {82}},\ \bibinfo {pages} {085311} (\bibinfo {year}
  {2010}{\natexlab{b}})}\BibitemShut {NoStop}%
\bibitem [{\citenamefont {{Bulnes Cuetara}}\ \emph {et~al.}(2011)\citenamefont
  {{Bulnes Cuetara}}, \citenamefont {Esposito},\ and\ \citenamefont
  {Gaspard}}]{PhysRevB.84.165114}%
  \BibitemOpen
  \bibfield  {author} {\bibinfo {author} {\bibfnamefont {G.}~\bibnamefont
  {{Bulnes Cuetara}}}, \bibinfo {author} {\bibfnamefont {M.}~\bibnamefont
  {Esposito}}, \ and\ \bibinfo {author} {\bibfnamefont {P.}~\bibnamefont
  {Gaspard}},\ }\href {\doibase 10.1103/PhysRevB.84.165114} {\bibfield
  {journal} {\bibinfo  {journal} {Phys. Rev. B}\ }\textbf {\bibinfo {volume}
  {84}},\ \bibinfo {pages} {165114} (\bibinfo {year} {2011})}\BibitemShut
  {NoStop}%
\bibitem [{\citenamefont {Harbola}\ \emph {et~al.}(2006)\citenamefont
  {Harbola}, \citenamefont {Esposito},\ and\ \citenamefont
  {Mukamel}}]{Harbola}%
  \BibitemOpen
  \bibfield  {author} {\bibinfo {author} {\bibfnamefont {U.}~\bibnamefont
  {Harbola}}, \bibinfo {author} {\bibfnamefont {M.}~\bibnamefont {Esposito}}, \
  and\ \bibinfo {author} {\bibfnamefont {S.}~\bibnamefont {Mukamel}},\ }\href
  {\doibase 10.1103/PhysRevB.74.235309} {\bibfield  {journal} {\bibinfo
  {journal} {Phys. Rev. B}\ }\textbf {\bibinfo {volume} {74}},\ \bibinfo
  {pages} {235309} (\bibinfo {year} {2006})}\BibitemShut {NoStop}%
\bibitem [{\citenamefont {Cohen}\ and\ \citenamefont
  {Rabani}(2011)}]{PhysRevB.84.075150}%
  \BibitemOpen
  \bibfield  {author} {\bibinfo {author} {\bibfnamefont {G.}~\bibnamefont
  {Cohen}}\ and\ \bibinfo {author} {\bibfnamefont {E.}~\bibnamefont {Rabani}},\
  }\href {\doibase 10.1103/PhysRevB.84.075150} {\bibfield  {journal} {\bibinfo
  {journal} {Phys. Rev. B}\ }\textbf {\bibinfo {volume} {84}},\ \bibinfo
  {pages} {075150} (\bibinfo {year} {2011})}\BibitemShut {NoStop}%
\bibitem [{\citenamefont {H\"artle}\ \emph {et~al.}(2013)\citenamefont
  {H\"artle}, \citenamefont {Cohen}, \citenamefont {Reichman},\ and\
  \citenamefont {Millis}}]{PhysRevB.88.235426}%
  \BibitemOpen
  \bibfield  {author} {\bibinfo {author} {\bibfnamefont {R.}~\bibnamefont
  {H\"artle}}, \bibinfo {author} {\bibfnamefont {G.}~\bibnamefont {Cohen}},
  \bibinfo {author} {\bibfnamefont {D.~R.}\ \bibnamefont {Reichman}}, \ and\
  \bibinfo {author} {\bibfnamefont {A.~J.}\ \bibnamefont {Millis}},\ }\href
  {\doibase 10.1103/PhysRevB.88.235426} {\bibfield  {journal} {\bibinfo
  {journal} {Phys. Rev. B}\ }\textbf {\bibinfo {volume} {88}},\ \bibinfo
  {pages} {235426} (\bibinfo {year} {2013})}\BibitemShut {NoStop}%
\bibitem [{\citenamefont {Esposito}\ and\ \citenamefont
  {Galperin}(2009)}]{PhysRevB.79.205303}%
  \BibitemOpen
  \bibfield  {author} {\bibinfo {author} {\bibfnamefont {M.}~\bibnamefont
  {Esposito}}\ and\ \bibinfo {author} {\bibfnamefont {M.}~\bibnamefont
  {Galperin}},\ }\href {\doibase 10.1103/PhysRevB.79.205303} {\bibfield
  {journal} {\bibinfo  {journal} {Phys. Rev. B}\ }\textbf {\bibinfo {volume}
  {79}},\ \bibinfo {pages} {205303} (\bibinfo {year} {2009})}\BibitemShut
  {NoStop}%
\bibitem [{\citenamefont {Galperin}\ \emph {et~al.}(2008)\citenamefont
  {Galperin}, \citenamefont {Nitzan},\ and\ \citenamefont
  {Ratner}}]{PhysRevB.78.125320}%
  \BibitemOpen
  \bibfield  {author} {\bibinfo {author} {\bibfnamefont {M.}~\bibnamefont
  {Galperin}}, \bibinfo {author} {\bibfnamefont {A.}~\bibnamefont {Nitzan}}, \
  and\ \bibinfo {author} {\bibfnamefont {M.~A.}\ \bibnamefont {Ratner}},\
  }\href {\doibase 10.1103/PhysRevB.78.125320} {\bibfield  {journal} {\bibinfo
  {journal} {Phys. Rev. B}\ }\textbf {\bibinfo {volume} {78}},\ \bibinfo
  {pages} {125320} (\bibinfo {year} {2008})}\BibitemShut {NoStop}%
\bibitem [{\citenamefont {Breuer}\ \emph {et~al.}(2016)\citenamefont {Breuer},
  \citenamefont {Laine}, \citenamefont {Piilo},\ and\ \citenamefont
  {Vacchini}}]{RevModPhys.88.021002}%
  \BibitemOpen
  \bibfield  {author} {\bibinfo {author} {\bibfnamefont {H.-P.}\ \bibnamefont
  {Breuer}}, \bibinfo {author} {\bibfnamefont {E.-M.}\ \bibnamefont {Laine}},
  \bibinfo {author} {\bibfnamefont {J.}~\bibnamefont {Piilo}}, \ and\ \bibinfo
  {author} {\bibfnamefont {B.}~\bibnamefont {Vacchini}},\ }\href {\doibase
  10.1103/RevModPhys.88.021002} {\bibfield  {journal} {\bibinfo  {journal}
  {Rev. Mod. Phys.}\ }\textbf {\bibinfo {volume} {88}},\ \bibinfo {pages}
  {021002} (\bibinfo {year} {2016})}\BibitemShut {NoStop}%
\bibitem [{\citenamefont {Kulkarni}\ \emph {et~al.}(2014)\citenamefont
  {Kulkarni}, \citenamefont {Cotlet},\ and\ \citenamefont
  {T{\"u}reci}}]{Kulkarni}%
  \BibitemOpen
  \bibfield  {author} {\bibinfo {author} {\bibfnamefont {M.}~\bibnamefont
  {Kulkarni}}, \bibinfo {author} {\bibfnamefont {O.}~\bibnamefont {Cotlet}}, \
  and\ \bibinfo {author} {\bibfnamefont {H.~E.}\ \bibnamefont {T{\"u}reci}},\
  }\href {\doibase 10.1103/PhysRevB.90.125402} {\bibfield  {journal} {\bibinfo
  {journal} {Phys. Rev. B}\ }\textbf {\bibinfo {volume} {90}},\ \bibinfo
  {pages} {125402} (\bibinfo {year} {2014})}\BibitemShut {NoStop}%
\bibitem [{\citenamefont {Viennot}\ \emph {et~al.}(2014)\citenamefont
  {Viennot}, \citenamefont {Delbecq}, \citenamefont {Dartiailh}, \citenamefont
  {Cottet},\ and\ \citenamefont {Kontos}}]{Viennot}%
  \BibitemOpen
  \bibfield  {author} {\bibinfo {author} {\bibfnamefont {J.~J.}\ \bibnamefont
  {Viennot}}, \bibinfo {author} {\bibfnamefont {M.~R.}\ \bibnamefont
  {Delbecq}}, \bibinfo {author} {\bibfnamefont {M.~C.}\ \bibnamefont
  {Dartiailh}}, \bibinfo {author} {\bibfnamefont {A.}~\bibnamefont {Cottet}}, \
  and\ \bibinfo {author} {\bibfnamefont {T.}~\bibnamefont {Kontos}},\ }\href
  {\doibase 10.1103/PhysRevB.89.165404} {\bibfield  {journal} {\bibinfo
  {journal} {Phys. Rev. B}\ }\textbf {\bibinfo {volume} {89}},\ \bibinfo
  {pages} {165404} (\bibinfo {year} {2014})}\BibitemShut {NoStop}%
\bibitem [{\citenamefont {Liu}\ \emph {et~al.}(2015)\citenamefont {Liu},
  \citenamefont {Stehlik}, \citenamefont {Eichler}, \citenamefont {Gullans},
  \citenamefont {Taylor},\ and\ \citenamefont {Petta}}]{Liu}%
  \BibitemOpen
  \bibfield  {author} {\bibinfo {author} {\bibfnamefont {Y.-Y.}\ \bibnamefont
  {Liu}}, \bibinfo {author} {\bibfnamefont {J.}~\bibnamefont {Stehlik}},
  \bibinfo {author} {\bibfnamefont {C.}~\bibnamefont {Eichler}}, \bibinfo
  {author} {\bibfnamefont {M.~J.}\ \bibnamefont {Gullans}}, \bibinfo {author}
  {\bibfnamefont {J.~M.}\ \bibnamefont {Taylor}}, \ and\ \bibinfo {author}
  {\bibfnamefont {J.~R.}\ \bibnamefont {Petta}},\ }\href {\doibase
  10.1126/science.aaa2501} {\bibfield  {journal} {\bibinfo  {journal}
  {Science}\ }\textbf {\bibinfo {volume} {347}},\ \bibinfo {pages} {285}
  (\bibinfo {year} {2015})}\BibitemShut {NoStop}%
\bibitem [{\citenamefont {Viennot}\ \emph {et~al.}(2015)\citenamefont
  {Viennot}, \citenamefont {Dartiailh}, \citenamefont {Cottet},\ and\
  \citenamefont {Kontos}}]{Viennot408}%
  \BibitemOpen
  \bibfield  {author} {\bibinfo {author} {\bibfnamefont {J.~J.}\ \bibnamefont
  {Viennot}}, \bibinfo {author} {\bibfnamefont {M.~C.}\ \bibnamefont
  {Dartiailh}}, \bibinfo {author} {\bibfnamefont {A.}~\bibnamefont {Cottet}}, \
  and\ \bibinfo {author} {\bibfnamefont {T.}~\bibnamefont {Kontos}},\ }\href
  {\doibase 10.1126/science.aaa3786} {\bibfield  {journal} {\bibinfo  {journal}
  {Science}\ }\textbf {\bibinfo {volume} {349}},\ \bibinfo {pages} {408}
  (\bibinfo {year} {2015})}\BibitemShut {NoStop}%
\bibitem [{\citenamefont {Liu}\ \emph {et~al.}(2018)\citenamefont {Liu},
  \citenamefont {Stehlik}, \citenamefont {Mi}, \citenamefont {Hartke},
  \citenamefont {Gullans},\ and\ \citenamefont
  {Petta}}]{PhysRevApplied.9.014030}%
  \BibitemOpen
  \bibfield  {author} {\bibinfo {author} {\bibfnamefont {Y.-Y.}\ \bibnamefont
  {Liu}}, \bibinfo {author} {\bibfnamefont {J.}~\bibnamefont {Stehlik}},
  \bibinfo {author} {\bibfnamefont {X.}~\bibnamefont {Mi}}, \bibinfo {author}
  {\bibfnamefont {T.~R.}\ \bibnamefont {Hartke}}, \bibinfo {author}
  {\bibfnamefont {M.~J.}\ \bibnamefont {Gullans}}, \ and\ \bibinfo {author}
  {\bibfnamefont {J.~R.}\ \bibnamefont {Petta}},\ }\href {\doibase
  10.1103/PhysRevApplied.9.014030} {\bibfield  {journal} {\bibinfo  {journal}
  {Phys. Rev. Applied}\ }\textbf {\bibinfo {volume} {9}},\ \bibinfo {pages}
  {014030} (\bibinfo {year} {2018})}\BibitemShut {NoStop}%
\bibitem [{\citenamefont {Beaudoin}\ \emph {et~al.}(2011)\citenamefont
  {Beaudoin}, \citenamefont {Gambetta},\ and\ \citenamefont
  {Blais}}]{PhysRevA.84.043832}%
  \BibitemOpen
  \bibfield  {author} {\bibinfo {author} {\bibfnamefont {F.}~\bibnamefont
  {Beaudoin}}, \bibinfo {author} {\bibfnamefont {J.~M.}\ \bibnamefont
  {Gambetta}}, \ and\ \bibinfo {author} {\bibfnamefont {A.}~\bibnamefont
  {Blais}},\ }\href {\doibase 10.1103/PhysRevA.84.043832} {\bibfield  {journal}
  {\bibinfo  {journal} {Phys. Rev. A}\ }\textbf {\bibinfo {volume} {84}},\
  \bibinfo {pages} {043832} (\bibinfo {year} {2011})}\BibitemShut {NoStop}%
\bibitem [{\citenamefont {Dinu}\ \emph {et~al.}(2018)\citenamefont {Dinu},
  \citenamefont {Moldoveanu},\ and\ \citenamefont
  {Gartner}}]{PhysRevB.97.195442}%
  \BibitemOpen
  \bibfield  {author} {\bibinfo {author} {\bibfnamefont {I.~V.}\ \bibnamefont
  {Dinu}}, \bibinfo {author} {\bibfnamefont {V.}~\bibnamefont {Moldoveanu}}, \
  and\ \bibinfo {author} {\bibfnamefont {P.}~\bibnamefont {Gartner}},\ }\href
  {\doibase 10.1103/PhysRevB.97.195442} {\bibfield  {journal} {\bibinfo
  {journal} {Phys. Rev. B}\ }\textbf {\bibinfo {volume} {97}},\ \bibinfo
  {pages} {195442} (\bibinfo {year} {2018})}\BibitemShut {NoStop}%
\bibitem [{\citenamefont {Gudmundsson}\ \emph {et~al.}(2012)\citenamefont
  {Gudmundsson}, \citenamefont {Jonasson}, \citenamefont {Tang}, \citenamefont
  {Goan},\ and\ \citenamefont {Manolescu}}]{Gudmundsson12:1109.4728}%
  \BibitemOpen
  \bibfield  {author} {\bibinfo {author} {\bibfnamefont {V.}~\bibnamefont
  {Gudmundsson}}, \bibinfo {author} {\bibfnamefont {O.}~\bibnamefont
  {Jonasson}}, \bibinfo {author} {\bibfnamefont {C.-S.}\ \bibnamefont {Tang}},
  \bibinfo {author} {\bibfnamefont {H.-S.}\ \bibnamefont {Goan}}, \ and\
  \bibinfo {author} {\bibfnamefont {A.}~\bibnamefont {Manolescu}},\ }\href
  {\doibase 10.1103/PhysRevB.85.075306} {\bibfield  {journal} {\bibinfo
  {journal} {Phys. Rev. B}\ }\textbf {\bibinfo {volume} {85}},\ \bibinfo
  {pages} {075306} (\bibinfo {year} {2012})}\BibitemShut {NoStop}%
\bibitem [{\citenamefont {Gudmundsson}\ \emph {et~al.}(2015)\citenamefont
  {Gudmundsson}, \citenamefont {Sitek}, \citenamefont {yi~Lin}, \citenamefont
  {Abdullah}, \citenamefont {Tang},\ and\ \citenamefont
  {Manolescu}}]{doi:10.1021/acsphotonics.5b00115}%
  \BibitemOpen
  \bibfield  {author} {\bibinfo {author} {\bibfnamefont {V.}~\bibnamefont
  {Gudmundsson}}, \bibinfo {author} {\bibfnamefont {A.}~\bibnamefont {Sitek}},
  \bibinfo {author} {\bibfnamefont {P.}~\bibnamefont {yi~Lin}}, \bibinfo
  {author} {\bibfnamefont {N.~R.}\ \bibnamefont {Abdullah}}, \bibinfo {author}
  {\bibfnamefont {C.-S.}\ \bibnamefont {Tang}}, \ and\ \bibinfo {author}
  {\bibfnamefont {A.}~\bibnamefont {Manolescu}},\ }\href@noop {} {\bibfield
  {journal} {\bibinfo  {journal} {ACS Photonics}\ }\textbf {\bibinfo {volume}
  {2}},\ \bibinfo {pages} {930} (\bibinfo {year} {2015})}\BibitemShut {NoStop}%
\bibitem [{\citenamefont {Cirio}\ \emph {et~al.}(2016)\citenamefont {Cirio},
  \citenamefont {{De Liberato}}, \citenamefont {Lambert},\ and\ \citenamefont
  {Nori}}]{PhysRevLett.116.113601}%
  \BibitemOpen
  \bibfield  {author} {\bibinfo {author} {\bibfnamefont {M.}~\bibnamefont
  {Cirio}}, \bibinfo {author} {\bibfnamefont {S.}~\bibnamefont {{De
  Liberato}}}, \bibinfo {author} {\bibfnamefont {N.}~\bibnamefont {Lambert}}, \
  and\ \bibinfo {author} {\bibfnamefont {F.}~\bibnamefont {Nori}},\ }\href
  {\doibase 10.1103/PhysRevLett.116.113601} {\bibfield  {journal} {\bibinfo
  {journal} {Phys. Rev. Lett.}\ }\textbf {\bibinfo {volume} {116}},\ \bibinfo
  {pages} {113601} (\bibinfo {year} {2016})}\BibitemShut {NoStop}%
\bibitem [{\citenamefont {Cirio}\ \emph {et~al.}(2019)\citenamefont {Cirio},
  \citenamefont {Shammah}, \citenamefont {Lambert}, \citenamefont
  {De~Liberato},\ and\ \citenamefont {Nori}}]{PhysRevLett.122.190403}%
  \BibitemOpen
  \bibfield  {author} {\bibinfo {author} {\bibfnamefont {M.}~\bibnamefont
  {Cirio}}, \bibinfo {author} {\bibfnamefont {N.}~\bibnamefont {Shammah}},
  \bibinfo {author} {\bibfnamefont {N.}~\bibnamefont {Lambert}}, \bibinfo
  {author} {\bibfnamefont {S.}~\bibnamefont {De~Liberato}}, \ and\ \bibinfo
  {author} {\bibfnamefont {F.}~\bibnamefont {Nori}},\ }\href {\doibase
  10.1103/PhysRevLett.122.190403} {\bibfield  {journal} {\bibinfo  {journal}
  {Phys. Rev. Lett.}\ }\textbf {\bibinfo {volume} {122}},\ \bibinfo {pages}
  {190403} (\bibinfo {year} {2019})}\BibitemShut {NoStop}%
\bibitem [{\citenamefont {Schachenmayer}\ \emph {et~al.}(2015)\citenamefont
  {Schachenmayer}, \citenamefont {Genes}, \citenamefont {Tignone},\ and\
  \citenamefont {Pupillo}}]{Genes}%
  \BibitemOpen
  \bibfield  {author} {\bibinfo {author} {\bibfnamefont {J.}~\bibnamefont
  {Schachenmayer}}, \bibinfo {author} {\bibfnamefont {C.}~\bibnamefont
  {Genes}}, \bibinfo {author} {\bibfnamefont {E.}~\bibnamefont {Tignone}}, \
  and\ \bibinfo {author} {\bibfnamefont {G.}~\bibnamefont {Pupillo}},\ }\href
  {\doibase 10.1103/PhysRevLett.114.196403} {\bibfield  {journal} {\bibinfo
  {journal} {Phys. Rev. Lett.}\ }\textbf {\bibinfo {volume} {114}},\ \bibinfo
  {pages} {196403} (\bibinfo {year} {2015})}\BibitemShut {NoStop}%
\bibitem [{\citenamefont {Breuer}\ and\ \citenamefont
  {Petruccione}(2007)}]{Petruccione}%
  \BibitemOpen
  \bibfield  {author} {\bibinfo {author} {\bibfnamefont {H.-P.}\ \bibnamefont
  {Breuer}}\ and\ \bibinfo {author} {\bibfnamefont {F.}~\bibnamefont
  {Petruccione}},\ }\href@noop {} {\emph {\bibinfo {title} {{The Theory of Open
  Quantum Systems}}}}\ (\bibinfo  {publisher} {Oxford University Press},\
  \bibinfo {year} {2007})\BibitemShut {NoStop}%
\bibitem [{\citenamefont {Poot}\ and\ \citenamefont {van~der
  Zant}(2012)}]{POOT2012273}%
  \BibitemOpen
  \bibfield  {author} {\bibinfo {author} {\bibfnamefont {M.}~\bibnamefont
  {Poot}}\ and\ \bibinfo {author} {\bibfnamefont {H.~S.}\ \bibnamefont {van~der
  Zant}},\ }\href {\doibase 10.1016/j.physrep.2011.12.004} {\bibfield
  {journal} {\bibinfo  {journal} {Physics Reports}\ }\textbf {\bibinfo {volume}
  {511}},\ \bibinfo {pages} {273} (\bibinfo {year} {2012})},\ \bibinfo {note}
  {mechanical systems in the quantum regime}\BibitemShut {NoStop}%
\bibitem [{\citenamefont {Tanatar}\ \emph {et~al.}(2019)\citenamefont
  {Tanatar}, \citenamefont {Moldoveanu}, \citenamefont {Dragomir},\ and\
  \citenamefont {Stanciu}}]{doi:10.1002/pssb.201800443}%
  \BibitemOpen
  \bibfield  {author} {\bibinfo {author} {\bibfnamefont {B.}~\bibnamefont
  {Tanatar}}, \bibinfo {author} {\bibfnamefont {V.}~\bibnamefont {Moldoveanu}},
  \bibinfo {author} {\bibfnamefont {R.}~\bibnamefont {Dragomir}}, \ and\
  \bibinfo {author} {\bibfnamefont {S.}~\bibnamefont {Stanciu}},\ }\href
  {\doibase 10.1002/pssb.201800443} {\bibfield  {journal} {\bibinfo  {journal}
  {physica status solidi (b)}\ }\textbf {\bibinfo {volume} {256}},\ \bibinfo
  {pages} {1800443} (\bibinfo {year} {2019})}\BibitemShut {NoStop}%
\bibitem [{\citenamefont {Settineri}\ \emph {et~al.}(2018)\citenamefont
  {Settineri}, \citenamefont {Macr\'{\i}}, \citenamefont {Ridolfo},
  \citenamefont {Di~Stefano}, \citenamefont {Kockum}, \citenamefont {Nori},\
  and\ \citenamefont {Savasta}}]{PhysRevA.98.053834}%
  \BibitemOpen
  \bibfield  {author} {\bibinfo {author} {\bibfnamefont {A.}~\bibnamefont
  {Settineri}}, \bibinfo {author} {\bibfnamefont {V.}~\bibnamefont
  {Macr\'{\i}}}, \bibinfo {author} {\bibfnamefont {A.}~\bibnamefont {Ridolfo}},
  \bibinfo {author} {\bibfnamefont {O.}~\bibnamefont {Di~Stefano}}, \bibinfo
  {author} {\bibfnamefont {A.~F.}\ \bibnamefont {Kockum}}, \bibinfo {author}
  {\bibfnamefont {F.}~\bibnamefont {Nori}}, \ and\ \bibinfo {author}
  {\bibfnamefont {S.}~\bibnamefont {Savasta}},\ }\href {\doibase
  10.1103/PhysRevA.98.053834} {\bibfield  {journal} {\bibinfo  {journal} {Phys.
  Rev. A}\ }\textbf {\bibinfo {volume} {98}},\ \bibinfo {pages} {053834}
  (\bibinfo {year} {2018})}\BibitemShut {NoStop}%
\bibitem [{\citenamefont {Zueco}\ and\ \citenamefont
  {Garc\'{\i}a-Ripoll}(2019)}]{PhysRevA.99.013807}%
  \BibitemOpen
  \bibfield  {author} {\bibinfo {author} {\bibfnamefont {D.}~\bibnamefont
  {Zueco}}\ and\ \bibinfo {author} {\bibfnamefont {J.}~\bibnamefont
  {Garc\'{\i}a-Ripoll}},\ }\href {\doibase 10.1103/PhysRevA.99.013807}
  {\bibfield  {journal} {\bibinfo  {journal} {Phys. Rev. A}\ }\textbf {\bibinfo
  {volume} {99}},\ \bibinfo {pages} {013807} (\bibinfo {year}
  {2019})}\BibitemShut {NoStop}%
\bibitem [{\citenamefont {Jonasson}\ \emph
  {et~al.}(2012{\natexlab{a}})\citenamefont {Jonasson}, \citenamefont {Tang},
  \citenamefont {Goan}, \citenamefont {Manolescu},\ and\ \citenamefont
  {Gudmundsson}}]{PhysRevE.86.046701}%
  \BibitemOpen
  \bibfield  {author} {\bibinfo {author} {\bibfnamefont {O.}~\bibnamefont
  {Jonasson}}, \bibinfo {author} {\bibfnamefont {C.-S.}\ \bibnamefont {Tang}},
  \bibinfo {author} {\bibfnamefont {H.-S.}\ \bibnamefont {Goan}}, \bibinfo
  {author} {\bibfnamefont {A.}~\bibnamefont {Manolescu}}, \ and\ \bibinfo
  {author} {\bibfnamefont {V.}~\bibnamefont {Gudmundsson}},\ }\href {\doibase
  10.1103/PhysRevE.86.046701} {\bibfield  {journal} {\bibinfo  {journal} {Phys.
  Rev. E}\ }\textbf {\bibinfo {volume} {86}},\ \bibinfo {pages} {046701}
  (\bibinfo {year} {2012}{\natexlab{a}})}\BibitemShut {NoStop}%
\bibitem [{\citenamefont {Schinabeck}\ \emph {et~al.}(2016)\citenamefont
  {Schinabeck}, \citenamefont {Erpenbeck}, \citenamefont {H\"artle},\ and\
  \citenamefont {Thoss}}]{PhysRevB.94.201407}%
  \BibitemOpen
  \bibfield  {author} {\bibinfo {author} {\bibfnamefont {C.}~\bibnamefont
  {Schinabeck}}, \bibinfo {author} {\bibfnamefont {A.}~\bibnamefont
  {Erpenbeck}}, \bibinfo {author} {\bibfnamefont {R.}~\bibnamefont {H\"artle}},
  \ and\ \bibinfo {author} {\bibfnamefont {M.}~\bibnamefont {Thoss}},\ }\href
  {\doibase 10.1103/PhysRevB.94.201407} {\bibfield  {journal} {\bibinfo
  {journal} {Phys. Rev. B}\ }\textbf {\bibinfo {volume} {94}},\ \bibinfo
  {pages} {201407} (\bibinfo {year} {2016})}\BibitemShut {NoStop}%
\bibitem [{\citenamefont {Gudmundsson}\ \emph {et~al.}(2009)\citenamefont
  {Gudmundsson}, \citenamefont {Gainar}, \citenamefont {Tang}, \citenamefont
  {Moldoveanu},\ and\ \citenamefont {Manolescu}}]{Gudmundsson09:113007}%
  \BibitemOpen
  \bibfield  {author} {\bibinfo {author} {\bibfnamefont {V.}~\bibnamefont
  {Gudmundsson}}, \bibinfo {author} {\bibfnamefont {C.}~\bibnamefont {Gainar}},
  \bibinfo {author} {\bibfnamefont {C.-S.}\ \bibnamefont {Tang}}, \bibinfo
  {author} {\bibfnamefont {V.}~\bibnamefont {Moldoveanu}}, \ and\ \bibinfo
  {author} {\bibfnamefont {A.}~\bibnamefont {Manolescu}},\ }\href
  {http://stacks.iop.org/1367-2630/11/113007} {\bibfield  {journal} {\bibinfo
  {journal} {New Journal of Physics}\ }\textbf {\bibinfo {volume} {11}},\
  \bibinfo {pages} {113007} (\bibinfo {year} {2009})}\BibitemShut {NoStop}%
\bibitem [{\citenamefont {Gudmundsson}\ \emph {et~al.}(2013)\citenamefont
  {Gudmundsson}, \citenamefont {Jonasson}, \citenamefont {Arnold},
  \citenamefont {Tang}, \citenamefont {Goan},\ and\ \citenamefont
  {Manolescu}}]{Gudmundsson:2013.305}%
  \BibitemOpen
  \bibfield  {author} {\bibinfo {author} {\bibfnamefont {V.}~\bibnamefont
  {Gudmundsson}}, \bibinfo {author} {\bibfnamefont {O.}~\bibnamefont
  {Jonasson}}, \bibinfo {author} {\bibfnamefont {T.}~\bibnamefont {Arnold}},
  \bibinfo {author} {\bibfnamefont {C.-S.}\ \bibnamefont {Tang}}, \bibinfo
  {author} {\bibfnamefont {H.-S.}\ \bibnamefont {Goan}}, \ and\ \bibinfo
  {author} {\bibfnamefont {A.}~\bibnamefont {Manolescu}},\ }\href {\doibase
  10.1002/prop.201200053} {\bibfield  {journal} {\bibinfo  {journal}
  {Fortschritte der Physik}\ }\textbf {\bibinfo {volume} {61}},\ \bibinfo
  {pages} {305} (\bibinfo {year} {2013})}\BibitemShut {NoStop}%
\bibitem [{\citenamefont {Arnold}\ \emph {et~al.}(2013)\citenamefont {Arnold},
  \citenamefont {Tang}, \citenamefont {Manolescu},\ and\ \citenamefont
  {Gudmundsson}}]{PhysRevB.87.035314}%
  \BibitemOpen
  \bibfield  {author} {\bibinfo {author} {\bibfnamefont {T.}~\bibnamefont
  {Arnold}}, \bibinfo {author} {\bibfnamefont {C.-S.}\ \bibnamefont {Tang}},
  \bibinfo {author} {\bibfnamefont {A.}~\bibnamefont {Manolescu}}, \ and\
  \bibinfo {author} {\bibfnamefont {V.}~\bibnamefont {Gudmundsson}},\ }\href
  {\doibase 10.1103/PhysRevB.87.035314} {\bibfield  {journal} {\bibinfo
  {journal} {Phys. Rev. B}\ }\textbf {\bibinfo {volume} {87}},\ \bibinfo
  {pages} {035314} (\bibinfo {year} {2013})}\BibitemShut {NoStop}%
\bibitem [{\citenamefont {Arnold}\ \emph {et~al.}(2015)\citenamefont {Arnold},
  \citenamefont {Tang}, \citenamefont {Manolescu},\ and\ \citenamefont
  {Gudmundsson}}]{2040-8986-17-1-015201}%
  \BibitemOpen
  \bibfield  {author} {\bibinfo {author} {\bibfnamefont {T.}~\bibnamefont
  {Arnold}}, \bibinfo {author} {\bibfnamefont {C.-S.}\ \bibnamefont {Tang}},
  \bibinfo {author} {\bibfnamefont {A.}~\bibnamefont {Manolescu}}, \ and\
  \bibinfo {author} {\bibfnamefont {V.}~\bibnamefont {Gudmundsson}},\ }\href
  {http://stacks.iop.org/2040-8986/17/i=1/a=015201} {\bibfield  {journal}
  {\bibinfo  {journal} {Journal of Optics}\ }\textbf {\bibinfo {volume} {17}},\
  \bibinfo {pages} {015201} (\bibinfo {year} {2015})}\BibitemShut {NoStop}%
\bibitem [{\citenamefont {Abdullah}\ \emph {et~al.}(2010)\citenamefont
  {Abdullah}, \citenamefont {Tang},\ and\ \citenamefont
  {Gudmundsson}}]{Abdullah10:195325}%
  \BibitemOpen
  \bibfield  {author} {\bibinfo {author} {\bibfnamefont {N.~R.}\ \bibnamefont
  {Abdullah}}, \bibinfo {author} {\bibfnamefont {C.-S.}\ \bibnamefont {Tang}},
  \ and\ \bibinfo {author} {\bibfnamefont {V.}~\bibnamefont {Gudmundsson}},\
  }\href {\doibase 10.1103/PhysRevB.82.195325} {\bibfield  {journal} {\bibinfo
  {journal} {Phys. Rev. B}\ }\textbf {\bibinfo {volume} {82}},\ \bibinfo
  {pages} {195325} (\bibinfo {year} {2010})}\BibitemShut {NoStop}%
\bibitem [{\citenamefont {Abdullah}\ \emph {et~al.}(2014)\citenamefont
  {Abdullah}, \citenamefont {Tang}, \citenamefont {Manolescu},\ and\
  \citenamefont {Gudmundsson}}]{Abdullah2014254}%
  \BibitemOpen
  \bibfield  {author} {\bibinfo {author} {\bibfnamefont {N.~R.}\ \bibnamefont
  {Abdullah}}, \bibinfo {author} {\bibfnamefont {C.-S.}\ \bibnamefont {Tang}},
  \bibinfo {author} {\bibfnamefont {A.}~\bibnamefont {Manolescu}}, \ and\
  \bibinfo {author} {\bibfnamefont {V.}~\bibnamefont {Gudmundsson}},\ }\href
  {\doibase 10.1016/j.physe.2014.07.030} {\bibfield  {journal} {\bibinfo
  {journal} {Physica E: Low-dimensional Systems and Nanostructures}\ }\textbf
  {\bibinfo {volume} {64}},\ \bibinfo {pages} {254} (\bibinfo {year}
  {2014})}\BibitemShut {NoStop}%
\bibitem [{\citenamefont {Gudmundsson}\ \emph {et~al.}(2016)\citenamefont
  {Gudmundsson}, \citenamefont {Sitek}, \citenamefont {Abdullah}, \citenamefont
  {Tang},\ and\ \citenamefont {Manolescu}}]{Gudmundsson16:AdP}%
  \BibitemOpen
  \bibfield  {author} {\bibinfo {author} {\bibfnamefont {V.}~\bibnamefont
  {Gudmundsson}}, \bibinfo {author} {\bibfnamefont {A.}~\bibnamefont {Sitek}},
  \bibinfo {author} {\bibfnamefont {N.~R.}\ \bibnamefont {Abdullah}}, \bibinfo
  {author} {\bibfnamefont {C.-S.}\ \bibnamefont {Tang}}, \ and\ \bibinfo
  {author} {\bibfnamefont {A.}~\bibnamefont {Manolescu}},\ }\href {\doibase
  10.1002/andp.201500298} {\bibfield  {journal} {\bibinfo  {journal} {Ann.
  Phys.}\ }\textbf {\bibinfo {volume} {528}},\ \bibinfo {pages} {394} (\bibinfo
  {year} {2016})}\BibitemShut {NoStop}%
\bibitem [{\citenamefont {Vargiamidis}\ \emph {et~al.}(2003)\citenamefont
  {Vargiamidis}, \citenamefont {Valassiades},\ and\ \citenamefont
  {Kyriakos}}]{Vargiamidis03:597}%
  \BibitemOpen
  \bibfield  {author} {\bibinfo {author} {\bibfnamefont {V.}~\bibnamefont
  {Vargiamidis}}, \bibinfo {author} {\bibfnamefont {O.}~\bibnamefont
  {Valassiades}}, \ and\ \bibinfo {author} {\bibfnamefont {D.~S.}\ \bibnamefont
  {Kyriakos}},\ }\href@noop {} {\bibfield  {journal} {\bibinfo  {journal}
  {Phys. stat. sol.}\ }\textbf {\bibinfo {volume} {236}},\ \bibinfo {pages}
  {597} (\bibinfo {year} {2003})}\BibitemShut {NoStop}%
\bibitem [{\citenamefont {Bardarson}\ \emph {et~al.}(2004)\citenamefont
  {Bardarson}, \citenamefont {Magnusdottir}, \citenamefont {Gudmundsdottir},
  \citenamefont {Tang}, \citenamefont {Manolescu},\ and\ \citenamefont
  {Gudmundsson}}]{PhysRevB.70.245308}%
  \BibitemOpen
  \bibfield  {author} {\bibinfo {author} {\bibfnamefont {J.~H.}\ \bibnamefont
  {Bardarson}}, \bibinfo {author} {\bibfnamefont {I.}~\bibnamefont
  {Magnusdottir}}, \bibinfo {author} {\bibfnamefont {G.}~\bibnamefont
  {Gudmundsdottir}}, \bibinfo {author} {\bibfnamefont {C.-S.}\ \bibnamefont
  {Tang}}, \bibinfo {author} {\bibfnamefont {A.}~\bibnamefont {Manolescu}}, \
  and\ \bibinfo {author} {\bibfnamefont {V.}~\bibnamefont {Gudmundsson}},\
  }\href {\doibase 10.1103/PhysRevB.70.245308} {\bibfield  {journal} {\bibinfo
  {journal} {Phys. Rev. B}\ }\textbf {\bibinfo {volume} {70}},\ \bibinfo
  {pages} {245308} (\bibinfo {year} {2004})}\BibitemShut {NoStop}%
\bibitem [{\citenamefont {Arnold}(2014)}]{ThorstenPhD}%
  \BibitemOpen
  \bibfield  {author} {\bibinfo {author} {\bibfnamefont {T.}~\bibnamefont
  {Arnold}},\ }\emph {\bibinfo {title} {{The influence of cavity photons on the
  transient transport of correlated electrons through a quantum ring with
  magnetic field and spin-orbit interaction}}},\ \href
  {http://hdl.handle.net/1946/19358} {Ph.D. thesis},\ \bibinfo  {school}
  {University of Iceland, http://hdl.handle.net/1946/19358} (\bibinfo {year}
  {2014})\BibitemShut {NoStop}%
\bibitem [{\citenamefont {Fujisawa}\ \emph {et~al.}(2003)\citenamefont
  {Fujisawa}, \citenamefont {Austing}, \citenamefont {Tokura}, \citenamefont
  {Hirayama},\ and\ \citenamefont {Tarucha}}]{Fujisawa_2003}%
  \BibitemOpen
  \bibfield  {author} {\bibinfo {author} {\bibfnamefont {T.}~\bibnamefont
  {Fujisawa}}, \bibinfo {author} {\bibfnamefont {D.~G.}\ \bibnamefont
  {Austing}}, \bibinfo {author} {\bibfnamefont {Y.}~\bibnamefont {Tokura}},
  \bibinfo {author} {\bibfnamefont {Y.}~\bibnamefont {Hirayama}}, \ and\
  \bibinfo {author} {\bibfnamefont {S.}~\bibnamefont {Tarucha}},\ }\href
  {\doibase 10.1088/0953-8984/15/33/201} {\bibfield  {journal} {\bibinfo
  {journal} {Journal of Physics: Condensed Matter}\ }\textbf {\bibinfo {volume}
  {15}},\ \bibinfo {pages} {R1395} (\bibinfo {year} {2003})}\BibitemShut
  {NoStop}%
\bibitem [{\citenamefont {Naser}\ \emph {et~al.}(2006)\citenamefont {Naser},
  \citenamefont {Ferry}, \citenamefont {Heeren}, \citenamefont {Reno},\ and\
  \citenamefont {Bird}}]{Naser}%
  \BibitemOpen
  \bibfield  {author} {\bibinfo {author} {\bibfnamefont {B.}~\bibnamefont
  {Naser}}, \bibinfo {author} {\bibfnamefont {D.~K.}\ \bibnamefont {Ferry}},
  \bibinfo {author} {\bibfnamefont {J.}~\bibnamefont {Heeren}}, \bibinfo
  {author} {\bibfnamefont {J.~L.}\ \bibnamefont {Reno}}, \ and\ \bibinfo
  {author} {\bibfnamefont {J.~P.}\ \bibnamefont {Bird}},\ }\href {\doibase
  10.1063/1.2337865} {\bibfield  {journal} {\bibinfo  {journal} {Applied
  Physics Letters}\ }\textbf {\bibinfo {volume} {89}},\ \bibinfo {pages}
  {083103} (\bibinfo {year} {2006})}\BibitemShut {NoStop}%
\bibitem [{\citenamefont {Lai}\ \emph {et~al.}(2009)\citenamefont {Lai},
  \citenamefont {Kuo},\ and\ \citenamefont {Li}}]{LAI2009886}%
  \BibitemOpen
  \bibfield  {author} {\bibinfo {author} {\bibfnamefont {W.-T.}\ \bibnamefont
  {Lai}}, \bibinfo {author} {\bibfnamefont {D.~M.}\ \bibnamefont {Kuo}}, \ and\
  \bibinfo {author} {\bibfnamefont {P.-W.}\ \bibnamefont {Li}},\ }\href
  {\doibase https://doi.org/10.1016/j.physe.2008.12.023} {\bibfield  {journal}
  {\bibinfo  {journal} {Physica E: Low-dimensional Systems and Nanostructures}\
  }\textbf {\bibinfo {volume} {41}},\ \bibinfo {pages} {886 } (\bibinfo {year}
  {2009})}\BibitemShut {NoStop}%
\bibitem [{\citenamefont {Kaestner}\ and\ \citenamefont
  {Kashcheyevs}(2015)}]{Kaestner_2015}%
  \BibitemOpen
  \bibfield  {author} {\bibinfo {author} {\bibfnamefont {B.}~\bibnamefont
  {Kaestner}}\ and\ \bibinfo {author} {\bibfnamefont {V.}~\bibnamefont
  {Kashcheyevs}},\ }\href {\doibase 10.1088/0034-4885/78/10/103901} {\bibfield
  {journal} {\bibinfo  {journal} {Reports on Progress in Physics}\ }\textbf
  {\bibinfo {volume} {78}},\ \bibinfo {pages} {103901} (\bibinfo {year}
  {2015})}\BibitemShut {NoStop}%
\bibitem [{\citenamefont {Moldoveanu}\ \emph
  {et~al.}(2009{\natexlab{b}})\citenamefont {Moldoveanu}, \citenamefont
  {Manolescu},\ and\ \citenamefont {Gudmundsson}}]{PhysRevB.80.205325}%
  \BibitemOpen
  \bibfield  {author} {\bibinfo {author} {\bibfnamefont {V.}~\bibnamefont
  {Moldoveanu}}, \bibinfo {author} {\bibfnamefont {A.}~\bibnamefont
  {Manolescu}}, \ and\ \bibinfo {author} {\bibfnamefont {V.}~\bibnamefont
  {Gudmundsson}},\ }\href {\doibase 10.1103/PhysRevB.80.205325} {\bibfield
  {journal} {\bibinfo  {journal} {Phys. Rev. B}\ }\textbf {\bibinfo {volume}
  {80}},\ \bibinfo {pages} {205325} (\bibinfo {year}
  {2009}{\natexlab{b}})}\BibitemShut {NoStop}%
\bibitem [{\citenamefont {Moldoveanu}\ \emph {et~al.}(2015)\citenamefont
  {Moldoveanu}, \citenamefont {Dinu}, \citenamefont {Tanatar},\ and\
  \citenamefont {Moca}}]{Moldoveanu_2015}%
  \BibitemOpen
  \bibfield  {author} {\bibinfo {author} {\bibfnamefont {V.}~\bibnamefont
  {Moldoveanu}}, \bibinfo {author} {\bibfnamefont {I.~V.}\ \bibnamefont
  {Dinu}}, \bibinfo {author} {\bibfnamefont {B.}~\bibnamefont {Tanatar}}, \
  and\ \bibinfo {author} {\bibfnamefont {C.~P.}\ \bibnamefont {Moca}},\ }\href
  {\doibase 10.1088/1367-2630/17/8/083020} {\bibfield  {journal} {\bibinfo
  {journal} {New Journal of Physics}\ }\textbf {\bibinfo {volume} {17}},\
  \bibinfo {pages} {083020} (\bibinfo {year} {2015})}\BibitemShut {NoStop}%
\bibitem [{\citenamefont {Moldoveanu}\ \emph {et~al.}(2011)\citenamefont
  {Moldoveanu}, \citenamefont {Cornean},\ and\ \citenamefont
  {Pillet}}]{PhysRevB.84.075464}%
  \BibitemOpen
  \bibfield  {author} {\bibinfo {author} {\bibfnamefont {V.}~\bibnamefont
  {Moldoveanu}}, \bibinfo {author} {\bibfnamefont {H.~D.}\ \bibnamefont
  {Cornean}}, \ and\ \bibinfo {author} {\bibfnamefont {C.-A.}\ \bibnamefont
  {Pillet}},\ }\href {\doibase 10.1103/PhysRevB.84.075464} {\bibfield
  {journal} {\bibinfo  {journal} {Phys. Rev. B}\ }\textbf {\bibinfo {volume}
  {84}},\ \bibinfo {pages} {075464} (\bibinfo {year} {2011})}\BibitemShut
  {NoStop}%
\bibitem [{\citenamefont {Cornean}\ \emph {et~al.}(2009)\citenamefont
  {Cornean}, \citenamefont {Neidhardt},\ and\ \citenamefont
  {Zagrebnov}}]{Cornean2009}%
  \BibitemOpen
  \bibfield  {author} {\bibinfo {author} {\bibfnamefont {H.~D.}\ \bibnamefont
  {Cornean}}, \bibinfo {author} {\bibfnamefont {H.}~\bibnamefont {Neidhardt}},
  \ and\ \bibinfo {author} {\bibfnamefont {V.~A.}\ \bibnamefont {Zagrebnov}},\
  }\href {\doibase 10.1007/s00023-009-0400-5} {\bibfield  {journal} {\bibinfo
  {journal} {Annales Henri Poincar{\'e}}\ }\textbf {\bibinfo {volume} {10}},\
  \bibinfo {pages} {61} (\bibinfo {year} {2009})}\BibitemShut {NoStop}%
\bibitem [{\citenamefont {Narozhny}\ and\ \citenamefont
  {Levchenko}(2016)}]{drag-RMP}%
  \BibitemOpen
  \bibfield  {author} {\bibinfo {author} {\bibfnamefont {B.~N.}\ \bibnamefont
  {Narozhny}}\ and\ \bibinfo {author} {\bibfnamefont {A.}~\bibnamefont
  {Levchenko}},\ }\href {\doibase 10.1103/RevModPhys.88.025003} {\bibfield
  {journal} {\bibinfo  {journal} {Rev. Mod. Phys.}\ }\textbf {\bibinfo {volume}
  {88}},\ \bibinfo {pages} {025003} (\bibinfo {year} {2016})}\BibitemShut
  {NoStop}%
\bibitem [{\citenamefont {S{\'a}nchez}\ \emph {et~al.}(2010)\citenamefont
  {S{\'a}nchez}, \citenamefont {L{\'o}pez}, \citenamefont {S{\'a}nchez},\ and\
  \citenamefont {B{\"u}ttiker}}]{PhysRevLett.104.076801}%
  \BibitemOpen
  \bibfield  {author} {\bibinfo {author} {\bibfnamefont {R.}~\bibnamefont
  {S{\'a}nchez}}, \bibinfo {author} {\bibfnamefont {R.}~\bibnamefont
  {L{\'o}pez}}, \bibinfo {author} {\bibfnamefont {D.}~\bibnamefont
  {S{\'a}nchez}}, \ and\ \bibinfo {author} {\bibfnamefont {M.}~\bibnamefont
  {B{\"u}ttiker}},\ }\href {\doibase 10.1103/PhysRevLett.104.076801} {\bibfield
   {journal} {\bibinfo  {journal} {Phys. Rev. Lett.}\ }\textbf {\bibinfo
  {volume} {104}},\ \bibinfo {pages} {076801} (\bibinfo {year}
  {2010})}\BibitemShut {NoStop}%
\bibitem [{\citenamefont {Kaasbjerg}\ and\ \citenamefont
  {Jauho}(2016)}]{APJauho-drag}%
  \BibitemOpen
  \bibfield  {author} {\bibinfo {author} {\bibfnamefont {K.}~\bibnamefont
  {Kaasbjerg}}\ and\ \bibinfo {author} {\bibfnamefont {A.-P.}\ \bibnamefont
  {Jauho}},\ }\href {\doibase 10.1103/PhysRevLett.116.196801} {\bibfield
  {journal} {\bibinfo  {journal} {Phys. Rev. Lett.}\ }\textbf {\bibinfo
  {volume} {116}},\ \bibinfo {pages} {196801} (\bibinfo {year}
  {2016})}\BibitemShut {NoStop}%
\bibitem [{\citenamefont {Lim}\ \emph {et~al.}(2018)\citenamefont {Lim},
  \citenamefont {S{\'a}nchez},\ and\ \citenamefont {L{\'o}pez}}]{Lim_2018}%
  \BibitemOpen
  \bibfield  {author} {\bibinfo {author} {\bibfnamefont {J.~S.}\ \bibnamefont
  {Lim}}, \bibinfo {author} {\bibfnamefont {D.}~\bibnamefont {S{\'a}nchez}}, \
  and\ \bibinfo {author} {\bibfnamefont {R.}~\bibnamefont {L{\'o}pez}},\ }\href
  {\doibase 10.1088/1367-2630/aaac0e} {\bibfield  {journal} {\bibinfo
  {journal} {New Journal of Physics}\ }\textbf {\bibinfo {volume} {20}},\
  \bibinfo {pages} {023038} (\bibinfo {year} {2018})}\BibitemShut {NoStop}%
\bibitem [{\citenamefont {Zhou}\ and\ \citenamefont
  {Guo}(2019)}]{PhysRevB.99.035423}%
  \BibitemOpen
  \bibfield  {author} {\bibinfo {author} {\bibfnamefont {C.}~\bibnamefont
  {Zhou}}\ and\ \bibinfo {author} {\bibfnamefont {H.}~\bibnamefont {Guo}},\
  }\href {\doibase 10.1103/PhysRevB.99.035423} {\bibfield  {journal} {\bibinfo
  {journal} {Phys. Rev. B}\ }\textbf {\bibinfo {volume} {99}},\ \bibinfo
  {pages} {035423} (\bibinfo {year} {2019})}\BibitemShut {NoStop}%
\bibitem [{\citenamefont {S{\'a}nchez}\ \emph {et~al.}(2017)\citenamefont
  {S{\'a}nchez}, \citenamefont {Thierschmann},\ and\ \citenamefont
  {Molenkamp}}]{Sanchez_2017}%
  \BibitemOpen
  \bibfield  {author} {\bibinfo {author} {\bibfnamefont {R.}~\bibnamefont
  {S{\'a}nchez}}, \bibinfo {author} {\bibfnamefont {H.}~\bibnamefont
  {Thierschmann}}, \ and\ \bibinfo {author} {\bibfnamefont {L.~W.}\
  \bibnamefont {Molenkamp}},\ }\href {\doibase 10.1088/1367-2630/aa8b94}
  {\bibfield  {journal} {\bibinfo  {journal} {New Journal of Physics}\ }\textbf
  {\bibinfo {volume} {19}},\ \bibinfo {pages} {113040} (\bibinfo {year}
  {2017})}\BibitemShut {NoStop}%
\bibitem [{\citenamefont {Bhandari}\ \emph {et~al.}(2018)\citenamefont
  {Bhandari}, \citenamefont {Chiriac{\`o}}, \citenamefont {Erdman},
  \citenamefont {Fazio},\ and\ \citenamefont {Taddei}}]{Th-drag}%
  \BibitemOpen
  \bibfield  {author} {\bibinfo {author} {\bibfnamefont {B.}~\bibnamefont
  {Bhandari}}, \bibinfo {author} {\bibfnamefont {G.}~\bibnamefont
  {Chiriac{\`o}}}, \bibinfo {author} {\bibfnamefont {P.~A.}\ \bibnamefont
  {Erdman}}, \bibinfo {author} {\bibfnamefont {R.}~\bibnamefont {Fazio}}, \
  and\ \bibinfo {author} {\bibfnamefont {F.}~\bibnamefont {Taddei}},\ }\href
  {\doibase 10.1103/PhysRevB.98.035415} {\bibfield  {journal} {\bibinfo
  {journal} {Phys. Rev. B}\ }\textbf {\bibinfo {volume} {98}},\ \bibinfo
  {pages} {035415} (\bibinfo {year} {2018})}\BibitemShut {NoStop}%
\bibitem [{\citenamefont {S{\'a}nchez}\ and\ \citenamefont
  {L{\'o}pez}(2016)}]{Sanchez16}%
  \BibitemOpen
  \bibfield  {author} {\bibinfo {author} {\bibfnamefont {D.}~\bibnamefont
  {S{\'a}nchez}}\ and\ \bibinfo {author} {\bibfnamefont {R.}~\bibnamefont
  {L{\'o}pez}},\ }\href {\doibase 10.1016/j.crhy.2016.08.005} {\bibfield
  {journal} {\bibinfo  {journal} {Comptes Rendus Physique}\ }\textbf {\bibinfo
  {volume} {17}},\ \bibinfo {pages} {1060} (\bibinfo {year}
  {2016})}\BibitemShut {NoStop}%
\bibitem [{\citenamefont {Sierra}\ and\ \citenamefont
  {S{\'a}nchez}(2014)}]{Sierra14}%
  \BibitemOpen
  \bibfield  {author} {\bibinfo {author} {\bibfnamefont {M.~A.}\ \bibnamefont
  {Sierra}}\ and\ \bibinfo {author} {\bibfnamefont {D.}~\bibnamefont
  {S{\'a}nchez}},\ }\href {\doibase 10.1103/PhysRevB.90.115313} {\bibfield
  {journal} {\bibinfo  {journal} {Phys. Rev. B}\ }\textbf {\bibinfo {volume}
  {90}},\ \bibinfo {pages} {115313} (\bibinfo {year} {2014})}\BibitemShut
  {NoStop}%
\bibitem [{\citenamefont {Torfason}\ \emph {et~al.}(2013)\citenamefont
  {Torfason}, \citenamefont {Manolescu}, \citenamefont {Erlingsson},\ and\
  \citenamefont {Gudmundsson}}]{Torfason13}%
  \BibitemOpen
  \bibfield  {author} {\bibinfo {author} {\bibfnamefont {K.}~\bibnamefont
  {Torfason}}, \bibinfo {author} {\bibfnamefont {A.}~\bibnamefont {Manolescu}},
  \bibinfo {author} {\bibfnamefont {S.~I.}\ \bibnamefont {Erlingsson}}, \ and\
  \bibinfo {author} {\bibfnamefont {V.}~\bibnamefont {Gudmundsson}},\ }\href
  {\doibase 10.1016/j.physe.2013.05.005} {\bibfield  {journal} {\bibinfo
  {journal} {Physica E}\ }\textbf {\bibinfo {volume} {53}},\ \bibinfo {pages}
  {178} (\bibinfo {year} {2013})}\BibitemShut {NoStop}%
\bibitem [{NJP(2014)}]{NJP2014}%
  \BibitemOpen
  \href@noop {} {\bibfield  {journal} {\bibinfo  {journal} {New Journal of
  Physics}\ }\textbf {\bibinfo {volume} {16}} (\bibinfo {year} {2014})},\
  \bibinfo {note} {edited by D. S{\'a}nchez and H. Linke}\BibitemShut {NoStop}%
\bibitem [{\citenamefont {Beenakker}\ and\ \citenamefont
  {Staring}(1992)}]{Beenakker92}%
  \BibitemOpen
  \bibfield  {author} {\bibinfo {author} {\bibfnamefont {C.~W.~J.}\
  \bibnamefont {Beenakker}}\ and\ \bibinfo {author} {\bibfnamefont {A.~A.~M.}\
  \bibnamefont {Staring}},\ }\href {\doibase 10.1103/PhysRevB.46.9667}
  {\bibfield  {journal} {\bibinfo  {journal} {Phys. Rev. B}\ }\textbf {\bibinfo
  {volume} {46}},\ \bibinfo {pages} {9667} (\bibinfo {year}
  {1992})}\BibitemShut {NoStop}%
\bibitem [{\citenamefont {Staring}\ \emph {et~al.}(1993)\citenamefont
  {Staring}, \citenamefont {Molenkamp}, \citenamefont {Alphenaar},
  \citenamefont {van Houten}, \citenamefont {Buyk}, \citenamefont {Mabesoone},
  \citenamefont {Beenakker},\ and\ \citenamefont {Foxon}}]{Staring93}%
  \BibitemOpen
  \bibfield  {author} {\bibinfo {author} {\bibfnamefont {A.~A.~M.}\
  \bibnamefont {Staring}}, \bibinfo {author} {\bibfnamefont {L.~W.}\
  \bibnamefont {Molenkamp}}, \bibinfo {author} {\bibfnamefont {B.~W.}\
  \bibnamefont {Alphenaar}}, \bibinfo {author} {\bibfnamefont {H.}~\bibnamefont
  {van Houten}}, \bibinfo {author} {\bibfnamefont {O.~J.~A.}\ \bibnamefont
  {Buyk}}, \bibinfo {author} {\bibfnamefont {M.~A.~A.}\ \bibnamefont
  {Mabesoone}}, \bibinfo {author} {\bibfnamefont {C.~W.~J.}\ \bibnamefont
  {Beenakker}}, \ and\ \bibinfo {author} {\bibfnamefont {C.~T.}\ \bibnamefont
  {Foxon}},\ }\href {\doibase 10.1209/0295-5075/22/1/011} {\bibfield  {journal}
  {\bibinfo  {journal} {Europhys. Lett.}\ }\textbf {\bibinfo {volume} {22}},\
  \bibinfo {pages} {57} (\bibinfo {year} {1993})}\BibitemShut {NoStop}%
\bibitem [{\citenamefont {Dzurak}\ \emph {et~al.}(1993)\citenamefont {Dzurak},
  \citenamefont {Smith}, \citenamefont {Pepper}, \citenamefont {Ritchie},
  \citenamefont {Frost}, \citenamefont {Jones},\ and\ \citenamefont
  {Hasko}}]{Dzurak93}%
  \BibitemOpen
  \bibfield  {author} {\bibinfo {author} {\bibfnamefont {A.}~\bibnamefont
  {Dzurak}}, \bibinfo {author} {\bibfnamefont {C.}~\bibnamefont {Smith}},
  \bibinfo {author} {\bibfnamefont {M.}~\bibnamefont {Pepper}}, \bibinfo
  {author} {\bibfnamefont {D.}~\bibnamefont {Ritchie}}, \bibinfo {author}
  {\bibfnamefont {J.}~\bibnamefont {Frost}}, \bibinfo {author} {\bibfnamefont
  {G.}~\bibnamefont {Jones}}, \ and\ \bibinfo {author} {\bibfnamefont
  {D.}~\bibnamefont {Hasko}},\ }\href {\doibase 10.1016/0038-1098(93)90819-9}
  {\bibfield  {journal} {\bibinfo  {journal} {Solid State Commun.}\ }\textbf
  {\bibinfo {volume} {87}},\ \bibinfo {pages} {1145} (\bibinfo {year}
  {1993})}\BibitemShut {NoStop}%
\bibitem [{\citenamefont {Svensson}\ \emph {et~al.}(2012)\citenamefont
  {Svensson}, \citenamefont {Persson}, \citenamefont {Hoffmann}, \citenamefont
  {Nakpathomkun}, \citenamefont {Nilsson}, \citenamefont {Xu}, \citenamefont
  {Samuelson},\ and\ \citenamefont {Linke}}]{Svensson12}%
  \BibitemOpen
  \bibfield  {author} {\bibinfo {author} {\bibfnamefont {S.~F.}\ \bibnamefont
  {Svensson}}, \bibinfo {author} {\bibfnamefont {A.~I.}\ \bibnamefont
  {Persson}}, \bibinfo {author} {\bibfnamefont {E.~A.}\ \bibnamefont
  {Hoffmann}}, \bibinfo {author} {\bibfnamefont {N.}~\bibnamefont
  {Nakpathomkun}}, \bibinfo {author} {\bibfnamefont {H.~A.}\ \bibnamefont
  {Nilsson}}, \bibinfo {author} {\bibfnamefont {H.~Q.}\ \bibnamefont {Xu}},
  \bibinfo {author} {\bibfnamefont {L.}~\bibnamefont {Samuelson}}, \ and\
  \bibinfo {author} {\bibfnamefont {H.}~\bibnamefont {Linke}},\ }\href
  {\doibase 10.1088/1367-2630/14/3/033041} {\bibfield  {journal} {\bibinfo
  {journal} {New Journal of Physics}\ }\textbf {\bibinfo {volume} {14}},\
  \bibinfo {pages} {033041} (\bibinfo {year} {2012})}\BibitemShut {NoStop}%
\bibitem [{\citenamefont {Reddy}\ \emph {et~al.}(2007)\citenamefont {Reddy},
  \citenamefont {Jang}, \citenamefont {Segalman},\ and\ \citenamefont
  {Majumdar}}]{Reddy1568}%
  \BibitemOpen
  \bibfield  {author} {\bibinfo {author} {\bibfnamefont {P.}~\bibnamefont
  {Reddy}}, \bibinfo {author} {\bibfnamefont {S.-Y.}\ \bibnamefont {Jang}},
  \bibinfo {author} {\bibfnamefont {R.~A.}\ \bibnamefont {Segalman}}, \ and\
  \bibinfo {author} {\bibfnamefont {A.}~\bibnamefont {Majumdar}},\ }\href
  {\doibase 10.1126/science.1137149} {\bibfield  {journal} {\bibinfo  {journal}
  {Science}\ }\textbf {\bibinfo {volume} {315}},\ \bibinfo {pages} {1568}
  (\bibinfo {year} {2007})}\BibitemShut {NoStop}%
\bibitem [{\citenamefont {Svensson}\ \emph {et~al.}(2013)\citenamefont
  {Svensson}, \citenamefont {Hoffmann}, \citenamefont {Nakpathomkun},
  \citenamefont {Wu}, \citenamefont {Xu}, \citenamefont {Nilsson},
  \citenamefont {S{\'a}nchez}, \citenamefont {Kashcheyevs},\ and\ \citenamefont
  {Linke}}]{Svensson13}%
  \BibitemOpen
  \bibfield  {author} {\bibinfo {author} {\bibfnamefont {S.~F.}\ \bibnamefont
  {Svensson}}, \bibinfo {author} {\bibfnamefont {E.~A.}\ \bibnamefont
  {Hoffmann}}, \bibinfo {author} {\bibfnamefont {N.}~\bibnamefont
  {Nakpathomkun}}, \bibinfo {author} {\bibfnamefont {P.~M.}\ \bibnamefont
  {Wu}}, \bibinfo {author} {\bibfnamefont {H.~Q.}\ \bibnamefont {Xu}}, \bibinfo
  {author} {\bibfnamefont {H.~A.}\ \bibnamefont {Nilsson}}, \bibinfo {author}
  {\bibfnamefont {D.}~\bibnamefont {S{\'a}nchez}}, \bibinfo {author}
  {\bibfnamefont {V.}~\bibnamefont {Kashcheyevs}}, \ and\ \bibinfo {author}
  {\bibfnamefont {H.}~\bibnamefont {Linke}},\ }\href@noop {} {\bibfield
  {journal} {\bibinfo  {journal} {New Journal of Physics}\ }\textbf {\bibinfo
  {volume} {15}},\ \bibinfo {pages} {105011} (\bibinfo {year}
  {2013})}\BibitemShut {NoStop}%
\bibitem [{\citenamefont {Zimbovskaya}(2015)}]{Zimbovskaya15}%
  \BibitemOpen
  \bibfield  {author} {\bibinfo {author} {\bibfnamefont {N.~A.}\ \bibnamefont
  {Zimbovskaya}},\ }\href@noop {} {\bibfield  {journal} {\bibinfo  {journal}
  {J. Chem. Phys.}\ }\textbf {\bibinfo {volume} {142}},\ \bibinfo {pages}
  {244310} (\bibinfo {year} {2015})}\BibitemShut {NoStop}%
\bibitem [{\citenamefont {Stanciu}\ \emph {et~al.}(2015)\citenamefont
  {Stanciu}, \citenamefont {Nemnes},\ and\ \citenamefont
  {Manolescu}}]{Stanciu15}%
  \BibitemOpen
  \bibfield  {author} {\bibinfo {author} {\bibfnamefont {A.~E.}\ \bibnamefont
  {Stanciu}}, \bibinfo {author} {\bibfnamefont {G.~A.}\ \bibnamefont {Nemnes}},
  \ and\ \bibinfo {author} {\bibfnamefont {A.}~\bibnamefont {Manolescu}},\
  }\href@noop {} {\bibfield  {journal} {\bibinfo  {journal} {Romanian J.
  Phys.}\ }\textbf {\bibinfo {volume} {60}},\ \bibinfo {pages} {716} (\bibinfo
  {year} {2015})}\BibitemShut {NoStop}%
\bibitem [{\citenamefont {Erlingsson}\ \emph {et~al.}(2017)\citenamefont
  {Erlingsson}, \citenamefont {Manolescu}, \citenamefont {Nemnes},
  \citenamefont {Bardarson},\ and\ \citenamefont {Sanchez}}]{Erlingsson17}%
  \BibitemOpen
  \bibfield  {author} {\bibinfo {author} {\bibfnamefont {S.}~\bibnamefont
  {Erlingsson}}, \bibinfo {author} {\bibfnamefont {A.}~\bibnamefont
  {Manolescu}}, \bibinfo {author} {\bibfnamefont {G.}~\bibnamefont {Nemnes}},
  \bibinfo {author} {\bibfnamefont {J.}~\bibnamefont {Bardarson}}, \ and\
  \bibinfo {author} {\bibfnamefont {D.}~\bibnamefont {Sanchez}},\ }\href
  {\doibase 10.1103/PhysRevLett.119.036804} {\bibfield  {journal} {\bibinfo
  {journal} {Phys. Rev. Lett.}\ }\textbf {\bibinfo {volume} {119}},\ \bibinfo
  {pages} {036804} (\bibinfo {year} {2017})}\BibitemShut {NoStop}%
\bibitem [{\citenamefont {Thorgilsson}\ \emph {et~al.}(2017)\citenamefont
  {Thorgilsson}, \citenamefont {Erlingsson},\ and\ \citenamefont
  {Manolescu}}]{Thorgilsson17}%
  \BibitemOpen
  \bibfield  {author} {\bibinfo {author} {\bibfnamefont {G.}~\bibnamefont
  {Thorgilsson}}, \bibinfo {author} {\bibfnamefont {S.~I.}\ \bibnamefont
  {Erlingsson}}, \ and\ \bibinfo {author} {\bibfnamefont {A.}~\bibnamefont
  {Manolescu}},\ }\href {\doibase 10.1088/1742-6596/906/1/012021} {\bibfield
  {journal} {\bibinfo  {journal} {Journal of Physics: Conference Series}\
  }\textbf {\bibinfo {volume} {906}},\ \bibinfo {pages} {012021} (\bibinfo
  {year} {2017})}\BibitemShut {NoStop}%
\bibitem [{\citenamefont {Erlingsson}\ \emph {et~al.}(2018)\citenamefont
  {Erlingsson}, \citenamefont {Bardarson},\ and\ \citenamefont
  {Manolescu}}]{Erlingsson18}%
  \BibitemOpen
  \bibfield  {author} {\bibinfo {author} {\bibfnamefont {S.~I.}\ \bibnamefont
  {Erlingsson}}, \bibinfo {author} {\bibfnamefont {J.~H.}\ \bibnamefont
  {Bardarson}}, \ and\ \bibinfo {author} {\bibfnamefont {A.}~\bibnamefont
  {Manolescu}},\ }\href {\doibase 10.3762/bjnano.9.107} {\bibfield  {journal}
  {\bibinfo  {journal} {Beilstein J. Nanotechnol}\ }\textbf {\bibinfo {volume}
  {9}},\ \bibinfo {pages} {1156} (\bibinfo {year} {2018})}\BibitemShut
  {NoStop}%
\bibitem [{\citenamefont {Zhang}\ \emph {et~al.}(2016)\citenamefont {Zhang},
  \citenamefont {Lou}, \citenamefont {Li}, \citenamefont {Reno}, \citenamefont
  {Pan}, \citenamefont {Watson}, \citenamefont {Manfra},\ and\ \citenamefont
  {Kono}}]{Zhang1005:2016}%
  \BibitemOpen
  \bibfield  {author} {\bibinfo {author} {\bibfnamefont {Q.}~\bibnamefont
  {Zhang}}, \bibinfo {author} {\bibfnamefont {M.}~\bibnamefont {Lou}}, \bibinfo
  {author} {\bibfnamefont {X.}~\bibnamefont {Li}}, \bibinfo {author}
  {\bibfnamefont {J.~L.}\ \bibnamefont {Reno}}, \bibinfo {author}
  {\bibfnamefont {W.}~\bibnamefont {Pan}}, \bibinfo {author} {\bibfnamefont
  {J.~D.}\ \bibnamefont {Watson}}, \bibinfo {author} {\bibfnamefont {M.~J.}\
  \bibnamefont {Manfra}}, \ and\ \bibinfo {author} {\bibfnamefont
  {J.}~\bibnamefont {Kono}},\ }\href {\doibase 10.1038/nphys3850;
  10.1038/nphys3850} {\bibfield  {journal} {\bibinfo  {journal} {Nature
  Physics}\ }\textbf {\bibinfo {volume} {12}},\ \bibinfo {pages} {1005}
  (\bibinfo {year} {2016})}\BibitemShut {NoStop}%
\bibitem [{\citenamefont {Gudmundsson}\ \emph {et~al.}(2017)\citenamefont
  {Gudmundsson}, \citenamefont {Abdullah}, \citenamefont {Sitek}, \citenamefont
  {Goan}, \citenamefont {Tang},\ and\ \citenamefont
  {Manolescu}}]{2016arXiv161109453G}%
  \BibitemOpen
  \bibfield  {author} {\bibinfo {author} {\bibfnamefont {V.}~\bibnamefont
  {Gudmundsson}}, \bibinfo {author} {\bibfnamefont {N.~R.}\ \bibnamefont
  {Abdullah}}, \bibinfo {author} {\bibfnamefont {A.}~\bibnamefont {Sitek}},
  \bibinfo {author} {\bibfnamefont {H.-S.}\ \bibnamefont {Goan}}, \bibinfo
  {author} {\bibfnamefont {C.-S.}\ \bibnamefont {Tang}}, \ and\ \bibinfo
  {author} {\bibfnamefont {A.}~\bibnamefont {Manolescu}},\ }\href {\doibase
  10.1103/PhysRevB.95.195307} {\bibfield  {journal} {\bibinfo  {journal} {Phys.
  Rev. B}\ }\textbf {\bibinfo {volume} {95}},\ \bibinfo {pages} {195307}
  (\bibinfo {year} {2017})}\BibitemShut {NoStop}%
\bibitem [{\citenamefont {Arnold}\ \emph
  {et~al.}(2014{\natexlab{a}})\citenamefont {Arnold}, \citenamefont {Tang},
  \citenamefont {Manolescu},\ and\ \citenamefont
  {Gudmundsson}}]{ARNOLD2014170}%
  \BibitemOpen
  \bibfield  {author} {\bibinfo {author} {\bibfnamefont {T.}~\bibnamefont
  {Arnold}}, \bibinfo {author} {\bibfnamefont {C.-S.}\ \bibnamefont {Tang}},
  \bibinfo {author} {\bibfnamefont {A.}~\bibnamefont {Manolescu}}, \ and\
  \bibinfo {author} {\bibfnamefont {V.}~\bibnamefont {Gudmundsson}},\ }\href
  {\doibase 10.1016/j.physe.2014.02.024} {\bibfield  {journal} {\bibinfo
  {journal} {Physica E: Low-dimensional Systems and Nanostructures}\ }\textbf
  {\bibinfo {volume} {60}},\ \bibinfo {pages} {170} (\bibinfo {year}
  {2014}{\natexlab{a}})}\BibitemShut {NoStop}%
\bibitem [{\citenamefont {Feranchuk}\ \emph {et~al.}(1996)\citenamefont
  {Feranchuk}, \citenamefont {Komarov},\ and\ \citenamefont
  {Ulyanenkov}}]{Feranchuk96:4035}%
  \BibitemOpen
  \bibfield  {author} {\bibinfo {author} {\bibfnamefont {I.~D.}\ \bibnamefont
  {Feranchuk}}, \bibinfo {author} {\bibfnamefont {L.~I.}\ \bibnamefont
  {Komarov}}, \ and\ \bibinfo {author} {\bibfnamefont {A.~P.}\ \bibnamefont
  {Ulyanenkov}},\ }\href {http://iopscience.iop.org/0305-4470/29/14/026}
  {\bibfield  {journal} {\bibinfo  {journal} {J. Phys. A: Math. Gen.}\ }\textbf
  {\bibinfo {volume} {29}},\ \bibinfo {pages} {4035} (\bibinfo {year}
  {1996})}\BibitemShut {NoStop}%
\bibitem [{\citenamefont {Li}\ \emph {et~al.}(2009)\citenamefont {Li},
  \citenamefont {Wang},\ and\ \citenamefont {Liu}}]{Li09:044212}%
  \BibitemOpen
  \bibfield  {author} {\bibinfo {author} {\bibfnamefont {X.-H.}\ \bibnamefont
  {Li}}, \bibinfo {author} {\bibfnamefont {K.-L.}\ \bibnamefont {Wang}}, \ and\
  \bibinfo {author} {\bibfnamefont {T.}~\bibnamefont {Liu}},\ }\href
  {http://iopscience.iop.org/0256-307X/26/4/044212} {\bibfield  {journal}
  {\bibinfo  {journal} {Chin. Phys. Lett.}\ }\textbf {\bibinfo {volume} {26}},\
  \bibinfo {pages} {044212} (\bibinfo {year} {2009})}\BibitemShut {NoStop}%
\bibitem [{\citenamefont {Jonasson}\ \emph
  {et~al.}(2012{\natexlab{b}})\citenamefont {Jonasson}, \citenamefont {Tang},
  \citenamefont {Goan}, \citenamefont {Manolescu},\ and\ \citenamefont
  {Gudmundsson}}]{Jonasson2011:01}%
  \BibitemOpen
  \bibfield  {author} {\bibinfo {author} {\bibfnamefont {O.}~\bibnamefont
  {Jonasson}}, \bibinfo {author} {\bibfnamefont {C.-S.}\ \bibnamefont {Tang}},
  \bibinfo {author} {\bibfnamefont {H.-S.}\ \bibnamefont {Goan}}, \bibinfo
  {author} {\bibfnamefont {A.}~\bibnamefont {Manolescu}}, \ and\ \bibinfo
  {author} {\bibfnamefont {V.}~\bibnamefont {Gudmundsson}},\ }\href
  {http://stacks.iop.org/1367-2630/14/i=1/a=013036} {\bibfield  {journal}
  {\bibinfo  {journal} {New Journal of Physics}\ }\textbf {\bibinfo {volume}
  {14}},\ \bibinfo {pages} {013036} (\bibinfo {year}
  {2012}{\natexlab{b}})}\BibitemShut {NoStop}%
\bibitem [{\citenamefont {Arnold}\ \emph
  {et~al.}(2014{\natexlab{b}})\citenamefont {Arnold}, \citenamefont {Tang},
  \citenamefont {Manolescu},\ and\ \citenamefont {Gudmundsson}}]{Arnold2014}%
  \BibitemOpen
  \bibfield  {author} {\bibinfo {author} {\bibfnamefont {T.}~\bibnamefont
  {Arnold}}, \bibinfo {author} {\bibfnamefont {C.-S.}\ \bibnamefont {Tang}},
  \bibinfo {author} {\bibfnamefont {A.}~\bibnamefont {Manolescu}}, \ and\
  \bibinfo {author} {\bibfnamefont {V.}~\bibnamefont {Gudmundsson}},\ }\href
  {\doibase 10.1140/epjb/e2014-50144-y} {\bibfield  {journal} {\bibinfo
  {journal} {The European Physical Journal B}\ }\textbf {\bibinfo {volume}
  {87}},\ \bibinfo {pages} {113} (\bibinfo {year}
  {2014}{\natexlab{b}})}\BibitemShut {NoStop}%
\bibitem [{\citenamefont {Abdullah}\ \emph {et~al.}(2013)\citenamefont
  {Abdullah}, \citenamefont {Tang}, \citenamefont {Manolescu},\ and\
  \citenamefont {Gudmundsson}}]{0953-8984-25-46-465302}%
  \BibitemOpen
  \bibfield  {author} {\bibinfo {author} {\bibfnamefont {N.~R.}\ \bibnamefont
  {Abdullah}}, \bibinfo {author} {\bibfnamefont {C.-S.}\ \bibnamefont {Tang}},
  \bibinfo {author} {\bibfnamefont {A.}~\bibnamefont {Manolescu}}, \ and\
  \bibinfo {author} {\bibfnamefont {V.}~\bibnamefont {Gudmundsson}},\ }\href
  {http://stacks.iop.org/0953-8984/25/i=46/a=465302} {\bibfield  {journal}
  {\bibinfo  {journal} {Journal of Physics: Condensed Matter}\ }\textbf
  {\bibinfo {volume} {25}},\ \bibinfo {pages} {465302} (\bibinfo {year}
  {2013})}\BibitemShut {NoStop}%
\bibitem [{\citenamefont {Abdullah}\ \emph
  {et~al.}(2016{\natexlab{a}})\citenamefont {Abdullah}, \citenamefont {Tang},
  \citenamefont {Manolescu},\ and\ \citenamefont
  {Gudmundsson}}]{ABDULLAH2016280}%
  \BibitemOpen
  \bibfield  {author} {\bibinfo {author} {\bibfnamefont {N.~R.}\ \bibnamefont
  {Abdullah}}, \bibinfo {author} {\bibfnamefont {C.-S.}\ \bibnamefont {Tang}},
  \bibinfo {author} {\bibfnamefont {A.}~\bibnamefont {Manolescu}}, \ and\
  \bibinfo {author} {\bibfnamefont {V.}~\bibnamefont {Gudmundsson}},\ }\href
  {\doibase 10.1016/j.physe.2016.06.023} {\bibfield  {journal} {\bibinfo
  {journal} {Physica E: Low-dimensional Systems and Nanostructures}\ }\textbf
  {\bibinfo {volume} {84}},\ \bibinfo {pages} {280} (\bibinfo {year}
  {2016}{\natexlab{a}})}\BibitemShut {NoStop}%
\bibitem [{\citenamefont {Abdullah}\ \emph
  {et~al.}(2016{\natexlab{b}})\citenamefont {Abdullah}, \citenamefont {Tang},
  \citenamefont {Manolescu},\ and\ \citenamefont
  {Gudmundsson}}]{0953-8984-28-37-375301}%
  \BibitemOpen
  \bibfield  {author} {\bibinfo {author} {\bibfnamefont {N.~R.}\ \bibnamefont
  {Abdullah}}, \bibinfo {author} {\bibfnamefont {C.-S.}\ \bibnamefont {Tang}},
  \bibinfo {author} {\bibfnamefont {A.}~\bibnamefont {Manolescu}}, \ and\
  \bibinfo {author} {\bibfnamefont {V.}~\bibnamefont {Gudmundsson}},\ }\href
  {http://stacks.iop.org/0953-8984/28/i=37/a=375301} {\bibfield  {journal}
  {\bibinfo  {journal} {Journal of Physics: Condensed Matter}\ }\textbf
  {\bibinfo {volume} {28}},\ \bibinfo {pages} {375301} (\bibinfo {year}
  {2016}{\natexlab{b}})}\BibitemShut {NoStop}%
\bibitem [{\citenamefont {Abdullah}\ \emph
  {et~al.}(2018{\natexlab{a}})\citenamefont {Abdullah}, \citenamefont {Tang},
  \citenamefont {Manolescu},\ and\ \citenamefont
  {Gudmundsson}}]{ABDULLAH2018199}%
  \BibitemOpen
  \bibfield  {author} {\bibinfo {author} {\bibfnamefont {N.~R.}\ \bibnamefont
  {Abdullah}}, \bibinfo {author} {\bibfnamefont {C.-S.}\ \bibnamefont {Tang}},
  \bibinfo {author} {\bibfnamefont {A.}~\bibnamefont {Manolescu}}, \ and\
  \bibinfo {author} {\bibfnamefont {V.}~\bibnamefont {Gudmundsson}},\ }\href
  {\doibase 10.1016/j.physleta.2017.11.007} {\bibfield  {journal} {\bibinfo
  {journal} {Physics Letters A}\ }\textbf {\bibinfo {volume} {382}},\ \bibinfo
  {pages} {199} (\bibinfo {year} {2018}{\natexlab{a}})}\BibitemShut {NoStop}%
\bibitem [{\citenamefont {Abdullah}\ \emph
  {et~al.}(2016{\natexlab{c}})\citenamefont {Abdullah}, \citenamefont {Tang},
  \citenamefont {Manolescu},\ and\ \citenamefont
  {Gudmundsson}}]{doi:10.1021/acsphotonics.5b00532}%
  \BibitemOpen
  \bibfield  {author} {\bibinfo {author} {\bibfnamefont {N.~R.}\ \bibnamefont
  {Abdullah}}, \bibinfo {author} {\bibfnamefont {C.-S.}\ \bibnamefont {Tang}},
  \bibinfo {author} {\bibfnamefont {A.}~\bibnamefont {Manolescu}}, \ and\
  \bibinfo {author} {\bibfnamefont {V.}~\bibnamefont {Gudmundsson}},\ }\href
  {\doibase 10.1021/acsphotonics.5b00532} {\bibfield  {journal} {\bibinfo
  {journal} {ACS Photonics}\ }\textbf {\bibinfo {volume} {3}},\ \bibinfo
  {pages} {249} (\bibinfo {year} {2016}{\natexlab{c}})}\BibitemShut {NoStop}%
\bibitem [{\citenamefont {Abdullah}\ \emph
  {et~al.}(2018{\natexlab{b}})\citenamefont {Abdullah}, \citenamefont {Tang},
  \citenamefont {Manolescu},\ and\ \citenamefont
  {Gudmundsson}}]{ABDULLAH2018102}%
  \BibitemOpen
  \bibfield  {author} {\bibinfo {author} {\bibfnamefont {N.~R.}\ \bibnamefont
  {Abdullah}}, \bibinfo {author} {\bibfnamefont {C.-S.}\ \bibnamefont {Tang}},
  \bibinfo {author} {\bibfnamefont {A.}~\bibnamefont {Manolescu}}, \ and\
  \bibinfo {author} {\bibfnamefont {V.}~\bibnamefont {Gudmundsson}},\ }\href
  {\doibase 10.1016/j.physe.2017.09.011} {\bibfield  {journal} {\bibinfo
  {journal} {Physica E: Low-dimensional Systems and Nanostructures}\ }\textbf
  {\bibinfo {volume} {95}},\ \bibinfo {pages} {102} (\bibinfo {year}
  {2018}{\natexlab{b}})}\BibitemShut {NoStop}%
\bibitem [{\citenamefont {Abdullah}\ \emph
  {et~al.}(2018{\natexlab{c}})\citenamefont {Abdullah}, \citenamefont {Arnold},
  \citenamefont {Tang}, \citenamefont {Manolescu},\ and\ \citenamefont
  {Gudmundsson}}]{Abdullah_2018}%
  \BibitemOpen
  \bibfield  {author} {\bibinfo {author} {\bibfnamefont {N.~R.}\ \bibnamefont
  {Abdullah}}, \bibinfo {author} {\bibfnamefont {T.}~\bibnamefont {Arnold}},
  \bibinfo {author} {\bibfnamefont {C.-S.}\ \bibnamefont {Tang}}, \bibinfo
  {author} {\bibfnamefont {A.}~\bibnamefont {Manolescu}}, \ and\ \bibinfo
  {author} {\bibfnamefont {V.}~\bibnamefont {Gudmundsson}},\ }\href {\doibase
  10.1088/1361-648x/aab255} {\bibfield  {journal} {\bibinfo  {journal} {Journal
  of Physics: Condensed Matter}\ }\textbf {\bibinfo {volume} {30}},\ \bibinfo
  {pages} {145303} (\bibinfo {year} {2018}{\natexlab{c}})}\BibitemShut
  {NoStop}%
\bibitem [{\citenamefont {Abdullah}\ \emph {et~al.}(2019)\citenamefont
  {Abdullah}, \citenamefont {Tang}, \citenamefont {Manolescu},\ and\
  \citenamefont {Gudmundsson}}]{Nzar-2019-Rabi}%
  \BibitemOpen
  \bibfield  {author} {\bibinfo {author} {\bibfnamefont {N.~R.}\ \bibnamefont
  {Abdullah}}, \bibinfo {author} {\bibfnamefont {C.-S.}\ \bibnamefont {Tang}},
  \bibinfo {author} {\bibfnamefont {A.}~\bibnamefont {Manolescu}}, \ and\
  \bibinfo {author} {\bibfnamefont {V.}~\bibnamefont {Gudmundsson}},\
  }\href@noop {} {\bibfield  {journal} {\bibinfo  {journal} {arXiv:1905.07492}\
  } (\bibinfo {year} {2019})}\BibitemShut {NoStop}%
\bibitem [{\citenamefont {Jonsson}\ \emph {et~al.}(2017)\citenamefont
  {Jonsson}, \citenamefont {Manolescu}, \citenamefont {Goan}, \citenamefont
  {Abdullah}, \citenamefont {Sitek}, \citenamefont {Tang},\ and\ \citenamefont
  {Gudmundsson}}]{2016arXiv161003223J}%
  \BibitemOpen
  \bibfield  {author} {\bibinfo {author} {\bibfnamefont {T.~H.}\ \bibnamefont
  {Jonsson}}, \bibinfo {author} {\bibfnamefont {A.}~\bibnamefont {Manolescu}},
  \bibinfo {author} {\bibfnamefont {H.-S.}\ \bibnamefont {Goan}}, \bibinfo
  {author} {\bibfnamefont {N.~R.}\ \bibnamefont {Abdullah}}, \bibinfo {author}
  {\bibfnamefont {A.}~\bibnamefont {Sitek}}, \bibinfo {author} {\bibfnamefont
  {C.-S.}\ \bibnamefont {Tang}}, \ and\ \bibinfo {author} {\bibfnamefont
  {V.}~\bibnamefont {Gudmundsson}},\ }\href {\doibase
  10.1016/j.cpc.2017.06.018} {\bibfield  {journal} {\bibinfo  {journal}
  {Computer Physics Communications}\ }\textbf {\bibinfo {volume} {220}},\
  \bibinfo {pages} {81} (\bibinfo {year} {2017})}\BibitemShut {NoStop}%
\bibitem [{\citenamefont {Petersen}\ and\ \citenamefont
  {Pedersen}(2012)}]{IMM2012-03274}%
  \BibitemOpen
  \bibfield  {author} {\bibinfo {author} {\bibfnamefont {K.~B.}\ \bibnamefont
  {Petersen}}\ and\ \bibinfo {author} {\bibfnamefont {M.~S.}\ \bibnamefont
  {Pedersen}},\ }\href {http://www2.imm.dtu.dk/pubdb/p.php?3274} {\enquote
  {\bibinfo {title} {{The Matrix Cookbook}},}\ } (\bibinfo {year} {2012}),\
  \bibinfo {note} {version 20121115}\BibitemShut {NoStop}%
\bibitem [{\citenamefont {Weidlich}(1971)}]{Weidlich71:325}%
  \BibitemOpen
  \bibfield  {author} {\bibinfo {author} {\bibfnamefont {W.}~\bibnamefont
  {Weidlich}},\ }\href {http://link.springer.com/article/10.1007%2FBF01395429}
  {\bibfield  {journal} {\bibinfo  {journal} {Zeitschrift f{\"u}r Physik}\
  }\textbf {\bibinfo {volume} {241}},\ \bibinfo {pages} {325} (\bibinfo {year}
  {1971})}\BibitemShut {NoStop}%
\bibitem [{\citenamefont {Nakano}\ \emph {et~al.}(2010)\citenamefont {Nakano},
  \citenamefont {Hatano},\ and\ \citenamefont {Petrosky}}]{Nakano2010}%
  \BibitemOpen
  \bibfield  {author} {\bibinfo {author} {\bibfnamefont {R.}~\bibnamefont
  {Nakano}}, \bibinfo {author} {\bibfnamefont {N.}~\bibnamefont {Hatano}}, \
  and\ \bibinfo {author} {\bibfnamefont {T.}~\bibnamefont {Petrosky}},\ }\href
  {\doibase 10.1007/s10773-010-0606-9} {\bibfield  {journal} {\bibinfo
  {journal} {International Journal of Theoretical Physics}\ }\textbf {\bibinfo
  {volume} {50}},\ \bibinfo {pages} {1134} (\bibinfo {year}
  {2010})}\BibitemShut {NoStop}%
\bibitem [{\citenamefont {Petrosky}(2010)}]{Petrosky01032010}%
  \BibitemOpen
  \bibfield  {author} {\bibinfo {author} {\bibfnamefont {T.}~\bibnamefont
  {Petrosky}},\ }\href {\doibase 10.1143/PTP.123.395} {\bibfield  {journal}
  {\bibinfo  {journal} {Progress of Theoretical Physics}\ }\textbf {\bibinfo
  {volume} {123}},\ \bibinfo {pages} {395} (\bibinfo {year}
  {2010})}\BibitemShut {NoStop}%
\bibitem [{\citenamefont {Swain}(1981)}]{0305-4470-14-10-013}%
  \BibitemOpen
  \bibfield  {author} {\bibinfo {author} {\bibfnamefont {S.}~\bibnamefont
  {Swain}},\ }\href {http://stacks.iop.org/0305-4470/14/i=10/a=013} {\bibfield
  {journal} {\bibinfo  {journal} {Journal of Physics A: Mathematical and
  General}\ }\textbf {\bibinfo {volume} {14}},\ \bibinfo {pages} {2577}
  (\bibinfo {year} {1981})}\BibitemShut {NoStop}%
\bibitem [{\citenamefont {Walls}\ and\ \citenamefont
  {Milburn}(2008)}]{Wallis-QO}%
  \BibitemOpen
  \bibfield  {author} {\bibinfo {author} {\bibfnamefont {D.}~\bibnamefont
  {Walls}}\ and\ \bibinfo {author} {\bibfnamefont {G.~J.}\ \bibnamefont
  {Milburn}},\ }\href {\doibase 10.1007/978-3-540-28574-8} {\emph {\bibinfo
  {title} {{Quantum optics}}}}\ (\bibinfo  {publisher} {Springer-Verlag Berlin
  Heidelberg},\ \bibinfo {year} {2008})\BibitemShut {NoStop}%
\bibitem [{\citenamefont {Goan}\ \emph {et~al.}(2011)\citenamefont {Goan},
  \citenamefont {Chen},\ and\ \citenamefont {Jian}}]{doi:10.1063/1.3570581}%
  \BibitemOpen
  \bibfield  {author} {\bibinfo {author} {\bibfnamefont {H.-S.}\ \bibnamefont
  {Goan}}, \bibinfo {author} {\bibfnamefont {P.-W.}\ \bibnamefont {Chen}}, \
  and\ \bibinfo {author} {\bibfnamefont {C.-C.}\ \bibnamefont {Jian}},\ }\href
  {\doibase 10.1063/1.3570581} {\bibfield  {journal} {\bibinfo  {journal} {The
  J. of Chem. Phys.}\ }\textbf {\bibinfo {volume} {134}},\ \bibinfo {pages}
  {124112} (\bibinfo {year} {2011})}\BibitemShut {NoStop}%
\bibitem [{\citenamefont {Gudmundsson}\ \emph {et~al.}(2018)\citenamefont
  {Gudmundsson}, \citenamefont {Abdullah}, \citenamefont {Sitek}, \citenamefont
  {Goan}, \citenamefont {Tang},\ and\ \citenamefont
  {Manolescu}}]{GUDMUNDSSON20181672}%
  \BibitemOpen
  \bibfield  {author} {\bibinfo {author} {\bibfnamefont {V.}~\bibnamefont
  {Gudmundsson}}, \bibinfo {author} {\bibfnamefont {N.~R.}\ \bibnamefont
  {Abdullah}}, \bibinfo {author} {\bibfnamefont {A.}~\bibnamefont {Sitek}},
  \bibinfo {author} {\bibfnamefont {H.-S.}\ \bibnamefont {Goan}}, \bibinfo
  {author} {\bibfnamefont {C.-S.}\ \bibnamefont {Tang}}, \ and\ \bibinfo
  {author} {\bibfnamefont {A.}~\bibnamefont {Manolescu}},\ }\href {\doibase
  10.1016/j.physleta.2018.04.017} {\bibfield  {journal} {\bibinfo  {journal}
  {Physics Letters A}\ }\textbf {\bibinfo {volume} {382}},\ \bibinfo {pages}
  {1672} (\bibinfo {year} {2018})}\BibitemShut {NoStop}%
\bibitem [{\citenamefont {{Gudmundsson}}\ \emph {et~al.}(2016)\citenamefont
  {{Gudmundsson}}, \citenamefont {{Jonsson}}, \citenamefont {{Bernodusson}},
  \citenamefont {{Abdullah}}, \citenamefont {{Sitek}}, \citenamefont {{Goan}},
  \citenamefont {{Tang}},\ and\ \citenamefont
  {{Manolescu}}}]{Gudmundsson16:AdP_10}%
  \BibitemOpen
  \bibfield  {author} {\bibinfo {author} {\bibfnamefont {V.}~\bibnamefont
  {{Gudmundsson}}}, \bibinfo {author} {\bibfnamefont {T.~H.}\ \bibnamefont
  {{Jonsson}}}, \bibinfo {author} {\bibfnamefont {M.~L.}\ \bibnamefont
  {{Bernodusson}}}, \bibinfo {author} {\bibfnamefont {N.~R.}\ \bibnamefont
  {{Abdullah}}}, \bibinfo {author} {\bibfnamefont {A.}~\bibnamefont {{Sitek}}},
  \bibinfo {author} {\bibfnamefont {H.-S.}\ \bibnamefont {{Goan}}}, \bibinfo
  {author} {\bibfnamefont {C.-S.}\ \bibnamefont {{Tang}}}, \ and\ \bibinfo
  {author} {\bibfnamefont {A.}~\bibnamefont {{Manolescu}}},\ }\href@noop {}
  {\bibfield  {journal} {\bibinfo  {journal} {Ann. Phys.}\ }\textbf {\bibinfo
  {volume} {529}},\ \bibinfo {pages} {1600177} (\bibinfo {year}
  {2016})}\BibitemShut {NoStop}%
\bibitem [{\citenamefont {{Gudmundsson}}\ \emph {et~al.}(2018)\citenamefont
  {{Gudmundsson}}, \citenamefont {{Abdullah}}, \citenamefont {{Sitek}},
  \citenamefont {{Goan}}, \citenamefont {{Tang}},\ and\ \citenamefont
  {{Manolescu}}}]{2017arXiv170603483G}%
  \BibitemOpen
  \bibfield  {author} {\bibinfo {author} {\bibfnamefont {V.}~\bibnamefont
  {{Gudmundsson}}}, \bibinfo {author} {\bibfnamefont {N.~R.}\ \bibnamefont
  {{Abdullah}}}, \bibinfo {author} {\bibfnamefont {A.}~\bibnamefont {{Sitek}}},
  \bibinfo {author} {\bibfnamefont {H.-S.}\ \bibnamefont {{Goan}}}, \bibinfo
  {author} {\bibfnamefont {C.-S.}\ \bibnamefont {{Tang}}}, \ and\ \bibinfo
  {author} {\bibfnamefont {A.}~\bibnamefont {{Manolescu}}},\ }\href@noop {}
  {\bibfield  {journal} {\bibinfo  {journal} {Annalen der Physik}\ }\textbf
  {\bibinfo {volume} {530}},\ \bibinfo {pages} {1700334} (\bibinfo {year}
  {2018})}\BibitemShut {NoStop}%
\bibitem [{\citenamefont {{De Liberato}}\ \emph {et~al.}(2009)\citenamefont
  {{De Liberato}}, \citenamefont {Gerace}, \citenamefont {Carusotto},\ and\
  \citenamefont {Ciuti}}]{PhysRevA.80.053810}%
  \BibitemOpen
  \bibfield  {author} {\bibinfo {author} {\bibfnamefont {S.}~\bibnamefont {{De
  Liberato}}}, \bibinfo {author} {\bibfnamefont {D.}~\bibnamefont {Gerace}},
  \bibinfo {author} {\bibfnamefont {I.}~\bibnamefont {Carusotto}}, \ and\
  \bibinfo {author} {\bibfnamefont {C.}~\bibnamefont {Ciuti}},\ }\href
  {\doibase 10.1103/PhysRevA.80.053810} {\bibfield  {journal} {\bibinfo
  {journal} {Phys. Rev. A}\ }\textbf {\bibinfo {volume} {80}},\ \bibinfo
  {pages} {053810} (\bibinfo {year} {2009})}\BibitemShut {NoStop}%
\bibitem [{\citenamefont {{Gudmundsson}}\ \emph
  {et~al.}(2019{\natexlab{a}})\citenamefont {{Gudmundsson}}, \citenamefont
  {{Gestsson}}, \citenamefont {{Abdullah}}, \citenamefont {{Tang}},
  \citenamefont {{Manolescu}},\ and\ \citenamefont
  {{Moldoveanu}}}]{2018arXiv180906930G}%
  \BibitemOpen
  \bibfield  {author} {\bibinfo {author} {\bibfnamefont {V.}~\bibnamefont
  {{Gudmundsson}}}, \bibinfo {author} {\bibfnamefont {H.}~\bibnamefont
  {{Gestsson}}}, \bibinfo {author} {\bibfnamefont {N.~R.}\ \bibnamefont
  {{Abdullah}}}, \bibinfo {author} {\bibfnamefont {C.-S.}\ \bibnamefont
  {{Tang}}}, \bibinfo {author} {\bibfnamefont {A.}~\bibnamefont {{Manolescu}}},
  \ and\ \bibinfo {author} {\bibfnamefont {V.}~\bibnamefont {{Moldoveanu}}},\
  }\href {\doibase 10.3762/bjnano.10.61} {\bibfield  {journal} {\bibinfo
  {journal} {Beilstein J. Nanotechnol.}\ }\textbf {\bibinfo {volume} {10}},\
  \bibinfo {pages} {606–616} (\bibinfo {year}
  {2019}{\natexlab{a}})}\BibitemShut {NoStop}%
\bibitem [{\citenamefont {{Abdullah}}\ \emph
  {et~al.}(2019{\natexlab{a}})\citenamefont {{Abdullah}}, \citenamefont
  {{Tang}}, \citenamefont {{Manolescu}},\ and\ \citenamefont
  {{Gudmundsson}}}]{2019arXiv190303655A}%
  \BibitemOpen
  \bibfield  {author} {\bibinfo {author} {\bibfnamefont {N.~R.}\ \bibnamefont
  {{Abdullah}}}, \bibinfo {author} {\bibfnamefont {C.-S.}\ \bibnamefont
  {{Tang}}}, \bibinfo {author} {\bibfnamefont {A.}~\bibnamefont {{Manolescu}}},
  \ and\ \bibinfo {author} {\bibfnamefont {V.}~\bibnamefont {{Gudmundsson}}},\
  }\href@noop {} {\bibfield  {journal} {\bibinfo  {journal} {arXiv:1903.03655}\
  } (\bibinfo {year} {2019}{\natexlab{a}})}\BibitemShut {NoStop}%
\bibitem [{\citenamefont {{Abdullah}}\ \emph
  {et~al.}(2019{\natexlab{b}})\citenamefont {{Abdullah}}, \citenamefont
  {{Tang}}, \citenamefont {{Manolescu}},\ and\ \citenamefont
  {{Gudmundsson}}}]{2019arXiv190404888A}%
  \BibitemOpen
  \bibfield  {author} {\bibinfo {author} {\bibfnamefont {N.~R.}\ \bibnamefont
  {{Abdullah}}}, \bibinfo {author} {\bibfnamefont {C.-S.}\ \bibnamefont
  {{Tang}}}, \bibinfo {author} {\bibfnamefont {A.}~\bibnamefont {{Manolescu}}},
  \ and\ \bibinfo {author} {\bibfnamefont {V.}~\bibnamefont {{Gudmundsson}}},\
  }\href@noop {} {\bibfield  {journal} {\bibinfo  {journal} {arXiv:1904.04888}\
  } (\bibinfo {year} {2019}{\natexlab{b}})}\BibitemShut {NoStop}%
\bibitem [{\citenamefont {Purcell}(1946)}]{PhysRev.69.681}%
  \BibitemOpen
  \bibfield  {author} {\bibinfo {author} {\bibfnamefont {E.~M.}\ \bibnamefont
  {Purcell}},\ }\href {\doibase 10.1103/PhysRev.69.674.2} {\bibfield  {journal}
  {\bibinfo  {journal} {Phys. Rev.}\ }\textbf {\bibinfo {volume} {69}},\
  \bibinfo {pages} {681} (\bibinfo {year} {1946})}\BibitemShut {NoStop}%
\bibitem [{\citenamefont {{Gudmundsson}}\ \emph
  {et~al.}(2019{\natexlab{b}})\citenamefont {{Gudmundsson}}, \citenamefont
  {{Abdullah}}, \citenamefont {{Tang}}, \citenamefont {{Manolescu}},\ and\
  \citenamefont {{Moldoveanu}}}]{2019arXiv190510883G}%
  \BibitemOpen
  \bibfield  {author} {\bibinfo {author} {\bibfnamefont {V.}~\bibnamefont
  {{Gudmundsson}}}, \bibinfo {author} {\bibfnamefont {N.~R.}\ \bibnamefont
  {{Abdullah}}}, \bibinfo {author} {\bibfnamefont {C.-S.}\ \bibnamefont
  {{Tang}}}, \bibinfo {author} {\bibfnamefont {A.}~\bibnamefont {{Manolescu}}},
  \ and\ \bibinfo {author} {\bibfnamefont {V.}~\bibnamefont {{Moldoveanu}}},\
  }\href@noop {} {\bibfield  {journal} {\bibinfo  {journal} {arXiv:1905.10883}\
  } (\bibinfo {year} {2019}{\natexlab{b}})}\BibitemShut {NoStop}%
\bibitem [{\citenamefont {Abdullah}\ \emph {et~al.}(2019)\citenamefont
  {Abdullah}, \citenamefont {Tang}, \citenamefont {Manolescu},\ and\
  \citenamefont {Gudmundsson}}]{2018arXiv181205665A}%
  \BibitemOpen
  \bibfield  {author} {\bibinfo {author} {\bibfnamefont {N.~R.}\ \bibnamefont
  {Abdullah}}, \bibinfo {author} {\bibfnamefont {C.-S.}\ \bibnamefont {Tang}},
  \bibinfo {author} {\bibfnamefont {A.}~\bibnamefont {Manolescu}}, \ and\
  \bibinfo {author} {\bibfnamefont {V.}~\bibnamefont {Gudmundsson}},\
  }\href@noop {} {\bibfield  {journal} {\bibinfo  {journal} {Nanomaterials}\
  }\textbf {\bibinfo {volume} {9}},\ \bibinfo {pages} {741} (\bibinfo {year}
  {2019})}\BibitemShut {NoStop}%
\bibitem [{\citenamefont {Stefanucci}\ and\ \citenamefont {van
  Leeuwen}(2013)}]{StefanucciBook:2013}%
  \BibitemOpen
  \bibfield  {author} {\bibinfo {author} {\bibfnamefont {G.}~\bibnamefont
  {Stefanucci}}\ and\ \bibinfo {author} {\bibfnamefont {R.}~\bibnamefont {van
  Leeuwen}},\ }\href@noop {} {\emph {\bibinfo {title} {{Nonequilibrium
  Many-Body Theory of Quantum Systems: A Modern Introduction}}}}\ (\bibinfo
  {publisher} {Cambridge University Press},\ \bibinfo {year}
  {2013})\BibitemShut {NoStop}%
\bibitem [{\citenamefont {Deng}\ \emph {et~al.}(2015)\citenamefont {Deng},
  \citenamefont {Orgiazzi}, \citenamefont {Shen}, \citenamefont {Ashhab},\ and\
  \citenamefont {Lupascu}}]{PhysRevLett.115.133601}%
  \BibitemOpen
  \bibfield  {author} {\bibinfo {author} {\bibfnamefont {C.}~\bibnamefont
  {Deng}}, \bibinfo {author} {\bibfnamefont {J.-L.}\ \bibnamefont {Orgiazzi}},
  \bibinfo {author} {\bibfnamefont {F.}~\bibnamefont {Shen}}, \bibinfo {author}
  {\bibfnamefont {S.}~\bibnamefont {Ashhab}}, \ and\ \bibinfo {author}
  {\bibfnamefont {A.}~\bibnamefont {Lupascu}},\ }\href {\doibase
  10.1103/PhysRevLett.115.133601} {\bibfield  {journal} {\bibinfo  {journal}
  {Phys. Rev. Lett.}\ }\textbf {\bibinfo {volume} {115}},\ \bibinfo {pages}
  {133601} (\bibinfo {year} {2015})}\BibitemShut {NoStop}%
\bibitem [{\citenamefont {Koski}\ \emph {et~al.}(2018)\citenamefont {Koski},
  \citenamefont {Landig}, \citenamefont {P\'alyi}, \citenamefont {Scarlino},
  \citenamefont {Reichl}, \citenamefont {Wegscheider}, \citenamefont {Burkard},
  \citenamefont {Wallraff}, \citenamefont {Ensslin},\ and\ \citenamefont
  {Ihn}}]{PhysRevLett.121.043603}%
  \BibitemOpen
  \bibfield  {author} {\bibinfo {author} {\bibfnamefont {J.~V.}\ \bibnamefont
  {Koski}}, \bibinfo {author} {\bibfnamefont {A.~J.}\ \bibnamefont {Landig}},
  \bibinfo {author} {\bibfnamefont {A.}~\bibnamefont {P\'alyi}}, \bibinfo
  {author} {\bibfnamefont {P.}~\bibnamefont {Scarlino}}, \bibinfo {author}
  {\bibfnamefont {C.}~\bibnamefont {Reichl}}, \bibinfo {author} {\bibfnamefont
  {W.}~\bibnamefont {Wegscheider}}, \bibinfo {author} {\bibfnamefont
  {G.}~\bibnamefont {Burkard}}, \bibinfo {author} {\bibfnamefont
  {A.}~\bibnamefont {Wallraff}}, \bibinfo {author} {\bibfnamefont
  {K.}~\bibnamefont {Ensslin}}, \ and\ \bibinfo {author} {\bibfnamefont
  {T.}~\bibnamefont {Ihn}},\ }\href {\doibase 10.1103/PhysRevLett.121.043603}
  {\bibfield  {journal} {\bibinfo  {journal} {Phys. Rev. Lett.}\ }\textbf
  {\bibinfo {volume} {121}},\ \bibinfo {pages} {043603} (\bibinfo {year}
  {2018})}\BibitemShut {NoStop}%
\bibitem [{\citenamefont {Pagel}\ and\ \citenamefont
  {Fehske}(2017)}]{PhysRevA.96.041802}%
  \BibitemOpen
  \bibfield  {author} {\bibinfo {author} {\bibfnamefont {D.}~\bibnamefont
  {Pagel}}\ and\ \bibinfo {author} {\bibfnamefont {H.}~\bibnamefont {Fehske}},\
  }\href {\doibase 10.1103/PhysRevA.96.041802} {\bibfield  {journal} {\bibinfo
  {journal} {Phys. Rev. A}\ }\textbf {\bibinfo {volume} {96}},\ \bibinfo
  {pages} {041802} (\bibinfo {year} {2017})}\BibitemShut {NoStop}%
\bibitem [{\citenamefont {Szczygielski}\ and\ \citenamefont
  {Alicki}(2015)}]{PhysRevA.92.022349}%
  \BibitemOpen
  \bibfield  {author} {\bibinfo {author} {\bibfnamefont {K.}~\bibnamefont
  {Szczygielski}}\ and\ \bibinfo {author} {\bibfnamefont {R.}~\bibnamefont
  {Alicki}},\ }\href {\doibase 10.1103/PhysRevA.92.022349} {\bibfield
  {journal} {\bibinfo  {journal} {Phys. Rev. A}\ }\textbf {\bibinfo {volume}
  {92}},\ \bibinfo {pages} {022349} (\bibinfo {year} {2015})}\BibitemShut
  {NoStop}%
\end{thebibliography}%

\end{document}